%% file: IIB_MM_4D.tex
\documentclass[12pt]{article}

\usepackage{amsmath}
\usepackage[usenames]{color}
\usepackage{graphics}
\usepackage{hangcaption}

\allowdisplaybreaks

\begin{document}

\baselineskip=18.6pt plus 0.2pt minus 0.1pt

\makeatletter
\@addtoreset{equation}{section}
\renewcommand{\theequation}{\thesection.\arabic{equation}}


\newcommand{\der}[1]{\partial_{#1}}
\newcommand{\dder}[2]{\frac{\partial{#1}}{\partial{#2}}}
\newcommand{\inv}[1]{\frac{1}{#1}}
\newcommand{\Rint}[1]{\int \hspace{-0.2em} {\rm d}{#1}\;}
\newcommand{\Tr}{{\rm Tr}}
\newcommand{\tr}{{\rm tr}}

\newcommand{\ph}[1]{\phantom{#1}}
\newcommand{\nn}{\nonumber}

\newcommand{\pref}[1]{(\ref{#1})}
\newcommand{\vbar}[1][]{\Bigr|_{#1}}
\newcommand{\tm}{\mbox{$\times$}}
\newcommand{\gf}[1]{\langle \; #1 \; \rangle}
\newcommand{\uslash}{\slash\hspace{-0.55em}u}
\newcommand{\graph}[2]{\scalebox{#1}{\includegraphics{#2}}}
\makeatother

\begin{titlepage}
\title{
\hfill\parbox{4cm}
{\normalsize KUNS-1779\\{\tt hep-th/0204240}}\\
\vspace{1cm}
Mean Field Approximation of IIB Matrix Model\\[5pt]
and Emergence of Four Dimensional Space-Time
}

\author{
H.~{\sc Kawai}$^a$\thanks{{\tt hkawai@gauge.scphys.kyoto-u.ac.jp}}
,\hspace{5pt}
S.~{\sc Kawamoto}$^a$\thanks{{\tt kawamoto@gauge.scphys.kyoto-u.ac.jp}}
,\hspace{5pt}
T.~{\sc Kuroki}$^a$\thanks{{\tt kuroki@gauge.scphys.kyoto-u.ac.jp}} ,\\
T.~{\sc Matsuo}$^b$\thanks{{\tt tmatsuo@yukawa.kyoto-u.ac.jp}}
{}\hspace*{2pt} and \hspace*{0pt}
S.~{\sc Shinohara}$^a$\thanks{
{\tt shunichi@gauge.scphys.kyoto-u.ac.jp}}
\\[7pt]
$^a${\it Department of Physics, Kyoto University, Kyoto 606-8502, Japan} \\
$^b${\it Yukawa Institute for Theoretical Physics, Kyoto University, }\\
{\it Kyoto 606-8502, Japan}
}

\date{\normalsize April, 2002}
\maketitle
\thispagestyle{empty}
\begin{abstract}
\normalsize\noindent
For the purpose of analyzing non-perturbative dynamics of
string theory,
Nishimura and Sugino have applied
an improved mean field approximation (IMFA) to IIB matrix model.
We have extracted the essence of the IMFA
and obtained a general scheme, the improved Taylor expansion,
that can be applied to a wide class of series which is not necessarily
convergent.
This approximation scheme with the help of the 2PI free energy
enables us to perform higher order calculations.
We have shown that the value of the free energy is stable at higher orders,
which supports the validity of the approximation.
Moreover, the ratio between the extent of ``our'' space-time
and that of the internal space is found to increase rapidly
as we take the higher orders into account. Our results
suggest that the four dimensional space-time emerges
spontaneously in IIB matrix model.
\end{abstract}

\end{titlepage}

\section{Introduction and summary}
\label{sec:intro_summry}

\vspace*{0pt}
After the enormous success in revealing perturbative dynamics
of string theory with/without D-branes
\cite{Polchinski:1995mt,Polchinski:string_theory,Maldacena:1998re,
Aharony:2000ti},
our focus will go on to the full non-perturbative
dynamics, or to the theory which has predictive power for the real world.

In such a stage, both a model which is
to be the true theory and a method to extract non-perturbative
information from it are important.
Some models have already been
conjectured as the constructive definition
of string theory and/or quantum gravity
\cite{Banks:1997vh,Ishibashi:1997xs,Smolin:2000kc,Azuma:2001re,
Smolin:2001wc}.
Among others we are interested in IIB matrix
model \cite{Ishibashi:1997xs,Aoki:1999bq}
which would be the
most promising to attack the non-perturbative dynamics
of string theory.

As to the method, J.~Nishimura and F.~Sugino \cite{Nishimura:2001sx}
have applied an improved mean field
approximation (IMFA) \cite{Kabat:2000hp}
to IIB matrix model
with an excellent idea for breaking the Lorentz symmetry,
and suggested that the four dimensional space-time
is more stable than those of the other dimensions.
In sec.~\ref{sec:MFA},
we review the IMFA with an application to a toy model.
Although the IMFA is somewhat mysterious, we find that
the essential feature of the IMFA can be captured in terms of the
improved Taylor expansion (ITE) which is elucidated in sec.\ref{sec:ITE}.
In \cite{Nishimura:2001sx} the authors have calculated to the third order
and got a result which is both remarkable
and uneasy:
\begin{description}
\item[remarkable fact]  Even at the third order,
  non-perturbative aspects
  of IIB matrix model can be captured definitely.
\item[uneasy fact] Not knowing the stability of the result,
  we can not strongly believe it.
\end{description}
To confirm the validity of the IMFA for IIB matrix model,
we should investigate
the stability of the result, especially the free energy, as we increase
the order of approximation.
The aim of this paper is to deepen the remarkable aspect
and to remove the uneasy problem at a time.
So we proceed to the next order calculation.
Since Feynman graphs to be calculated
increase in number and become complicated,
we need to introduce a new method.
It turns out nice to use the two-particle irreducible (2PI)
free energy which is explained in sec.~\ref{sec:2PI}.
With the help of the 2PI free energy we have completed the fifth order
calculation and see how the above aim is achieved.
We find that the contribution of the fifth order to the free energy
is smaller than that of
the third order, which supports the validity of this approximation.

The ratio between the extent of the four dimensional space-time and
that of the extra internal space is found to
increase from three (the third order) to six (the fifth order)
as is shown in sec.~\ref{sec:result}.
This indicates that the ratio is infinite in the full theory and
the flat four dimensional space-time is generated.

\section{Improved mean field approximation}
\label{sec:MFA}

\subsection{general prescription}

Suppose we have some action function $S(x)$ and its
partition function
\begin{equation}
  Z = \Rint{x} e^{-S(x)}.
\end{equation}
In general, this integral can not be performed analytically,
and we need an approximation scheme to evaluate
the partition function.
In the mean field approximation, we add the mean field action $S_0(x;a)$
which can be integrated analytically,
and subtract it as an interaction term.
The mean field action contains a set of  parameters $a$ which
should be determined later according to the order of the approximation.
The original action $S(x)$ is also treated as an interaction term.
Here we introduce a coupling $g$ formally, which will be set to $1$ at
the end of the calculation.
To get the $k$-th order approximation,
we convert the expression to a formal power series with respect to $g$
and truncate it up to order $k$.
The prescription is as follows:
\begin{align}
  Z & = \Rint{x} e^{-S_0(x;a)
               - g \bigl(S(x)-S_0(x;a)\bigr)}\\
    & \Downarrow {\rm convert\ to\ the\ formal\ power\
        series\ w.r.t.\ }g\nn \\
  Z  & \sim \sum_{n=0}^{\infty} \frac{(-g)^n}{n!}
           \Rint{x} \bigl(S(x)-S_0(x;a)\bigr)^n
            e^{-S_0(x;a)}
\label{eq:fps} \\
    & \Downarrow {\rm truncate\ the\ summation}\nn \\
  Z_{k}(a)  & = \sum_{n=0}^{k} \frac{(-g)^n}{n!}
           \Rint{x} \bigl(S(x)-S_0(x;a)\bigr)^n
            e^{-S_0(x;a)}.
\end{align}

The truncated partition function $Z_{k}(a)$ is a function of the
parameters $a$, though the original partition function is independent
of them. Thus we can determine the values of the parameters so that
the truncated partition function is stationary with respect to $a$.
We call this procedure the improved mean field approximation (IMFA).

To approximate the correlation function, we use the same truncation:
\begin{equation}
  \gf{f(x)}_k =  \inv{Z_k(a)} \sum_{n=0}^{k} \frac{(-g)^n}{n!}
           \Rint{x} f(x) \bigl(S(x)-S_0(x;a)\bigr)^n
           e^{-S_0(x;a)}
           \Biggr|_{{\rm neglect\ }O(g^{k+1})} .
\end{equation}

Note that at the first order IMFA is the same
as the variational principle.
Note also that the power series (\ref{eq:fps}) is in general only an
asymptotic series if we fix the parameter $a$. Nevertheless by
determining $a$ order by order, we can approximate the original
partition function with high accuracy as we will see in the next
subsection.

\subsection{application to a toy model}

As an example, let us apply the IMFA to the zero dimensional $\phi^4$
model whose partition function can be integrated exactly.
The action is
\begin{equation}
  \label{action_phi4}
  S = \frac{m^2}{2}\phi^2+\frac{\lambda}{4!}\phi^4.
\end{equation}
The exact values of the partition function and
the two point function in the massless case ($m^2=0, \lambda=1$)
in which we are especially
interested are
\begin{align}
  Z_{\rm massless} & = \Rint{\phi} \; e^{-S}
       = 2\cdot 3! \cdot (4!)^{-3/4} \cdot \Gamma(1/4), \\
  F_{\rm massless} & = -\log Z \simeq -1.389388802, \\
  \gf{\phi^2}_{\rm massless} & = \inv{Z} \Rint{\phi} \; \phi^2 e^{-S}
       = (4!)^{1/2} \frac{\Gamma(3/4)}{\Gamma(1/4)}
       \simeq 1.655801764.
\end{align}
We adopt the following action as $S_0$,
\begin{equation}
  \label{S_0_phi4}
  S_0 = \frac{m_0^2}{2}\phi^2,
\end{equation}
where $m_0^2$ is a variational parameter which corresponds to $a$
in the previous subsection.
Even in the massless case, this action enables us to calculate
the free energy $F_{k}=-\log Z_{k}$ by using the ordinary
Feynman diagrams.

Graphically, $F_{k}$ is the sum of all connected vacuum graphs shown
as in eq.~(\ref{F_phi4}),
where a line and a cross with a circle represent a
propagator $1/m_0^2$ and a mass counter term $g(m_0^2-m^2)$ respectively.
The four-point vertex denotes $g\lambda$.

\begin{equation}
  \label{F_phi4}
  \begin{array}[c]{cccccccc}
F = &  \raisebox{- 9pt}{\scalebox{0.7}{\includegraphics{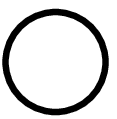}}} & + &
       \underbrace{
       \raisebox{-20pt}{\scalebox{0.7}{\includegraphics{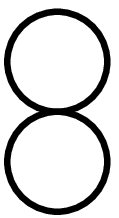}}} +
       \raisebox{-13pt}{\scalebox{0.7}{\includegraphics{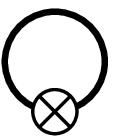}}} } &
       + &
       \underbrace{
       \raisebox{-25pt}{\scalebox{0.7}{\includegraphics{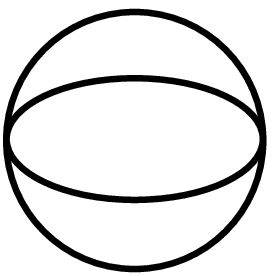}}} +
       \raisebox{-30pt}{\scalebox{0.7}{\includegraphics{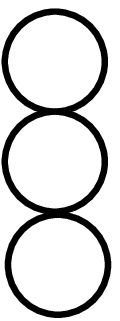}}} +
       \raisebox{-20pt}{\scalebox{0.7}{\includegraphics{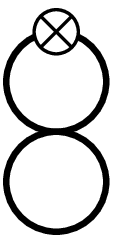}}} +
       \raisebox{-14pt}{\scalebox{0.7}{\includegraphics{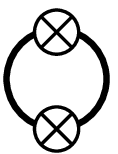}}} } &
       + & \cdots \\
   & O(g^0) &  & O(g^1) & & O(g^2) & & O(g^3)
\end{array}
\end{equation}

\begin{figure}[htbp]
  \begin{center}
    \leavevmode
    \scalebox{1}{\includegraphics{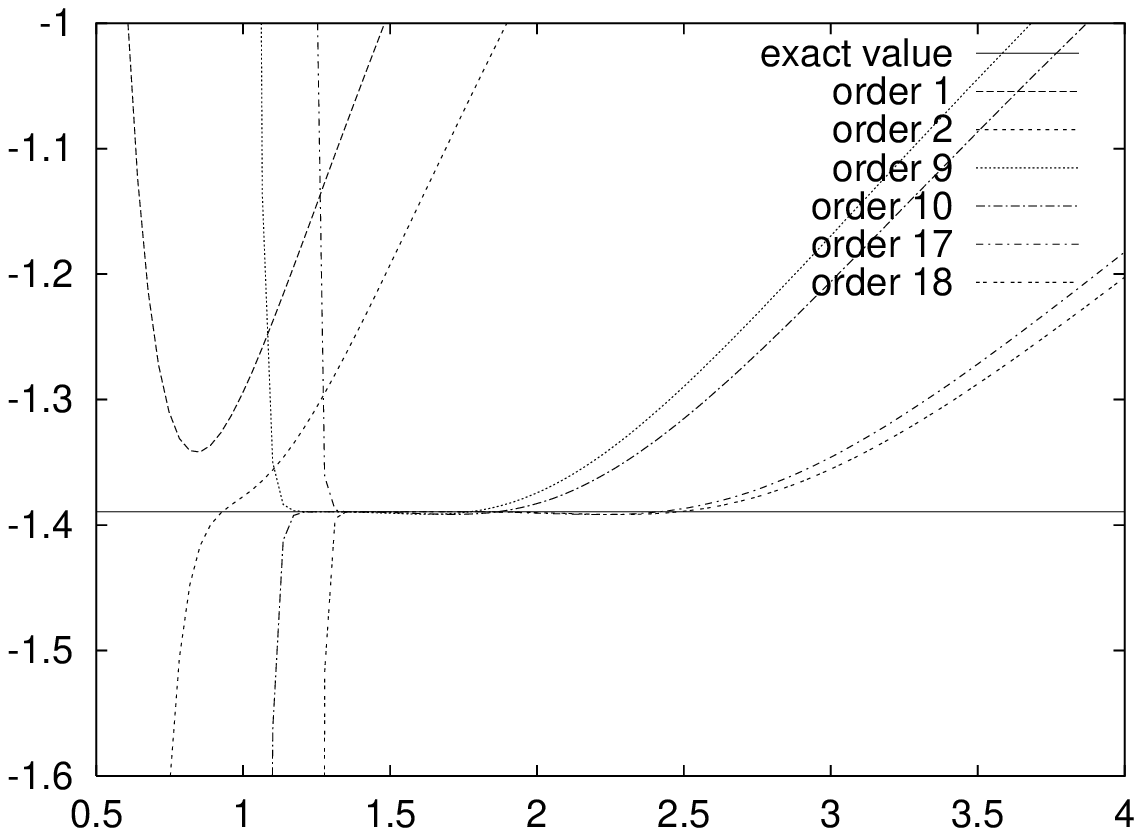}}
    \captionwidth = 30em
    \hangcaption{Truncated free energies in the IMFA scheme.
       The horizontal axis denotes $m_0^2$.}
    \label{fig:f_phi4_NS}
  \end{center}
\end{figure}

\begin{figure}[htbp]
  \begin{center}
    \leavevmode
    \scalebox{1}{\includegraphics{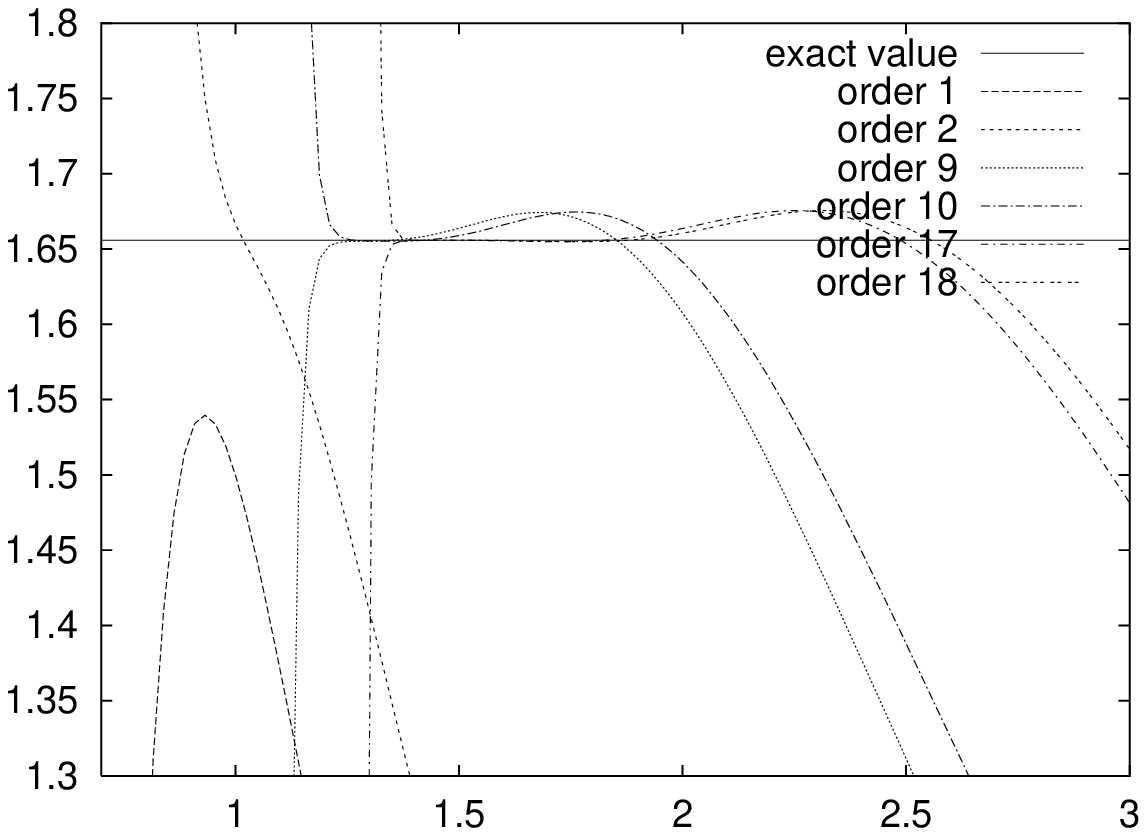}}
    \captionwidth = 30em
    \hangcaption{Truncated two point functions in the IMFA scheme.
       The horizontal axis is $m_0^2$.}
    \label{fig:2pt_phi4}
  \end{center}
\end{figure}

Figs.\ref{fig:f_phi4_NS} and \ref{fig:2pt_phi4} show
the approximated free energies and two point functions
as functions of $m_0^2$ in the massless case.
We find that the higher the order of the approximation,
the wider the plateau at the exact values in both cases.
This phenomenon is expected because the free energy $F$ should not
depend on the parameter $m_0^2$ which is introduced artificially.
Although the truncated free energy $F_k$ depends on $m_0^2$,
we expect $m_0^2$ dependence becomes weaker for larger values of $k$.
Since the truncated free energies and two point functions are merely
polynomials in $1/m_0^2$,
the existence of the plateau indicates that extremum
points of these polynomials concentrate on limited regions.
Once the plateau exists, small changes of $m_0^2$ do not affect the
resultant values.
Therefore if we have a plateau, we can regard the value in the region
as a good approximation of the exact value.


\section{Improved Taylor expansion}
\label{sec:ITE}

The IMFA is a rather curious approximation scheme.
It is interesting to understand the ground on which it works.
In this section, we see that the improved Taylor expansion (ITE)
explains some of the essential aspects of the IMFA.

Suppose we have a sequence of functions $\{ f_i(x) \}_{i=0, 1,
\ldots}$ which does not necessarily converge.
We define an improved sequence $\{ f_i^{\rm improved}(x;x_0) \}_{i=0,1,
\ldots}$ by 
\begin{align}
  \label{ITE}
  f^{\rm improved}_{i}(x;x_0) & = \sum_{n=0}^{i}
        \inv{n!} (x-x_0)^n f_{i-n}^{(n)}(x_0).
\end{align}
We call this prescription the improved Taylor expansion (ITE).
We emphasize that each $f_i^{\rm improved}$ is constructed from finite 
number of $f_j$'s, that is, $f_0, f_1, \ldots, f_i$.
This differs from the ordinary Taylor expansion in that
we use $f_j$ with smaller $j$'s for higher derivative terms.
As we see below, different values of $x_0$ make $\{ f_i^{\rm improved}
\}_{i=0,1,\ldots}$ converge in different regions.
This property enables us to obtain a series which converges 
around various values of $x$ by choosing proper $x_0$ correspondingly.

\begin{figure}[htbp]
  \begin{center}
    \leavevmode
    {\scalebox{1}{\includegraphics{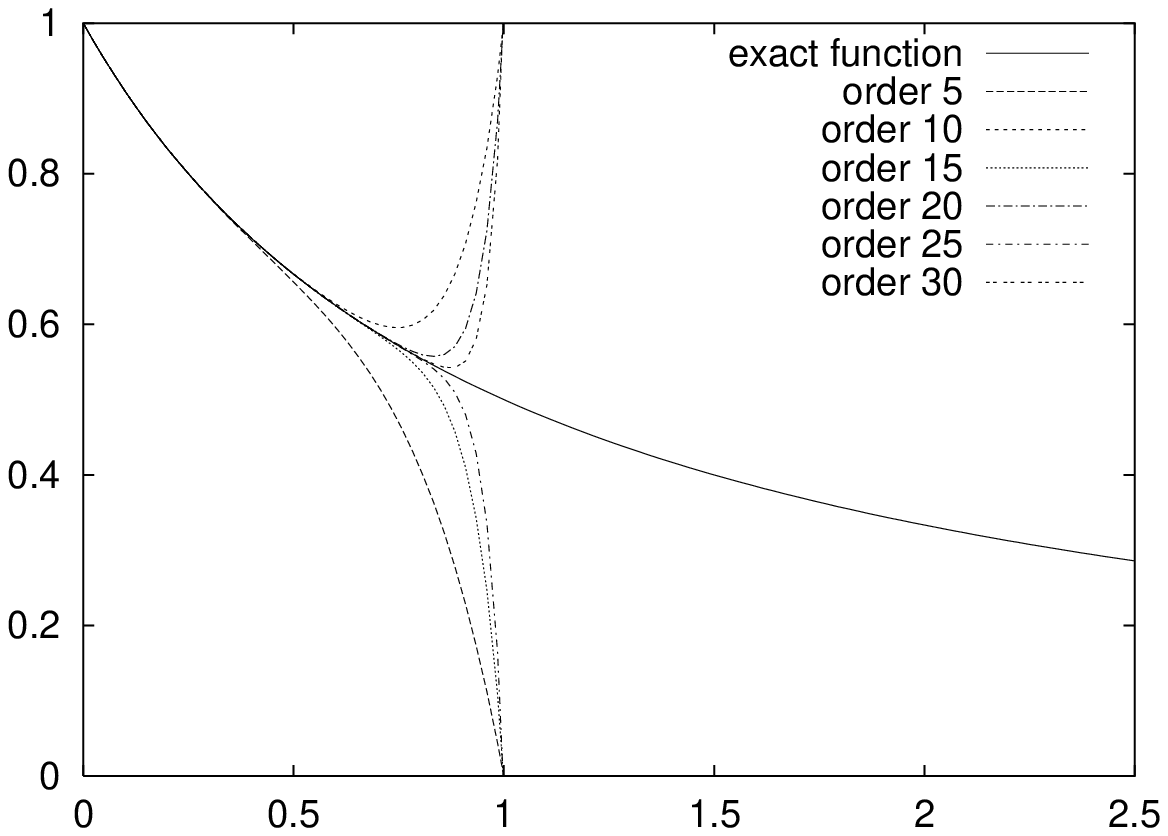}}}
    \hangcaption{Original series. The horizontal axis is $x$.}
    \label{fig:ori_1_x}
  \end{center}
\end{figure}

\begin{figure}[htbp]
  \begin{center}
    \leavevmode
    {\scalebox{1}{\includegraphics{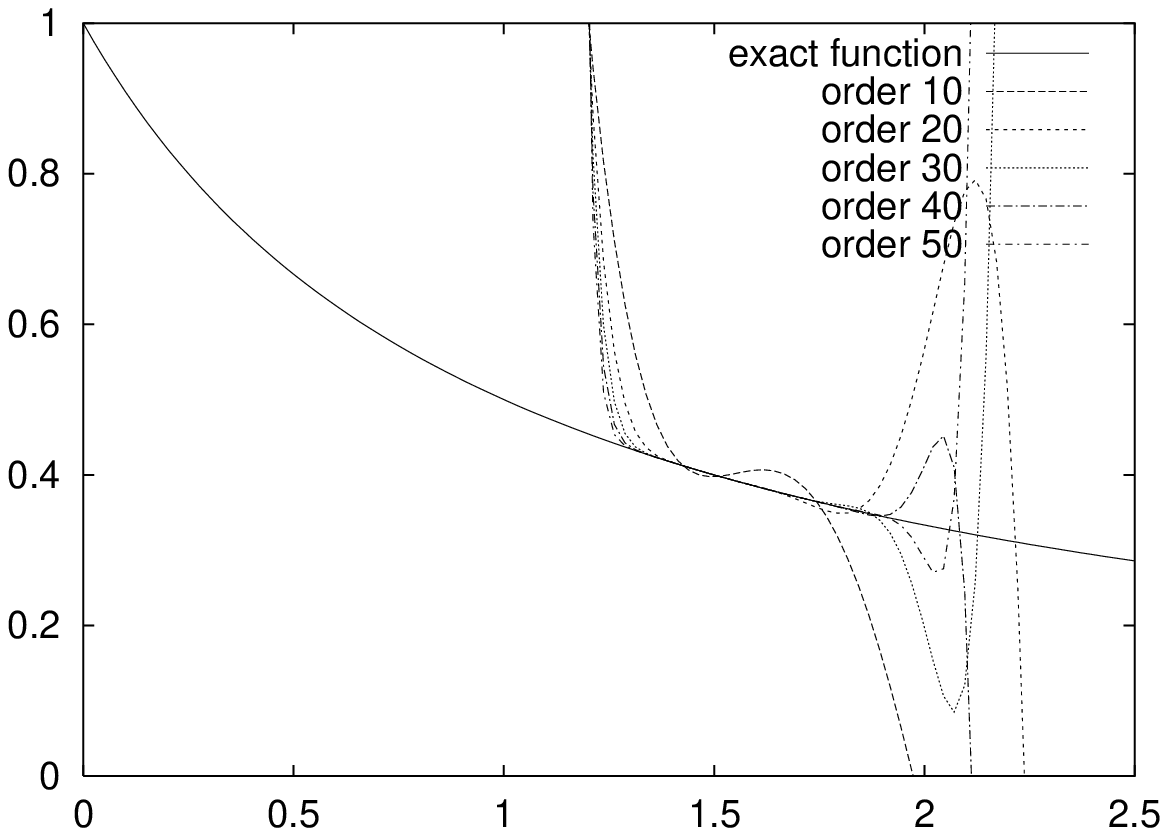}}}
    \hangcaption{ITE around $x_0=1.1$. The horizontal axis is $x$.}
    \label{fig:ITE}
  \end{center}
\end{figure}

For example, we consider a power series given by
\begin{equation}
  f(x) = \sum_{n=0}^{\infty} (-)^n x^n \ (=\frac{1}{1+x}),
  \label{original_expansion}
\end{equation}
and define $\{ f_i \}_{i=0,1,\ldots }$ by
\begin{equation}
    f_i  = \sum_{n=0}^{i} (-)^n x^n.
  \label{truncated_sum}
\end{equation}
The behavior of this sequence $\{ f_i \}_{i=0,1,\ldots }$ is shown in
Fig.\ref{fig:ori_1_x}, which
manifests that the convergence radius of the series is 1.
The improved sequence $\{f_i^{\rm improved}(x;x_0) \}_{i=0,1,\ldots }$
for $x_0=1.1$, for instance,
behaves as in Fig.\ref{fig:ITE}.
This sequence converges well around $x\sim 3/2$.
Thus the improved sequence has a totally different
convergence region from that of the original sequence.
As we can see from this example, the reason why we use $f_j$ with
smaller $j$'s in the definition of the ITE eq. (\ref{ITE}) is that
$f_j$ with the higher $j$'s make behavior of the sum worse for a
divergent series.

\begin{figure}[htbp]
  \begin{center}
    \leavevmode
    {\scalebox{1}{\includegraphics{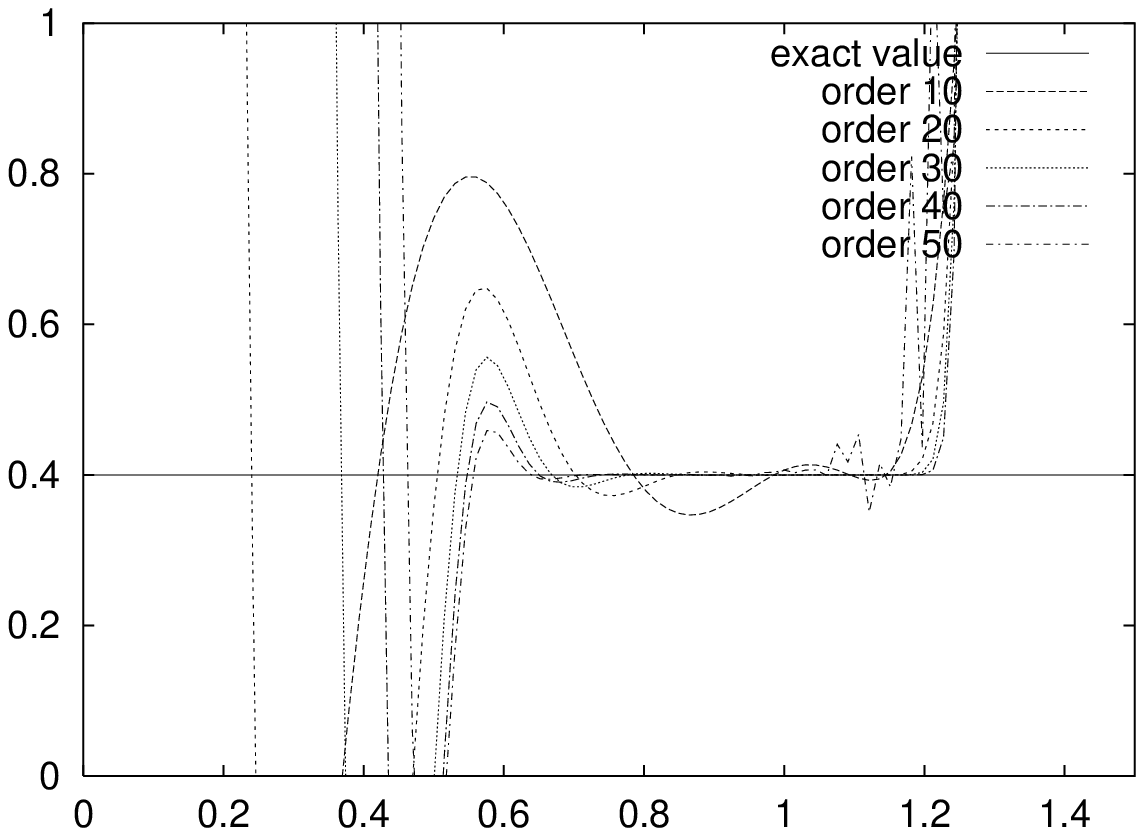}}}
    \hangcaption{The plateau in ITE scheme. The horizontal axis is $x_0$.}
    \label{fig:plateu_ITE}
  \end{center}
\end{figure}

To see the relation with the IMFA more clearly,
let us fix $x$ to 3/2, outside the
convergent radius ($x$ corresponds to $m^2$ in eq.~(\ref{action_phi4})).
Instead we regard $f^{\rm improved}_{i}(3/2;x_0)$ as a function of $x_0$.
Now $x_0$ corresponds to a parameter in the IMFA
($m_0^2$ in eq.~(\ref{S_0_phi4})).
As seen from Fig.\ref{fig:plateu_ITE}, there are plateaus on the
exact value $f(3/2)=0.4$. The plateau gets wider as we increase the
order of the approximation.
The approximation is nice and the improved
functions are almost independent of the variational parameter $x_0$
as expected.

The expansion (\ref{ITE}) can be written in a concise form.
First, we introduce a formal expansion parameter $g$, and define a function
from the sequence of functions $\{ f_i(x) \}_{i=0,1,\ldots}$,
\begin{equation}
  \label{f_g}
  f(x) = \sum_{i=0}^{\infty} g^i \, h_i(x),
\end{equation}
where $h_i(x) = f_i(x) - f_{i-1}(x) \ (i=1,2, \ldots)$ and $h_0(x) =f_0(x)$. 
For example, in the case of power series eq. (\ref{original_expansion})
we have $h_i(x) = (-)^i \, x^i$.
Next we substitute $x_0 + g(x-x_0)$ for $x$ in this $f(x)$ and
truncate it with respect to $g$:
\begin{equation}
  \label{f_g_truncated}
  f^{\rm improved}_{i}(x;x_0) = f(x_0 + g(x-x_0))\Big|_i.
\end{equation}
Here $|_i$ denotes neglecting $O(g^{i+1})$ terms and setting
$g$=$1$.
One can easily verify that this expression reproduces the functions in 
the improved sequence eq.~(\ref{ITE}).

In relation to the IMFA, the derivatives correspond to insertions of mass
counter terms.
As a result, the IMFA is included in a more general scheme, the ITE.
The abstract feature of the ITE
may be useful to find the essence of the IMFA.
Once we understand the essence, we could refine the IMFA in various ways
and use a refined method to grasp the non-perturbative properties
of IIB matrix model further.

For reference, some curious behavior of the ITE and an application
to an asymptotic series are examined
in appendices \ref{sec:curios_ITE} and
\ref{sec:ITE_asy}.

\section{2PI free energy and Legendre transformation}
\label{sec:2PI}

In the application of the IMFA to various models, we can drastically
simplify the computation by using the 2PI free energy.
In this paper, we apply the Legendre transformation to the 2PI free
energy to obtain the ordinary free energy, and search its extrema.

Although within this procedure the 2PI free energy seems just a technical
tool, we can give it a more fundamental meaning related to the
Schwinger-Dyson equations (SDE).
In fact, we can apply the ITE scheme directly to the 2PI free energy,
which we explain in appendix \ref{sec:ITE_2PI}.

\subsection{Schwinger-Dyson equations and 2PI free energy}

The SDE are the consistency conditions among
the Green functions and in some cases we can extract non-perturbative
information from them.
Again we consider the massless $\phi^4$ model.
For the full propagator $c = \gf{\phi^2}$, we can derive
the SDE as shown in eq.~(\ref{SDE}).
\begin{equation}
  \label{SDE}
  0 = \frac{1}{\scalebox{0.7}{\includegraphics{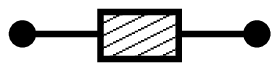}}}
      + \scalebox{0.7}{\includegraphics{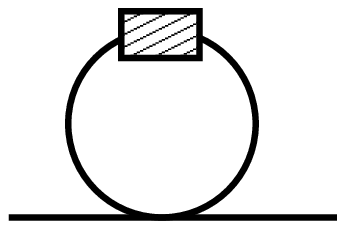}}
      + \raisebox{-24pt}{\scalebox{0.7}{\includegraphics{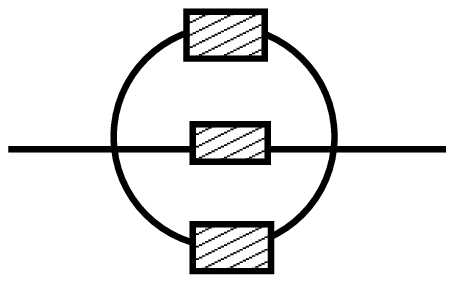}}}
      + \cdots.
\end{equation}
The derivation of this equation is given
in appendix \ref{sec:der_SDE}. Note that each graph has no self-energy
part when we regard each full propagator as a single line.

This equation is, in fact, the stationary condition of the two
particle irreducible (2PI) free energy which we call $G$
hereafter.
\begin{equation}
  \label{G}
  G = \ \raisebox{-17pt}{\scalebox{0.7}{\includegraphics{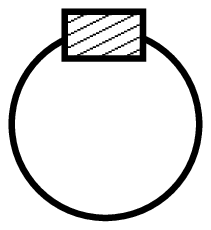}}}
      \ +\  \raisebox{-40pt}{\scalebox{0.7}{\includegraphics{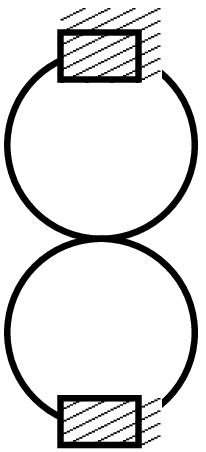}}}
      \ +\  \raisebox{-43pt}{\scalebox{0.7}{\includegraphics{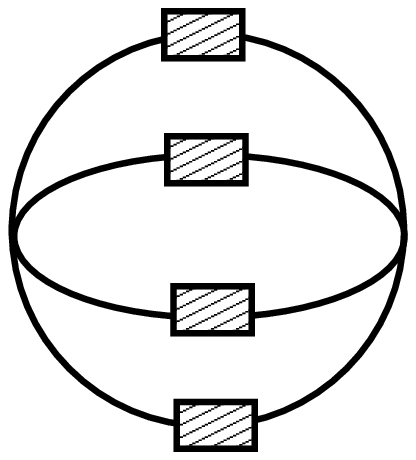}}}
      \ +\  \cdots.
\end{equation}
2PI graphs are the ones that remain connected
after cutting any two distinct propagators.
In other words, 2PI graphs do not contain any self-energy part
as a subgraph. 
This property comes from the fact that
the SDE do not contain any self-energy part.
The correspondence between the SDE for the full propagators
and the 2PI free energy holds generally in field theory.

The advantage of considering the 2PI free energy is
that we can greatly reduce the number of graphs.
Next, we will explain the relation between the 2PI free energy
and the ordinary free energy.

\subsection{Legendre transformation}

Let us start with the free energy,
\begin{align}
  Z(J) & = \Rint{\phi} e^{-S(\phi)-J\phi^2},\\
  F(J) & = -\log Z(J).
\end{align}
Here we introduce a source $J$ for $\phi^2$.
We apply the Legendre transformation to $F(J)$ with respect to $J$.
\begin{align}
  K(J) & = \dder{F(J)}{J},\\
  G(K) & = \left[ F(J) - J \dder{F(J)}{J} \right]_{J=J(K)},
\end{align}
where $K$ is the conjugate variable of $J$.
To investigate the original theory, we set $J=0$.
Then $K(J=0)$ is equal to the full two-point function $c$,
\begin{equation}
  K(J=0) = \gf{\phi^2}_{J=0} = c,
\end{equation}
and $G(K=c)$ is nothing but the sum of the 2PI vacuum graphs \cite{Fukuda:1995}.

For example, we use this procedure for the $\phi^4$ model,
eq.~(\ref{action_phi4}).
The free energy $F$ is given by the sum of the connected vacuum graphs,
\begin{equation}
  \label{F}
  F = \inv{2}\log (m^2+2J) + \frac{\lambda}{8}\inv{(m^2+2J)^2}
      -\lambda^2\left(\inv{48}+\inv{16}\right)\inv{(m^2+2J)^4}+ O(\lambda^3),
\end{equation}
where each term corresponds to the one in eq. (\ref{F_phi4}) other
than the graphs with mass counter terms.
The full propagator becomes as follows:
\begin{equation}
  \label{c_phi4}
  c = \frac{1}{m^2} - \frac{\lambda}{2}\inv{(m^2)^3} +
        \lambda^2\frac{2}{3}\inv{(m^2)^5} + O(\lambda^3).
\end{equation}
This equation can be solved with respect to $1/m^2$ iteratively,
and we get the 2PI free energy $G$.
\begin{align}
  \frac{1}{m^2} &
  = c + \frac{\lambda}{2}c^3 + \frac{\lambda^2}{12}c^5 + O(\lambda^3),\\
  G & = -\inv{2}-\inv{2}\log c + \inv{2}c \, m^2 +\frac{\lambda}{8}c^2
  -\frac{\lambda^2}{48}c^4    + O(\lambda^3),
\end{align}
which reproduces eq.~(\ref{G}) correctly when $m^2=0$
\footnote{Here an additive constant is not important.}.

In the practical calculations, it is easier to compute the 2PI free
energy $G(c)$ first and then obtain $F(m^2)$ by applying the
inverse Legendre transformation to $G(c)$.

Once we obtain the free energy $F(m^2)$, we replace $\lambda$ with $g\lambda$
and apply the ITE with respect to $m^2$.
For example if we consider the massless case corresponding to
IIB matrix model,
the improved free energies are
\begin{align}
  \label{imp_f}
  F_{k}^{\rm improved} & = F(m_0^2-g m_0^2) \bigr|_k\\
& = \left. -\inv{2} \log c_0 +g\left(\frac{\lambda}{8} c_0^2-\inv{2}\right)
         +g^2 \left(-\frac{\lambda^2}{48} c_0^4-\frac{\lambda^2}{16} c_0^4
                +\frac{\lambda}{4} c_0^2-\inv{4}\right) + \cdots \right|_k,
\end{align}
where $c_0$ stands for $1/m_0^2$.
Again, the last equation is the same as eq.~(\ref{F_phi4}) term by term.

\section{The ansatz and the results}
\label{sec:result}

\subsection{IIB matrix model}

Our aim is to analyze IIB matrix model \cite{Ishibashi:1997xs}.
The action is
\begin{equation}
  S = -\frac{1}{g_0^2}\Tr \Bigl(\; \frac{1}{4} [A_\mu,A_\nu]^2
      -\frac{1}{2}\, \bar{\psi} \Gamma^\mu [A_\mu, \psi] \;\Bigr),
\end{equation}
where $A_\mu(\mu =1,\ldots,10)$ and $\psi^\alpha(\alpha=1,\ldots,16)$
are all $N \times N$ hermitian matrices transforming as a vector and
a left-handed spinor representation under SO(10).
The symmetries of IIB matrix model are the matrix rotation U($N$),
ten dimensional Lorentz symmetry SO(10) and the type IIB supersymmetry.
As discussed in \cite{Aoki:1999bq}, the distribution of
the eigenvalues of $A_\mu$ is interpreted as the space-time itself.

It is difficult to integrate this action exactly.
So we apply the IMFA by adding and subtracting a quadratic term.
We specify the full propagators for bosons and fermions
instead of the quadratic term $S_0$.
Now we focus on the dimensionality of space-time or the ratio between
``our'' space and the ``internal'' one, and consider such ansatz that
the  U($N$) symmetry remains unbroken.
So gauge indices are set to preserve the U($N$) symmetry.
\begin{align}
  \gf{A_{\mu\;\phantom{i}j}^{\phantom{\mu}\;i}\;
      A_{\nu\;\phantom{k}l}^{\phantom{\nu}\;k}}
    & = g_0^2 \, C_{(\mu\nu)}\;
          \delta^{i}_{\phantom{i}l}\, \delta^{\phantom{j}k}_{j},\\
  \gf{\psi^{\alpha\; i}_{\phantom{\alpha\; i}j}\;
      \psi^{\beta\; k}_{\phantom{\beta\; k}l}}
    & = g_0^2 \, \frac{i}{3!} u_{[\mu\nu\rho]}
         (\Gamma^{\mu\nu\rho}{\cal C}^{-1})^{\alpha\beta}\;
        \delta^{i}_{\phantom{i}l}\, \delta^{\phantom{j}k}_{j},
\end{align}
where ${\cal C}$ is the charge conjugation matrix such that
\begin{align}
 {}^t{\cal C}=-{\cal C}, \hspace{1em}
   {\cal C}\Gamma^{\mu} = - {}^t \Gamma^{\mu} {\cal C}
    \hspace{0.5em} ( \mu =1,\ldots ,10 ).
\end{align}
Since the U($N$) index part is symmetric in each propagator,
the Lorentz index part is forced to be symmetric for bosons
and anti-symmetric for fermions.
We have converted the spinor indices $\alpha,\beta$ to
the rank three anti-symmetric tensor $u_{[\mu\nu\rho]}$
through the gamma matrices, which is equivalent to the antisymmetric
product of two left-handed spinor representations.

Here we give an overview. First, we calculate
the 2PI free energy $G$ by evaluating planar vacuum diagrams with the
above exact propagators.
We consider the large-$N$ limit with the 't~Hooft coupling
$g=g_0^2 N$ fixed to 1.
We use this $g$ as the formal expansion parameter in the ITE later.
Secondly,
in order to investigate in what shape the eigenvalues distribute and
to compare the free energies of various vacua
we try various ansatz with symmetries which are subgroups of SO(10).
With these symmetries we can
restrict the form of the propagators and reduce the number of
variational parameters.
Thirdly, for each ansatz we make the inverse Legendre transformation to the
2PI free energy $G$ to get the ordinary free energy $F$.
We further apply the ITE with respect to the conjugate parameters of $C$
and $u$ as in eq. (\ref{imp_f}) and obtain the improved free energy
$F^{\rm improved}$.
Finally, in each ansatz we extremize $F^{\rm  improved}$ and compute
the extent of the space-time from $\gf{ \Tr\; (A_\mu A_\nu)}$.

\subsection{2PI free energy}

We calculate the 2PI free energy $G$ to the fifth order.
The Feynman rules and all graphs with their symmetry factors
are shown in appendices
\ref{sec:Feynmann_rules} and \ref{sec:2PIgraphs}.
At a glance we notice that the number of graphs are much reduced
by the condition of 2PI.
For example, at the third order there are only four 2PI graphs as in
appendix \ref{third_order_graphs}, although thirty four connected
graphs contribute to the ordinary free energy.

The general form of the 2PI free energy is given by\footnote{We
adjust an additive constant to the definition of
\cite{{Nishimura:2001sx}}.}
\begin{equation}
\label{generalG}
\hspace*{-5pt}
   \frac{G(C,u)}{N^2}
         = 3(1\!+\!\log2)
          \!-\!\frac{1}{2}\tr\log C+\frac{1}{2}\tr\log\uslash
          \!-\!\frac{g}{2}
           \left(\tr C^2-(\tr C)^2
                \!+\!\tr(\uslash\Gamma^{\mu}\uslash\Gamma^{\nu})C_{\mu\nu}
           \right)\!+\!\cdots,
\end{equation}
where $\uslash^{\alpha}_{\phantom{\alpha}\beta}
=u_{\mu\nu\rho}(\Gamma^{\mu\nu\rho})^{\alpha}_{\phantom{\alpha}\beta}/3!$
and the traces are taken in the vector representation for $C's$
and in the left-handed spinor representation for $\uslash$'s.
Although we can write down the higher order terms without evaluating
the traces as in eq.~(\ref{generalG}), it is technically difficult
to evaluate these traces and obtain a concise expression of $G$.
Instead we evaluate the traces for each ansatz.
We present in appendix \ref{sec:calcG} the explicit forms
of $G$ to the fifth order for the SO(7) and the SO(4) ansatz
which are explained in the next subsection.

\subsection{symmetries and ansatz}

The guideline of making the ansatz is as follows:
\begin{itemize}
\item First of all, we fix a subgroup of SO(10) from SO(7) to SO(1),
  which will be interpreted as the rotational symmetry
  of ``our'' space-time. The component of the fermion two-point function
  $u$ is non-zero only when no indices are along
  ``our'' space-time by this symmetry. In fact, we shall see
  that the extent of ``our'' space is larger than that of
  the remaining ``internal'' space.
\item Because $u$ is a rank three anti-symmetric tensor, we
  decompose the internal space into the sum of three dimensional
  spaces and the rest, e.g. 4=3+1 or 8=3+3+2. We then assume the
  rotational symmetry for each space, e.g. SO(3)$\times$SO(1) or
  SO(3)$\times$SO(3)$\times$SO(2).
  This symmetry forces the component of $u$ to be zero unless all
  three indices lie in one of the three dimensional spaces.
\item Finally, we impose the permutation symmetry on the internal
  three dimensional spaces.
\end{itemize}

In this procedure, it is very important to include all parameters
permitted by the imposed symmetry.
Even if the free energy has an extremum in a randomly restricted
parameter space, it need not be a real one in the full parameter space.
On the other hand if we restrict the parameters by a symmetry,
the free energy is promised to be stationary in the directions of
the parameters which break the symmetry.

\underline{SO(7) ansatz}

The fermion propagator is represented
by the rank three anti-symmetric tensor $u$.
Therefore if we give a non-zero value
to only one element of $u$, SO(10) symmetry breaks down to
 SO(7)$\times$SO(3).
Under this symmetry group,
propagators are represented by three parameters, $V_1,V_2,u$.
\begin{align}
  C_{\mu\nu} & =
  \begin{array}{cc}
    \begin{array}{c}
       1 \\ \vdots \\ 7 \\\hline 8 \\ \vdots \\ 10 \\
    \end{array}
    \left(
      \begin{array}{ccc|ccc}
        V_1 &&&&&\\
        & \ddots &&&& \\
        & & V_1 &&&\\ \hline
        & & & V_2 \\
        & & & & \ddots \\
         & & & & & V_2
      \end{array}
    \right),
  \end{array}\\
  \uslash & = u \Gamma^{8,9,10}.
\end{align}

\underline{SO(6) ansatz}

Next, we consider the case where SO(6) symmetry survives.
$u_{[\mu\nu\rho]}$ has non-zero values only
in the extra four dimensional space.
A tensor $u_{\mu\nu\rho}$ of SO(4) is dual to a vector $\tilde u^{\mu}$.
Therefore up to SO(4) rotation, all configurations are identical.
As a result SO(3)$\,\subset\,$SO(4) transverse to $\tilde u^{\mu}$
remains unbroken.
Therefore this ansatz has SO(6)$\times$SO(3) symmetry.
\begin{align}
  \label{so(6)}
  C_{\mu\nu} & =
  \begin{array}{cc}
    \begin{array}{c}
       1 \\ \vdots \\ 6 \\ \hline 7 \\ \hline 8 \\ \vdots \\ 10 \\
    \end{array}
    \left(
      \begin{array}{ccc|c|ccc}
        V_1 &&&&&&\\
        & \ddots &&&& \\
        & & V_1 &&&&\\ \hline
        & & & V_2 &&&\\ \hline
        & & & & V_3 &&\\
        & & & & & \ddots  &\\
        & & & & & &V_3
      \end{array}
    \right),
  \end{array}\\
  \uslash & = u \Gamma^{8,9,10}.
\end{align}

\underline{SO(5) ansatz}

With five extra dimensions, the situation becomes a little complicated.
$u_{[\mu\nu\rho]}$ of SO(5) is now dual to a rank two anti-symmetric
tensor (adjoint representation) $\tilde u^{[\mu\nu]}$.
In general, $\tilde u^{[\mu\nu]}$ can be taken to the standard form
by SO(5), there exists two free parameters.
For simplicity (though it may not be natural), we set one of them to zero.
As to the fermion propagator, after all, we take the same form in these
three cases.
This ansatz has SO(5)$\times$SO(2)$\times$SO(3) symmetry.
\begin{align}
  C_{\mu\nu} & = {\rm diag}({\rm five}\ V_1{\rm 's},\
     {\rm two}\ V_2{\rm 's},\ {\rm three}\ V_3{\rm 's}), \\
  \uslash & =u \Gamma^{8,9,10}.
\end{align}

\underline{SO(4) ansatz}

Assuming SO(4) symmetry, the extra dimension is 6$\,=\,$3+3.
We can naturally impose simple symmetry
SO(3)$\times$SO(3)$\times Z_2$ where
$Z_2$ symmetry exchanges two SO(3) factors with
the reversion of the 1st direction (for no parity
change in total). Consequently we arrive at
the three parameter ansatz.
This ansatz has SO(4)$\times$SO(3)$\times$SO(3)$\times Z_2$ symmetry.
\begin{align}
  C_{\mu\nu} & =
  \begin{array}{cc}
    \begin{array}{c}
       1 \\ \vdots \\ 4 \\\hline 5 \\ \vdots \\ 10 \\
    \end{array}
    \left(
      \begin{array}{ccc|ccc}
        V_1 &&&&&\\
        & \ddots &&&& \\
        & & V_1 &&&\\ \hline
        & & & V_2 \\
        & & & & \ddots \\
         & & & & & V_2
      \end{array}
    \right),
  \end{array}\\
  \uslash & = \inv{\sqrt{2}}u \left(\Gamma^{5,6,7} +\Gamma^{8,9,10}\right).
\end{align}

\underline{SO(3) ansatz}

We decompose the extra $7$ dimensions to $3+3+1$ , and impose
SO(3)$\times$SO(3)$\times Z_2$ symmetry.
The fermion propagator is the same as the above case.
This ansatz has SO(3)$\times$SO(3)$\times$SO(3)$\times Z_2$ symmetry.
\begin{align}
  C_{\mu\nu} & = {\rm diag}({\rm three}\ V_1{\rm 's},\
     {\rm one}\ V_2,\ {\rm six}\ V_3{\rm 's}), \\
  \uslash & = \inv{\sqrt{2}}u \left(\Gamma^{5,6,7} +\Gamma^{8,9,10}\right).
\end{align}
In the first SO(3) factor (which will be interpreted as ``our''
space-time),
the reversion of 1st direction included in $Z_2$ forbids the
non-zero value of $u_{1,2,3}$.

\underline{SO(2) ansatz}

We decompose the extra $8$ dimensions to $2+3+3$ , and impose
SO(2)$\times$SO(2)$\times$SO(3)$\times$SO(3)$\times Z_2$ symmetry.
We then have the following form:
\begin{align}
  C_{\mu\nu} & = {\rm diag}({\rm two}\ V_1{\rm 's},\
     {\rm two}\ V_2{\rm 's},\ {\rm six}\ V_3{\rm 's}),\\
  \uslash & = \inv{\sqrt{2}}u \left(\Gamma^{5,6,7} +\Gamma^{8,9,10}\right).
\end{align}

\underline{SO(1) ansatz}\footnote{We use the term SO(1) to
  indicate one dimensional space-time.}

Finally we consider the one dimensional space-time.
We divide nine extra directions into three parts each of which
has three dimensions.
Three SO(3) symmetries and permutation $S_3$ of these factors
are imposed.
$S_3$ is combined with the reversion of the 1st direction
in order to be a subgroup of the original SO(10).
This ansatz has SO(3)$\times$SO(3)$\times$SO(3)$\times S_3$ symmetry.
\begin{align}
  C_{\mu\nu} & = {\rm diag}({\rm one}\ V_1,\ {\rm nine}\ V_2{\rm 's}),\\
  \uslash & = \inv{\sqrt{3}}u \left(
     \Gamma^{2,3,4} +\Gamma^{5,6,7} +\Gamma^{8,9,10}\right).
\end{align}

\subsection{free energy}

The ansatz listed above have three or four parameters. Namely
the boson propagator has two or three parameters $V_i$ and
the fermion propagator has one parameter $u$.
After substituting each ansatz into the general expression of $G$
as eq.~(\ref{generalG}),
we obtain the 2PI free energy $G(V_i,u)$.
The explicit forms of $G(V_i,u)$ are given in appendix \ref{sec:calcG}
for the SO(7) and the SO(4) ansatz. As we discuss later, these cases
turn out to be of particular importance.
Then we make the inverse Legendre transformation to $G(V_i,u)$
in the following way. We define conjugate variables to $V_i$ and $u$
as
\begin{align}
  M^i & = \dder{}{V_i}G(V,u),\\
    m & = \dder{}{u}G(V,u).
\end{align}
The ordinary free energy $F$ is
\begin{equation}
F(M^i,m)=G-\sum_iM^iV_i-mu.
\end{equation}

Then we apply the ITE to $F(M^i,m)$:
\begin{equation}
  F^{\rm improved}_{k}(M^i,m;M^i_0,m_0)
   = F(M^i_0 + g(M^i-M^i_0),m_0 +g(m-m_0))\Big|_k.
\end{equation}
Here $|_k$ denotes expanding with respect to $g$,
neglecting $O(g^{k+1})$ terms and setting $g=1$
as in eq.~(\ref{f_g_truncated}).
Since the action of IIB matrix model does not have a mass term,
the improved free energy is given by
\begin{equation}
\label{F_imp}
F^{\rm improved}_{k}(0,0;M^i_0,m_0)
= F(M^i_0 - gM^i_0,m_0 - gm_0)\Big|_k.
\end{equation}
The explicit forms of $F(M^i_0 - gM^i_0,m_0 - gm_0)$ for the SO(7)
and the SO(4) ansatz are given in appendix \ref{sec:calcF} .
We differentiate this free energy with respect to each parameter
and set the results to zero.
By solving these equations we get the extrema
of the improved free energy.
If there are more than one solutions, we select the one which gives the
smallest free energy among them. Such a solution is expected
to be on the plateau.

\begin{figure}[htbp]
  \begin{center}
    \leavevmode
    {\scalebox{0.9}{\includegraphics{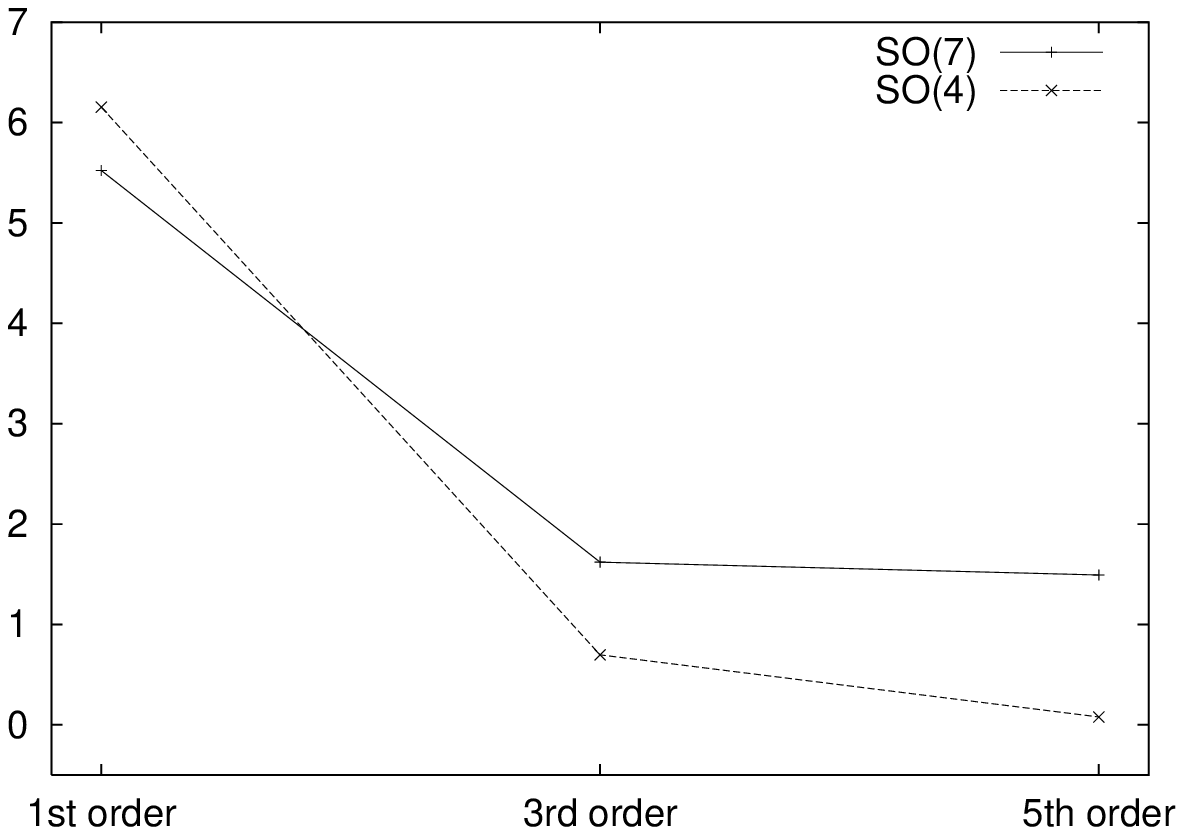}}}
    \hangcaption{Free energies.}
    \label{fig:free_energy}
  \end{center}
\end{figure}

Fig.\ref{fig:free_energy} shows the free energies for the
SO(7) and the SO(4) ansatz. As we explain below, the other cases are reduced
to these two cases. Their numerical values are shown in
Table \ref{tab:numerical_values}.
The corresponding values of $M_0^i$ and $m_0$ are shown in appendix
\ref{sec:Position_of_plateau}.
We find that the contribution of the fifth order
is smaller than that of the third order.
This fact suggests that the IMFA works well,
and we can trust these results as in
the case of the $\phi^4$ toy model.

\begin{table}[htbp]
  \begin{center}
    \leavevmode
    \begin{tabular}{c|c||l|l|l} \hline
      ansatz & quantities &
         \multicolumn{1}{c|}{1st order} &
         \multicolumn{1}{c|}{3rd order} &
         \multicolumn{1}{c}{5th order} \\
       \hline \hline
      SO(7) & $F$ &\hspace{3pt} 5.522722156  &\hspace{3pt} 1.620942535  &
              1.492429842\\ \cline{2-5}
            & $R$ &\hspace{3pt} 0.6407423748 &\hspace{3pt} 0.7670244718 &
              0.8051803940\\ \cline{2-5}
            & $r$ &\hspace{3pt} 0.3276944430 &\hspace{3pt} 0.3556261073 &
              0.3844627899\\ \cline{2-5}
         & $\rho$ &\hspace{3pt} 1.955304364  &\hspace{3pt} 2.156828390  &
              2.094299930\\ \hline \hline
      SO(4) & $F$ &\hspace{3pt} 6.153347114  &\hspace{3pt} 0.696885300  &
              0.077693647\\ \cline{2-5}
            & $R$ &\hspace{3pt} 0.7500535231 &\hspace{3pt} 1.161504610  &
               1.7090706440 \\ \cline{2-5}
            & $r$ &\hspace{3pt} 0.4038451083 & \hspace{3pt} 0.3796625830 &
               0.2691460598 \\ \cline{2-5}
         &$\rho$ &\hspace{3pt} 1.857280202  &\hspace{3pt} 3.059307619  &
               6.349974602 \\  \hline
    \end{tabular}
    \caption{Numerical values.}
    \label{tab:numerical_values}
  \end{center}
\end{table}

Some comments are in order.
\begin{itemize}
\item At the 2nd and the 4th order
  there is no reasonable extremum.
  As in the case of the $\phi^4$ model,
  extrema do not necessarily appear in the even and lower order levels.
  Therefore we consider only the odd orders.
\item For the SO(6) ansatz, the extremum solution reduces to that of
  the SO(7) ansatz,
  i.e. $V_1$ and $V_2$ in eq.~(\ref{so(6)}) take the same value
  and the SO(6) symmetry is enhanced to SO(7).
  In the same way the SO(5) ansatz reduces to the SO(7) ansatz.
  This is natural because the fermion propagators have the same form
  in these cases.
  On the other hand the SO(3) and the SO(2) ansatz reduce to the SO(4)
  ansatz.
  We thus conclude that the fermions play a crucial role
  in the process of the spontaneous breakdown of SO(10).
\item The SO(1) ansatz has no solution even at the 1st and the
  3rd order.
  This may indicate that one dimensional space is not realized
  as ``our'' space.
\end{itemize}

\subsection{extent of space-time}

\begin{figure}[htbp]
  \begin{center}
    \leavevmode
    {\scalebox{0.9}{\includegraphics{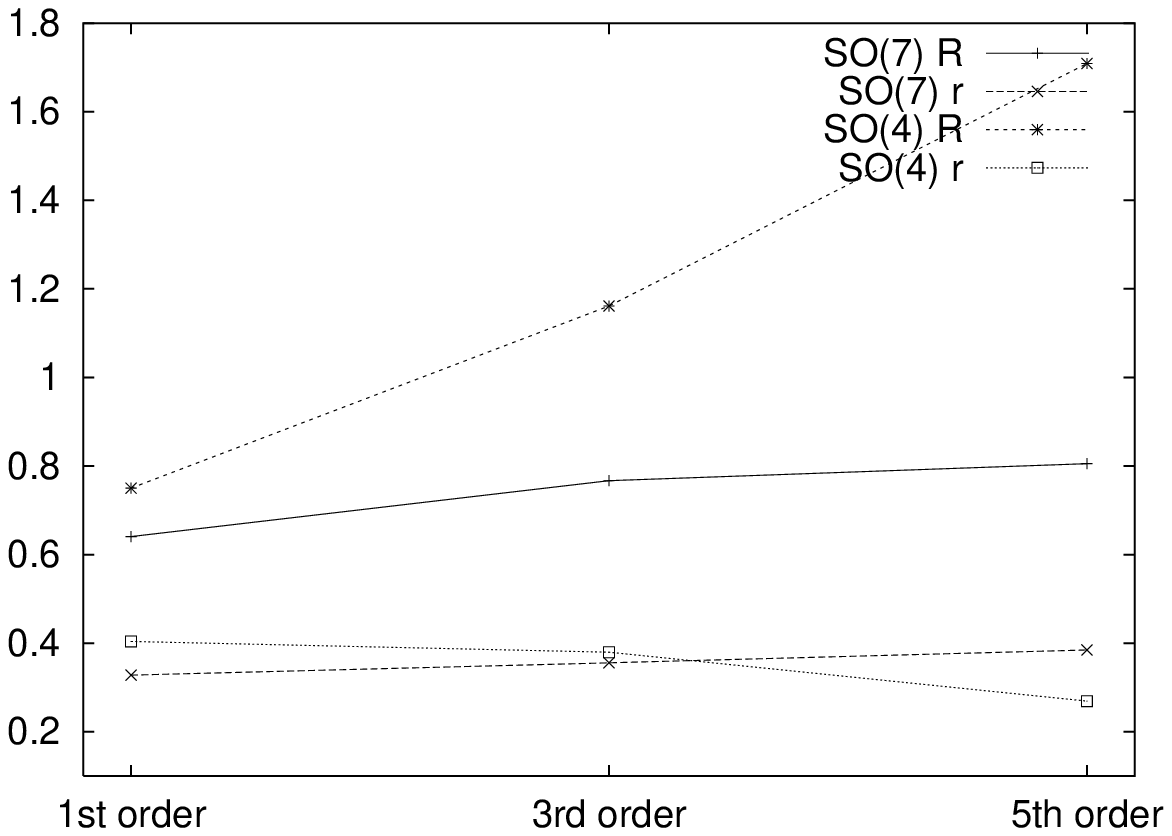}}}
    \hangcaption{Extent of the space-time.}
    \label{fig:extension}
  \end{center}
\end{figure}

\begin{figure}[htbp]
  \begin{center}
    \leavevmode
    {\scalebox{0.9}{\includegraphics{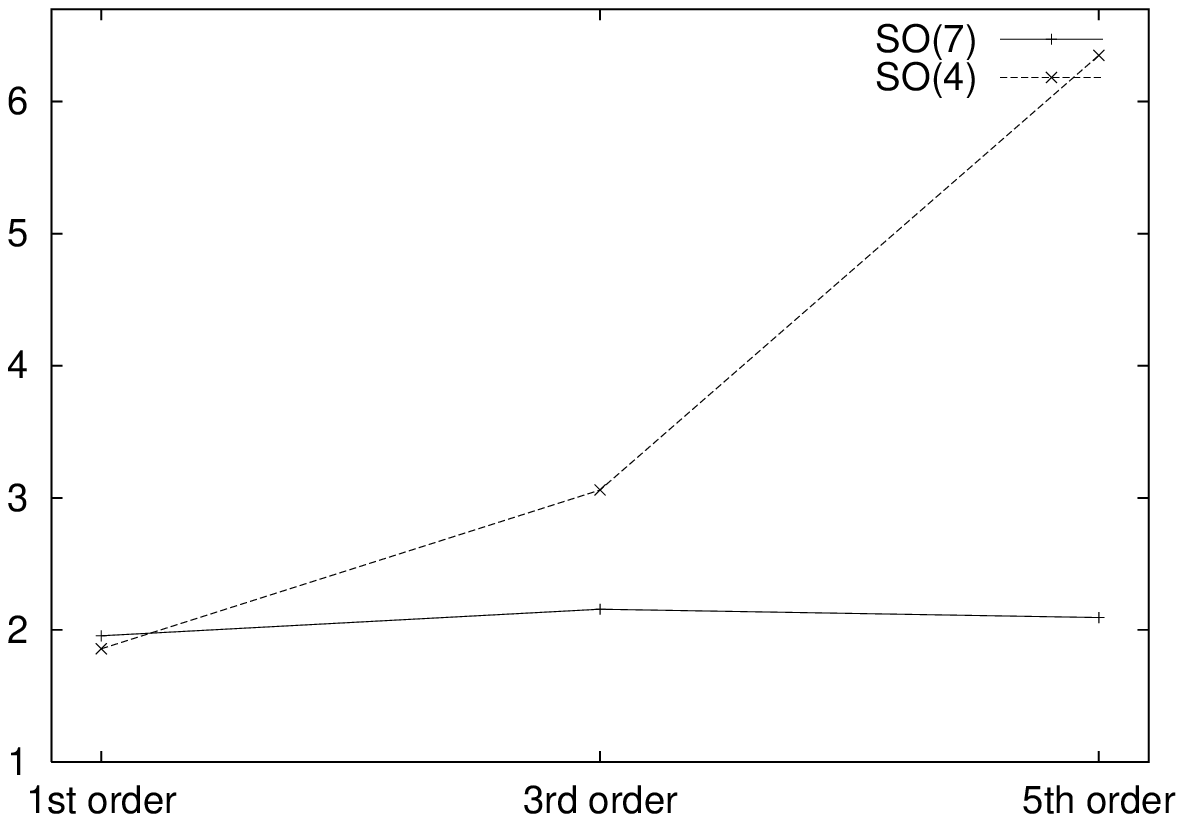}}}
    \hangcaption{Ratio between the extent of ``our''
    space-time and that of the ``internal'' space.}
    \label{fig:ratio}
  \end{center}
\end{figure}

Next, let us consider the extent of ``our''
space-time $R$ and that of the ``internal'' space $r$.
Here, for both the SO(7) and the SO(4) ansatz, $R$ and $r$ are defined
by
\begin{align}
  R^2 & = \gf{\frac{1}{N} \; \Tr A_1^{\phantom{1}2} }
  = - \dder{F(M^i,m)}{M^1}\ , \\
  r^2 & = \gf{\frac{1}{N} \; \Tr A_{10}^{\phantom{10}2} }
  = - \dder{F(M^i,m)}{M^2}.
\end{align}
These quantities can be also improved by the ITE, and obtained as
in eq.~(\ref{F_imp}):
\begin{align}
\gf{\frac{1}{N} \; \Tr A_1^{\phantom{1}2} }_k & =
 - \dder{F}{M^1}(M^i_0 - g M^i_0, m_0 -g m_0)\Big|_k,\\
\gf{\frac{1}{N} \; \Tr A_{10}^{\phantom{10}2} }_k & =
  - \dder{F}{M^2}(M^i_0 - g M^i_0, m_0 -g m_0)\Big|_k.
\end{align}
We evaluate these quantities using the values of $M_0^i$ and $m_0$
given in appendix \ref{sec:Position_of_plateau}.
$R$ and $r$ are shown in Fig.\ref{fig:extension},
and their ratio $\rho=R/r$ is plotted in Fig.\ref{fig:ratio}.

In contrast to the stability of the free energy,
both $R$ and $r$ for the SO(4) ansatz changes considerably.
It is striking that their ratio grows rapidly as
we increase the order of the approximation.

The fact that the SO(4) ansatz gives the lowest free energy
and the large value of the extent ratio
indicates that the path integral of IIB matrix model is dominated
by the nearly four dimensional distribution of the eigenvalues.

\section{Conclusions and discussions}
\label{sec:conc.}

Our main results are the following:
\begin{itemize}
\item Four dimensional configurations dominate in the path integral
  of IIB matrix model.
\item The fermionic field plays an important role in the
  breakdown of the Lorentz symmetry.
  Its two point function seems to determine the remaining symmetry.
\item
  For the SO(7) ansatz, the free energy and the extent ratio
  become stable at the 5th order. On the other hand, for the
  SO(4) ansatz, the free  energy still decreases at the 5th order, but
  it seems to approach to a definite value. On the other hand the
  ratio grows more
  rapidly as expected. Stability of the result requires still higher
  order analysis.
\end{itemize}

The IMFA is, in a rigorous sense, not an ``approximation'' scheme.
We do not know the true reason why a free energy converges to its
exact value.
But if we apply the IMFA to some models, the results are magically nice.
To find a trick behind the IMFA should be very important.

In any ansatz except the SO(1), the ``internal'' space shrinks
in such directions that the fermion full propagator has non-zero value.
We need to understand this phenomenon, shrinking by fermion,
at least qualitatively. From this point of view, it would be
interesting to clarify the relation between our result
and the mechanism of the symmetry breaking  by the phase of the fermion
determinant suggested in \cite{Nishimura:2000}.


In relation to the above connection between the ``compactification''
of the space and the fermion full propagator, we make a brief
comment on a possibility that the full SO(10) symmetry will be restored.
One might wonder that eigenvalue distribution having
the full SO(10) symmetry is likely to be realized. However, as far as
our approximation scheme is concerned, such distribution corresponds
to setting all of the components of $u$ to be zero, in which
both the ordinary free energy and the 2PI free energy apparently diverge
logarithmically. Therefore, it cannot be a minimum energy
configuration or a solution to the SDE. This is consistent with the
results in \cite{Nishimura:2000}.

Growth of the extent ratio for the SO(4) ansatz is quite
interesting.
To what value will it reach as going to the higher order, infinite
or finite? If it is approaching infinity, we can regard it
as an indication that the space-time
is spontaneously compactified to the four dimensional flat space.

At any rate,
the IMFA partly grasps the non-perturbative dynamics of
string theory.
We have extracted information on the dimension of the space-time
as the first step.
Probably the next step is to predict the gauge
symmetry.
Although some refinement to the IMFA is required,
we will become able to achieve it in the near future.

There may be some directions to refine the IMFA.
\begin{itemize}
\item Not only two point functions but three or more point functions
  can be included to the SDE.
  Also it may be interesting to include one-point functions.
\item In the present paper, we have used the 2PI free energy and
  the full propagators just as a calculational
  technique to reduce the number of graphs.
  But the notion of the full propagators, which has a direct connection
  with the SDE, may be more powerful in examining
  various aspects of the non-perturbative dynamics. In light of this,
  it would be important to investigate a direct application of the ITE
  to the 2PI free energy as done in appendix \ref{sec:ITE_2PI}.
\end{itemize}

Although there remain many unknown or ambiguous problems
such as the existence of gravity, chiral fermions, the connection
with the low energy effective field theory and so on,
we hope we can proceed step by step with this new method.

\section*{Acknowledgments}
The authors would like to thank T.~Azuma, J.~Nishimura, F.~Sugino
 and T.~Yokono for valuable discussions and useful comments.
This work is supported in part by Grant-in-Aid for Scientific Research
from Ministry of Education, Science, Sports and Culture of Japan
(\#1640290, \#03603, \#02846, \#01794 and \#03529).
The work of T.~K., T.~M., S.~K. and S.~S. is supported in part
by the Japan Society for the
Promotion of Science under the Post- and Pre-doctoral Research Program.

\appendix

\section{Some curious behavior of ITE}
\label{sec:curios_ITE}

In sec.\ref{sec:ITE}, we apply the ITE to the series $\sum (-)^n x^n$.
The value at $x=3/2$ is well approximated.
The method in sec.\ref{sec:ITE}, however, is not effective
when we are to evaluate the series farther from $x=1$.
Therefore we use the ITE in a different way.
The function we consider is
\begin{align}
  \label{exp_x}
  f(x) & = \sum_{n=0}^{\infty} g^n (-)^n x^{n} (= \inv{1+g x}),\\
  f_k(x) & = f(x)\big|_k.
\end{align}
As $x$ becomes large, the behavior of the power series in $x$ would be
worse. In such a case, a more proper variable is the inverse of $x$,
and we apply the ITE to
it, i.e.
\begin{align}
  \label{inv_x_ITE}
  \frac{1}{x} & \to \frac{1}{x_0}+g\left( \inv{x}-\inv{x_0} \right),\\
  f^{\rm improved}_k & =
         f\left(\inv{\frac{1}{x_0}+g\left( \inv{x}-\inv{x_0}
  \right)}\right) \Bigg|_k.
\end{align}
The original series of $f(x)$ without the ITE behave as shown in
Fig.\ref{fig:ori_1_x}.

\begin{figure}[htbp]
  \begin{center}
    \leavevmode
    {\scalebox{1}{\includegraphics{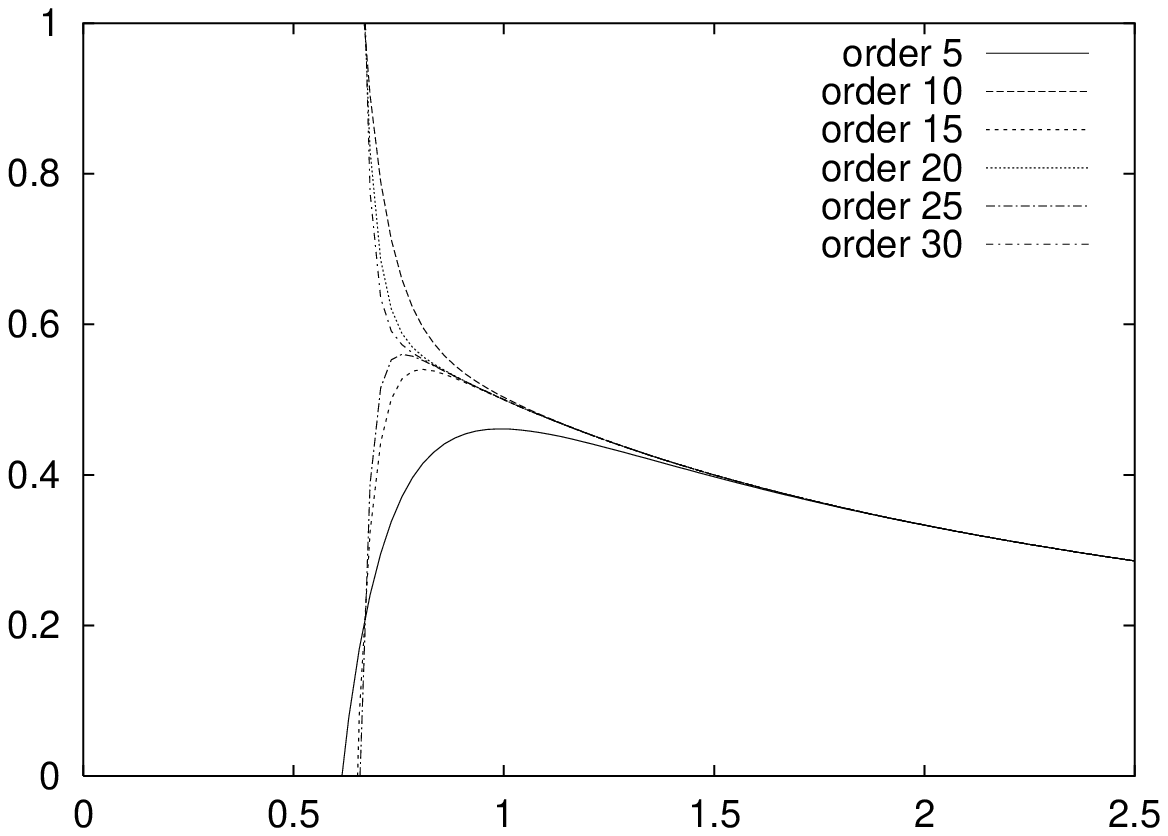}}}
    \hangcaption{Improved series. The horizontal axis is $x$.}
    \label{fig:ITE_1_x}
  \end{center}
\end{figure}

After applying the ITE, the behavior changes drastically.
Fig.\ref{fig:ITE_1_x} shows the
improved series at $x_0=0.8$.
The improved series behaves well in the region $x>1$
in which the original series diverges.
For example, if we evaluate the value at $x=3$, the
error\footnote{Error is defined here as
$\displaystyle{\frac{({\rm approximated\ value})-({\rm exact\ value})}
{({\rm exact\ value})}}$.}
is almost
$10^{-({\rm order\ of\ approximation})}$.

\section{Improved Taylor expansion for asymptotic series}
\label{sec:ITE_asy}

\begin{figure}[htbp]
  \begin{center}
    \leavevmode
    {\scalebox{1}{\includegraphics{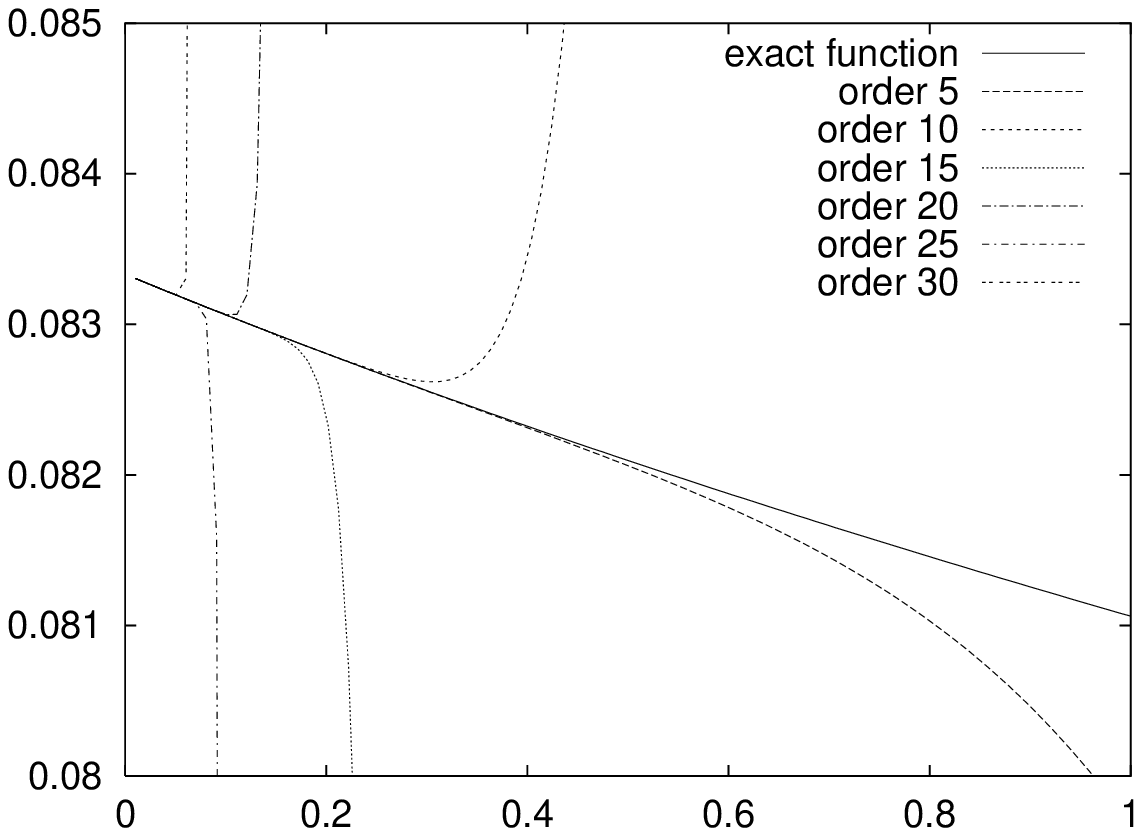}}}
    \hangcaption{Original truncated functions. The horizontal axis is $x$.}
    \label{fig:original_lngamma}
  \end{center}
\end{figure}

\begin{figure}[htbp]
  \begin{center}
    \leavevmode
    {\scalebox{1}{\includegraphics{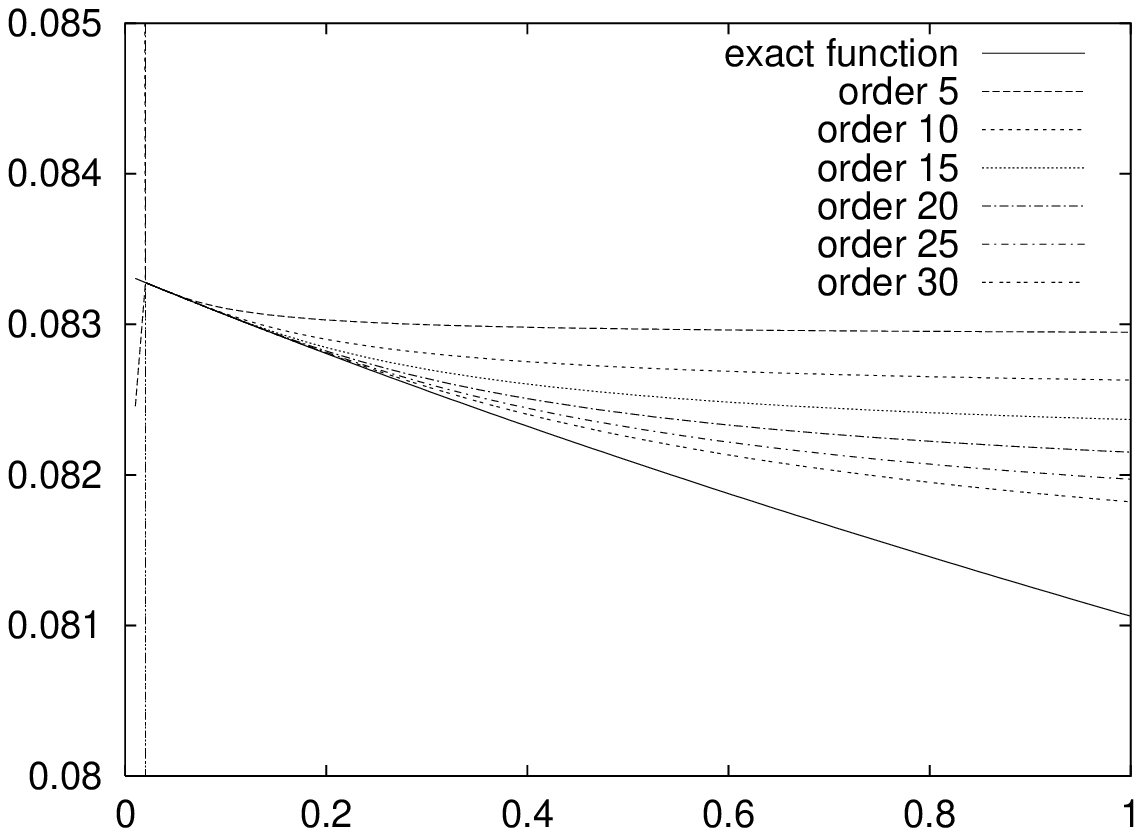}}}
    \hangcaption{Improved functions. The horizontal axis is $x$.}
    \label{fig:ITE_lngamma}
  \end{center}
\end{figure}

\begin{figure}[htbp]
  \begin{center}
    \leavevmode
    {\scalebox{1}{\includegraphics{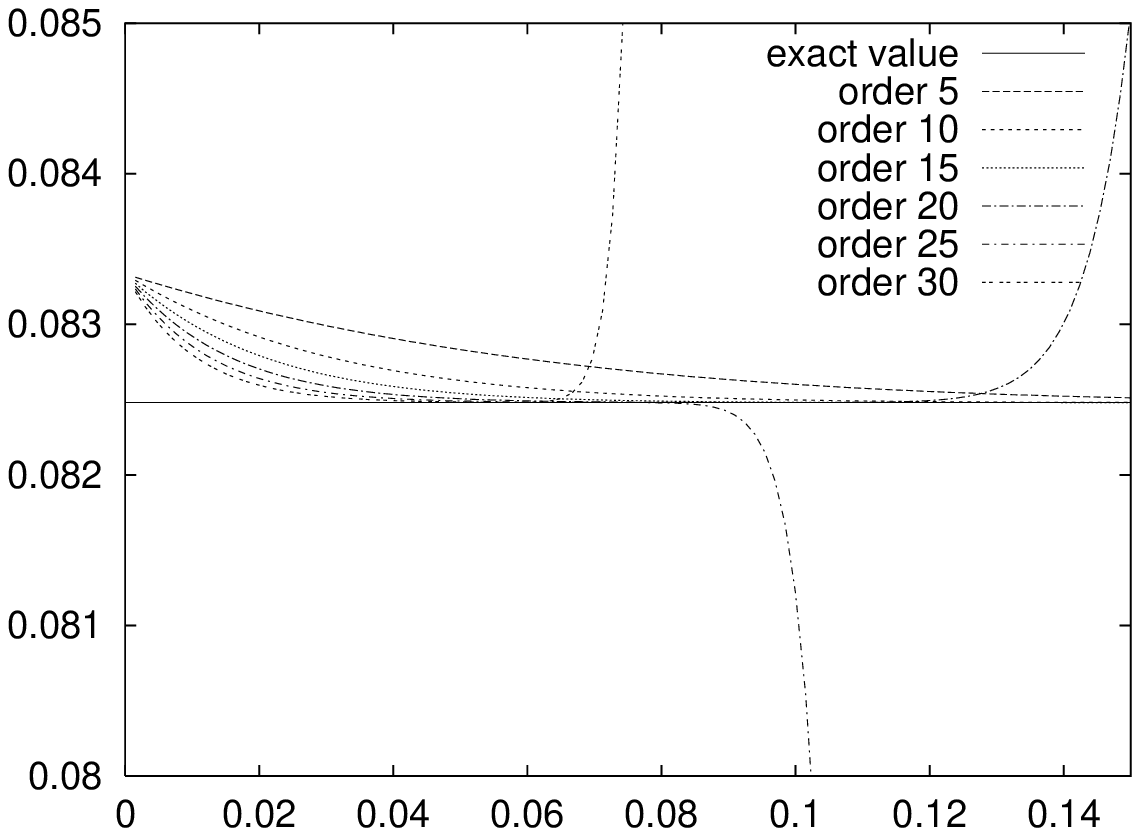}}}
    \hangcaption{Plateau for each improved function. The horizontal
    axis is $x_0$.}
    \label{fig:plateu_lngamma}
  \end{center}
\end{figure}

Although our actual interest is in analyzing IIB matrix model
and its planar sum has a finite convergence radius,
ITE can also be applied to asymptotic series.
We have already seen in sec.\ref{sec:MFA} that the IMFA for the
$\phi^4$ model works well.
The expansion in $g$ of this model is only asymptotic and not convergent.
As another example, we apply the ITE to $\Gamma$ function here.
\begin{align}
  \label{lngamma}
 J(x) & = x^{-1/2} \left( \log \Gamma(x^{-1/2})
         -x^{-1/2} (\log x^{-1/2}-1 )
         +\frac{1}{2} \log \frac{x^{-1/2}}{2 \pi} \right)\\
   & = \inv{12}-\inv{360}gx+\inv{1260}g^2 x^2
        -\inv{1680}g^3 x^3+\inv{1188}g^4 x^4-\frac{691}{360360} g^5
         x^5 \cdots,
\end{align}
which is also an asymptotic series.
As in the above example, we apply the ITE for the inverse variable.
Figs.\ref{fig:original_lngamma} and \ref{fig:ITE_lngamma}
show respectively the truncated functions without improvement and
with the ITE around $x_0=0.03$.
The behavior in the large $x$ region is smoothed out after the ITE.
If we fix $x$ to $1/3$, conversely, there is a plateau for each improved
function as shown in Fig.\ref{fig:plateu_lngamma}.
The errors from the exact values at the extremum
are about $10^{-5}$ for the improved functions of order between 6 and 26.
On the other hand the truncated functions without improvement have errors
of about $10^{-4}$, if the order of truncation lies between 3 and 7.
As going to larger values of $x$, errors grow in both cases.
At any rate the error becomes small and the oscillation becomes smooth by
the ITE.

\section{Direct application of ITE to 2PI free energy}
\label{sec:ITE_2PI}

The ITE can be applied not only to the ordinary free energy
but also to the 2PI free energy which has a direct
connection to the SDE.

Let us consider the massless matrix $\phi^4$ model \cite{Brezin:1978sv}.
\begin{equation}
S=N \frac{g^2}{4}\Tr \, \phi^4 .
\end{equation}
In the planar limit the perturbative expansion has a finite
convergence radius. Therefore we expect that this model is more
tractable than the usual $\phi^4$ model.

The 2PI free energy divided by $N^2$ is expanded as follows :
\begin{equation}
  G(x) = -\inv{2}\log(x)+\frac{g}{2}x^2-\frac{g^2}{8}x^4
       +\frac{g^3}{6}x^6-\frac{3g^4}{8} x^8 +\frac{11g^5}{10}x^{10}
       + O(g^6),
\end{equation}
where $x$ is the full propagator.
We can improve this function by replacing
$x$ with $x_0+g(x-x_0)$.

\begin{figure}[htbp]
  \begin{center}
    \leavevmode
    {\scalebox{1}{\includegraphics{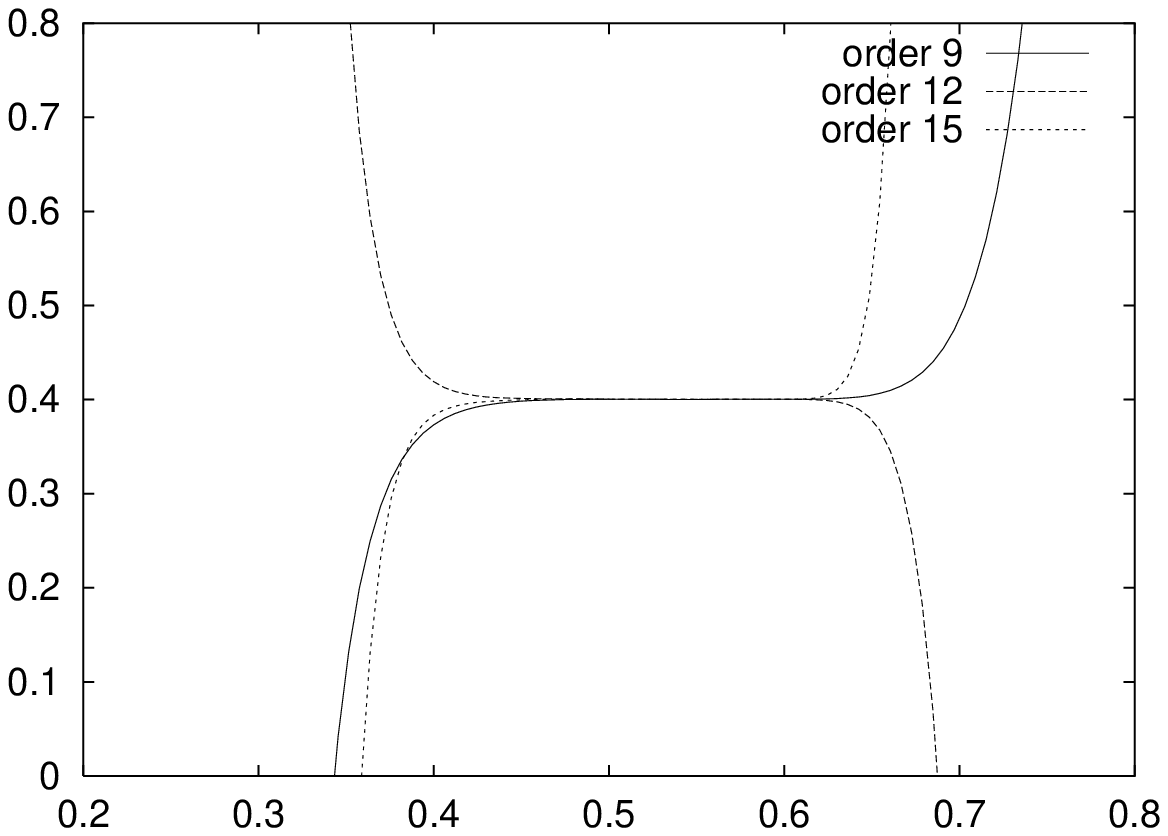}}}
    \hangcaption{Plateau of improved $G$ for $x=0.8$.
    The horizontal axis is $x_0$.}
    \label{fig:ITE_phi4_pl}
  \end{center}
\end{figure}

If we fix $x$ and regard $G$ as a function of
$x_0$, we see a clear plateau as shown in
Fig.\ref{fig:ITE_phi4_pl}.

\begin{figure}[htbp]
  \begin{center}
    \leavevmode
    {\scalebox{1}{\includegraphics{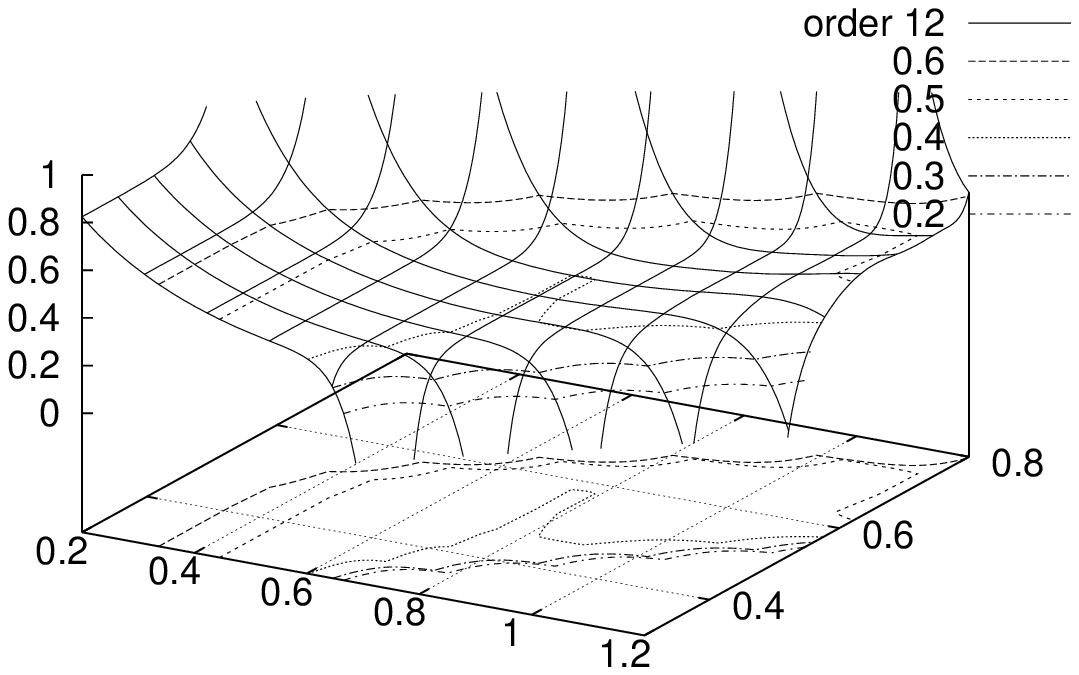}}}
    \hangcaption{3D plot of improved $G$ at the 12th order.
    ($0.2 \leq x \leq 1.2$,
                                          $0.8 \leq x_0 \leq 1.2$).}
    \label{fig:3d_phi4}
  \end{center}
\end{figure}

\begin{figure}[htbp]
  \begin{center}
    \leavevmode
    {\scalebox{1}{\includegraphics{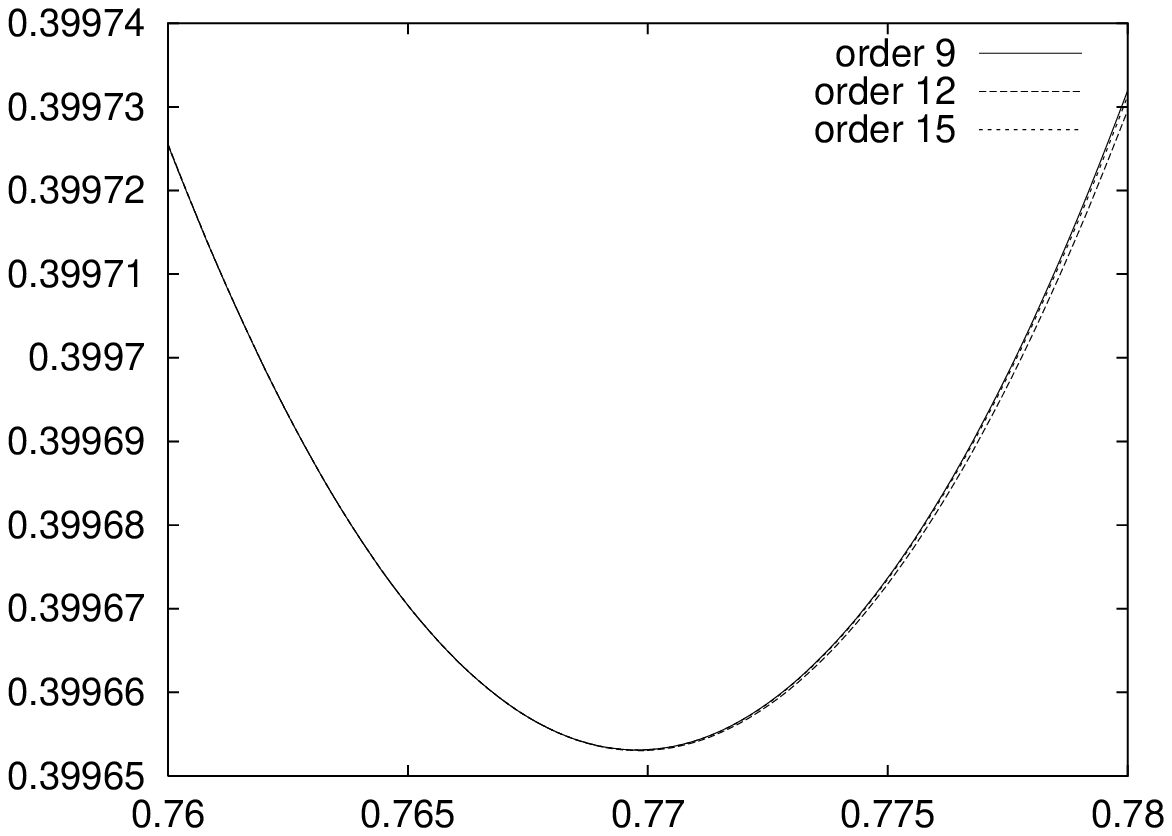}}}
    \hangcaption{Minimum of improved $G$. The horizontal axis is $x$.}
    \label{fig:ITE_phi4_min}
  \end{center}
\end{figure}

When we plot the improved 2PI free energy $G$ by two variables $x$ and
$x_0$ as in Fig.\ref{fig:3d_phi4},
the contour at $G=0.4$ runs in parallel with $x_0$ axis at $x\sim 0.77$.
Consequently, we have the plateau for $x_0$ and achieve a nice
approximation around there.
Here, in contrast to the case of the ordinary free energy,
it is the full propagator
$x$ which gives the minimum of $G$ that we are interested in,
because it is nothing but a solution to the SDE.
Fixing $x_0=0.585$ and plotting $G$ as a function of $x$,
we succeed to evaluate $x=0.77$ at the minimum of $G$,
Fig.\ref{fig:ITE_phi4_min}.
This result almost reproduces the exact value $4\sqrt{3}/9\sim
0.7698$.
In this way we are able to apply the ITE directly to the 2PI free energy.
There is no need for Legendre transformation, and
we have more direct connection with the SDE.

\section{Feynman rules}
\label{sec:Feynmann_rules}

We write down the Feynman rule for IIB matrix model.
\begin{itemize}
\item Propagators:
  \begin{itemize}
  \item boson \hspace{9pt}: \raisebox{-18pt}{\input{b_pro.latex}}
          \hspace{1em}= $\ \displaystyle{g_0^2} \; C_{\mu\nu}$\vspace*{10pt}\,.
  \item fermion : \raisebox{-18pt}{\input{f_pro.latex}}
          \hspace{1em}= $\ \displaystyle{g_0^2 \,\frac{1}{3!}}
          \; i u_{\mu\nu\rho}
               (\Gamma^{\mu\nu\rho}{\cal C}^{-1})^{\alpha\beta}$\,.
  \end{itemize}
\item Vertices:
  \begin{itemize}
  \item four-boson \hspace{19pt}: \hspace{3pt}
        \raisebox{-43pt}{\input{b_vert.latex}}
        \hspace{1em} =
        $\ \ \displaystyle{\frac{1}{g_0^2}}
        (2\delta^{\mu\rho}\delta^{\nu\lambda}
           -\delta^{\mu\nu}\delta^{\rho\lambda}
           -\delta^{\mu\lambda}\delta^{\nu\rho})$\,.
  \item boson-fermion : \hspace{3pt}
        \raisebox{-35pt}{\input{b-f_vert.latex}}
        \hspace{1em} = $\ \ \displaystyle{\frac{1}{g_0^2}} ({\cal C}\Gamma^\mu)_{\alpha\beta}$\,.
  \end{itemize}
\item Extra factors:
  \begin{itemize}
  \item $-1$ for each fermion loop.
  \item $N$ for each color loop.
  \end{itemize}
\end{itemize}

\section{Planar Feynman graphs}
\label{sec:2PIgraphs}
\setcounter{subsection}{-1}
Here we show all the Feynman graphs from the zeroth order to the
fifth order.
The number below each graph is the symmetry factor.
The value of each graph is given by
\begin{equation}
  (-1) \times \inv{\rm symmetry\ factor} \times
  ( {\rm factor\ from\ Feynman\ rules} ).
\end{equation}
The first factor of minus one originates from
the minus sign of the definition of the free energy,
$F=-\log Z$.

\subsection{zeroth order}
\begin{tabular}{cc}
  \graph{0.77}{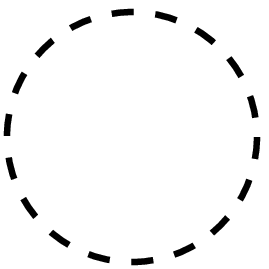} & \graph{0.77}{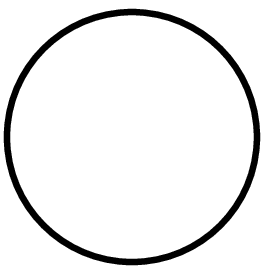} \\
    --- & ---
\end{tabular}

\subsection{first order}
\begin{tabular}{cccc}
  \graph{0.77}{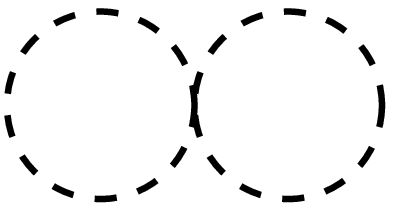} & \graph{0.77}{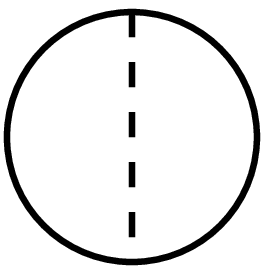} \\
   2 & 2
\end{tabular}

\subsection{second order}
\begin{tabular}{cc}
  \graph{0.77}{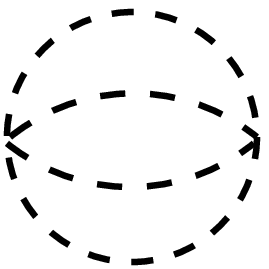} & \graph{0.77}{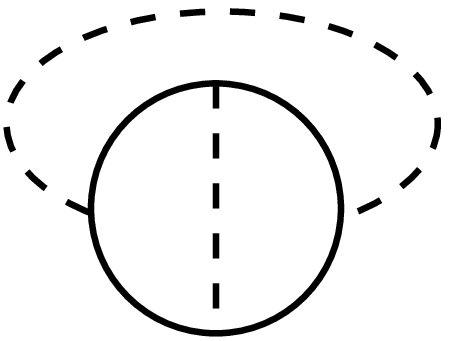}\\
    8 & 4
\end{tabular}

\subsection{third order}
\label{third_order_graphs}
\begin{tabular}{cccc}
  \graph{0.77}{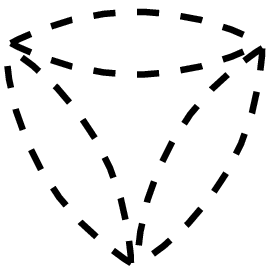} & \graph{0.77}{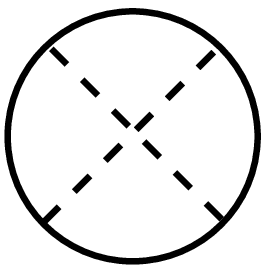} &
  \graph{0.77}{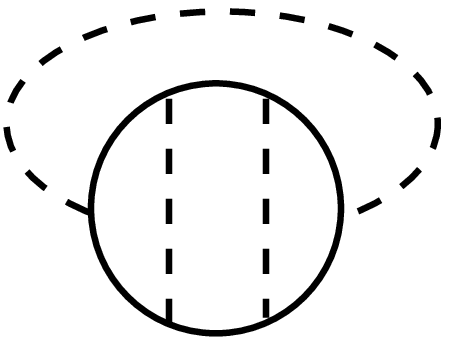} & \graph{0.77}{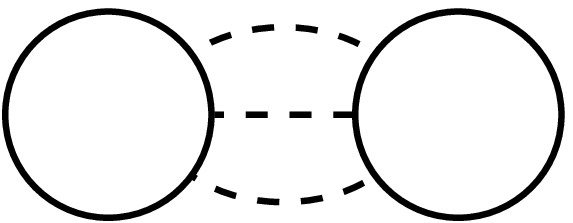} \\
    6 & 4 & 2 & 6
\end{tabular}

\subsection{fourth order}
\begin{tabular}{cccc}
  \graph{0.77}{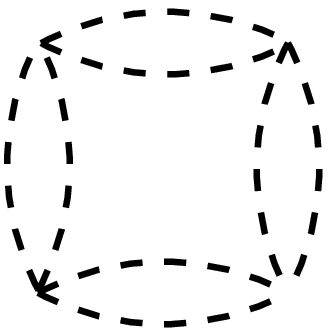} & \graph{0.77}{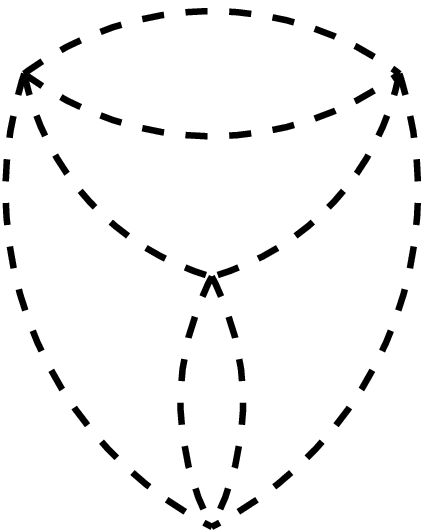} &
  \graph{0.77}{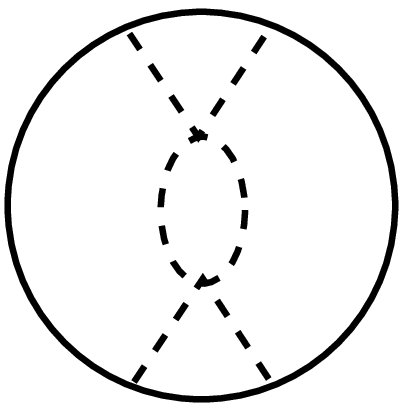} & \graph{0.77}{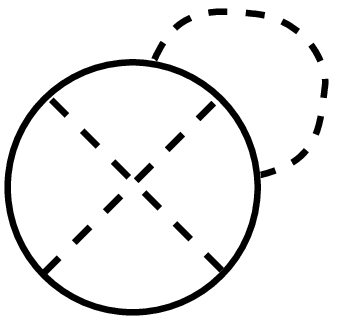} \\
   8 & 4 & 2 & 1
\end{tabular}

\begin{tabular}{cccc}
  \graph{0.77}{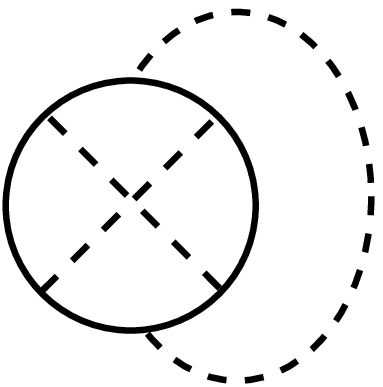} & \graph{0.77}{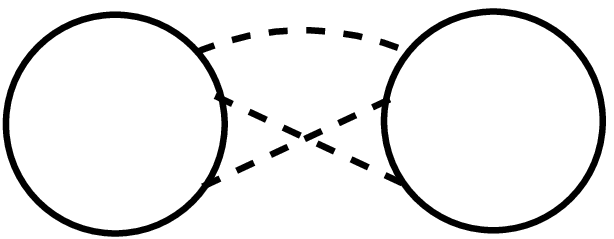} &
  \graph{0.77}{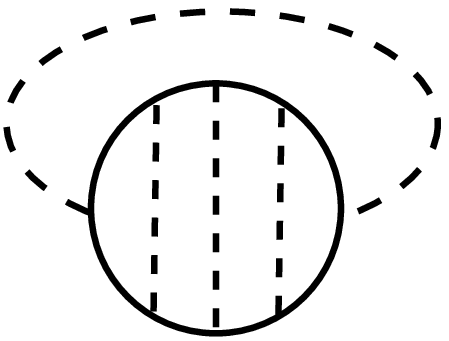} & \graph{0.77}{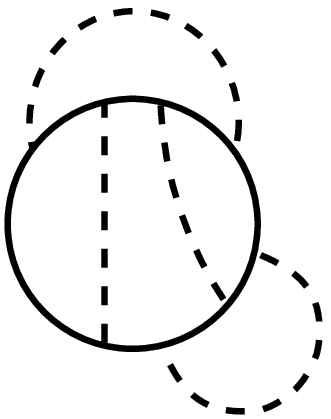} \\
   2 & 2 & 2 & 2
\end{tabular}

\begin{tabular}{cccc}
  \graph{0.77}{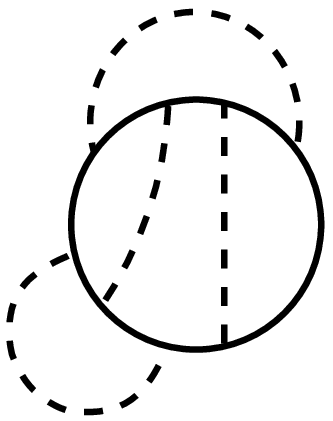} & \graph{0.77}{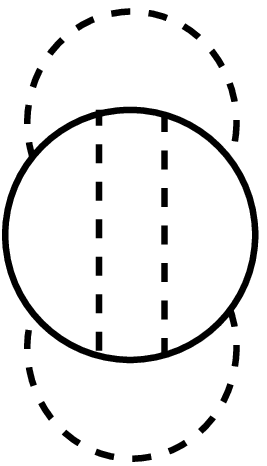} &
  \graph{0.77}{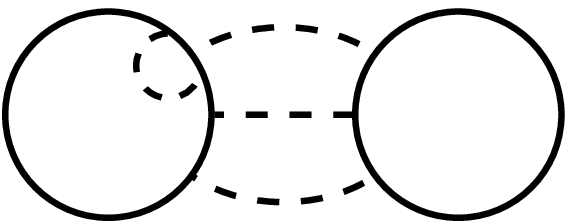} & \graph{0.77}{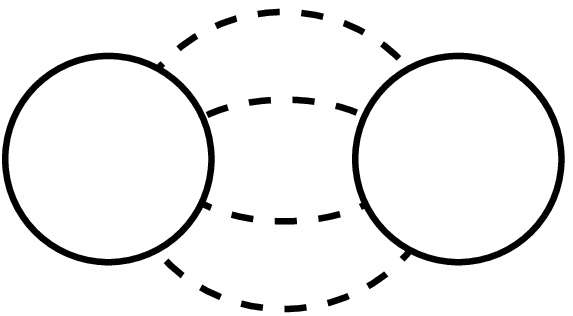} \\
   2  & 4 & 1 & 8
\end{tabular}

\subsection{fifth order}
\begin{tabular}{cccc}
  \graph{0.77}{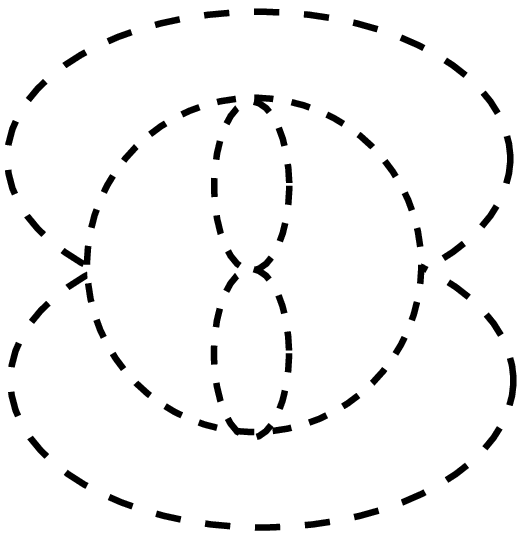} & \graph{0.77}{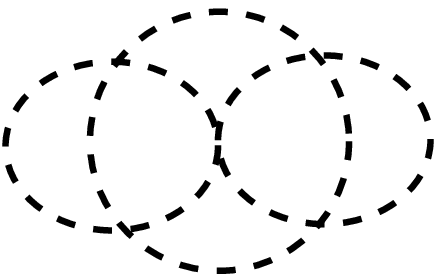} &
  \graph{0.77}{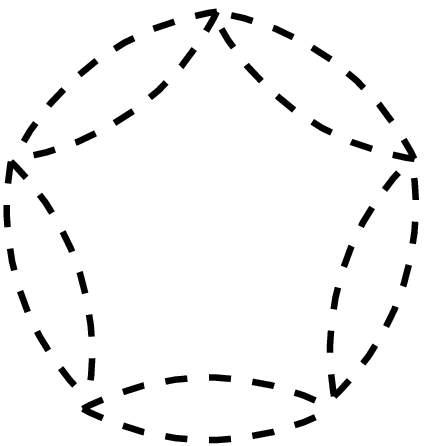} & \graph{0.77}{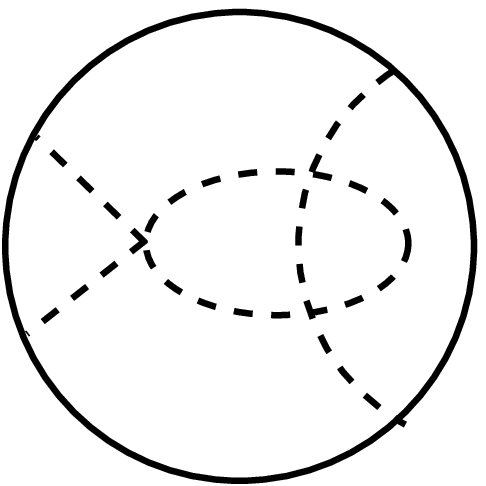} \\
   2 & 2 & 10 & 1
\end{tabular}

\begin{tabular}{cccc}
  \graph{0.77}{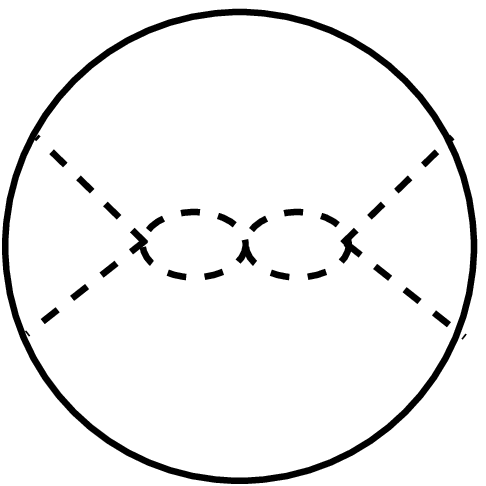} & \graph{0.77}{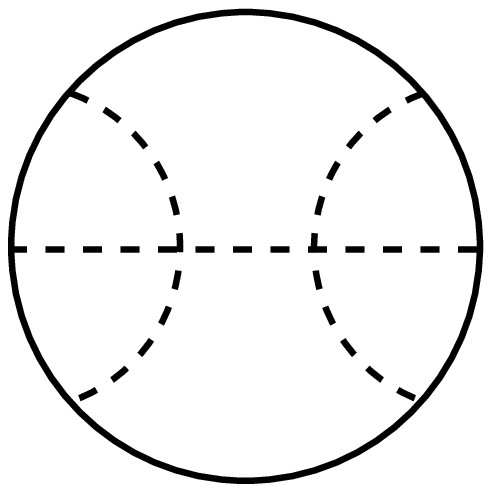} &
  \graph{0.77}{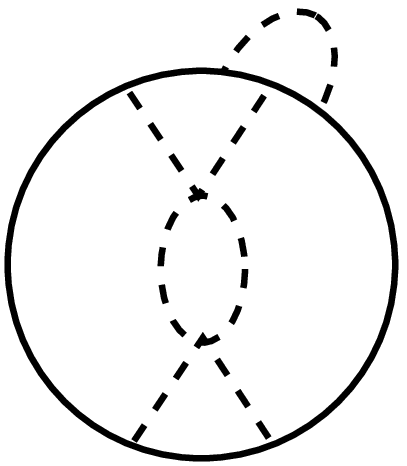} & \graph{0.77}{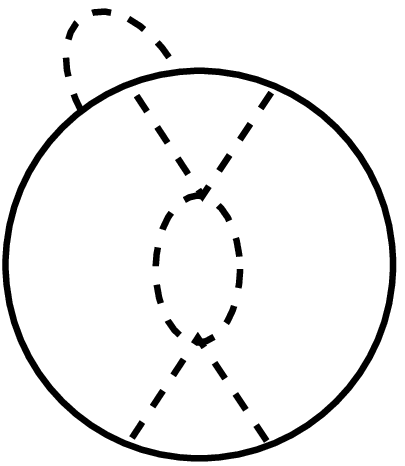} \\
   2 & 2 & 1 & 1
\end{tabular}

\begin{tabular}{ccc}
  \graph{0.77}{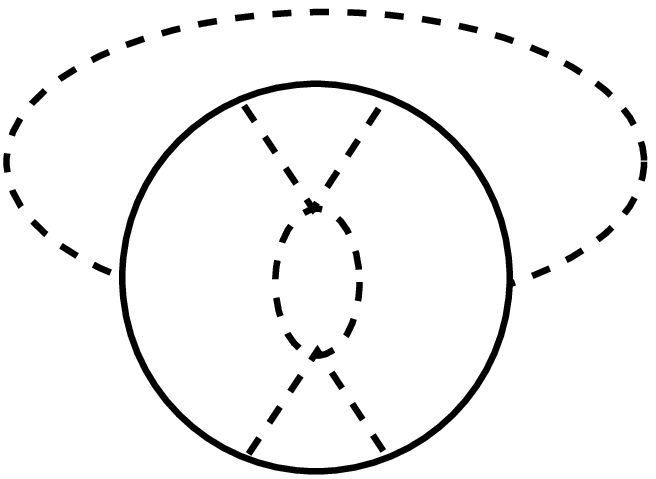} & \graph{0.77}{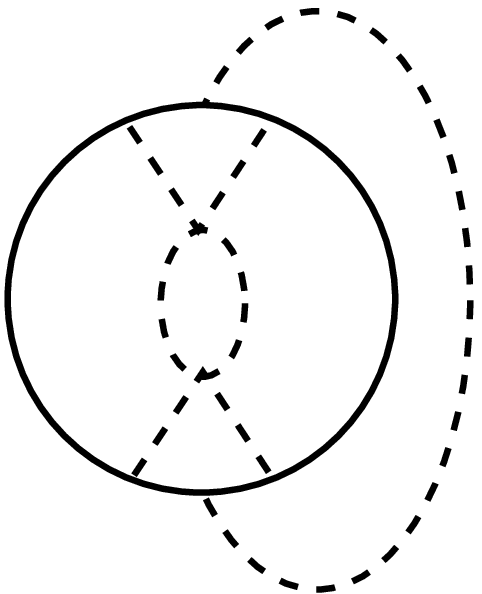} &
  \graph{0.77}{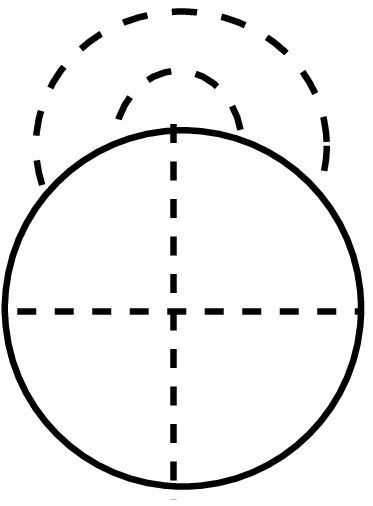} \\
   2  & 2 & 1
\end{tabular}

\begin{tabular}{ccc}
  \graph{0.77}{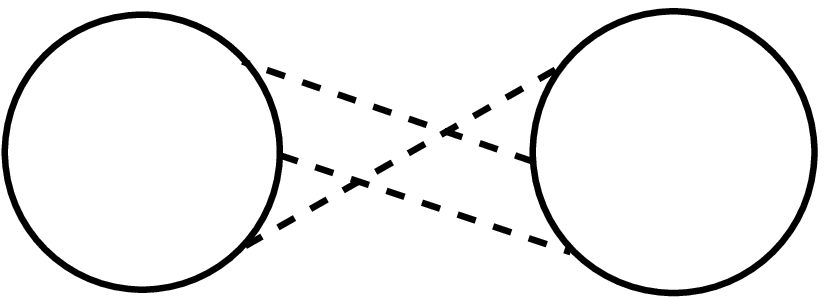} &
  \graph{0.77}{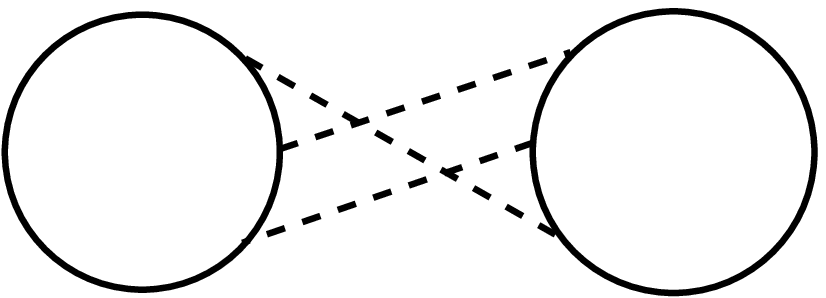} \\
   2  & 2
\end{tabular}

\begin{tabular}{cc}
  \graph{0.77}{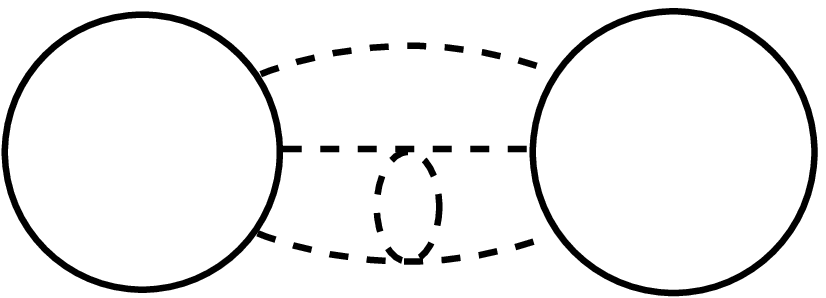} &
  \graph{0.77}{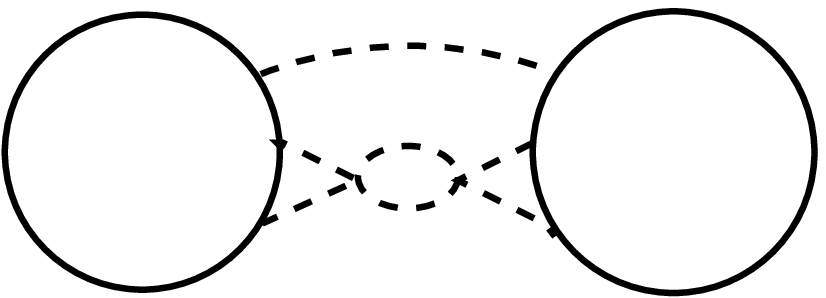} \\
   2 & 2
\end{tabular}

\begin{tabular}{cccc}
  \graph{0.77}{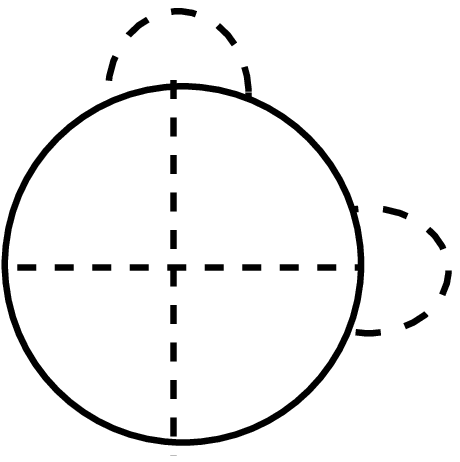} & \graph{0.77}{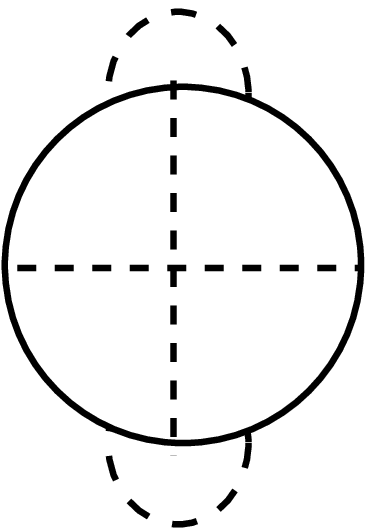} &
  \graph{0.77}{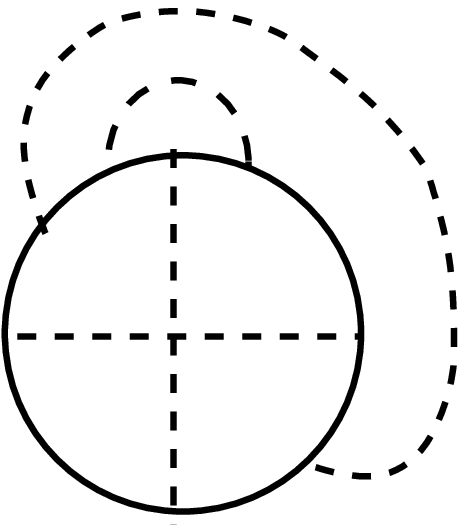} & \graph{0.77}{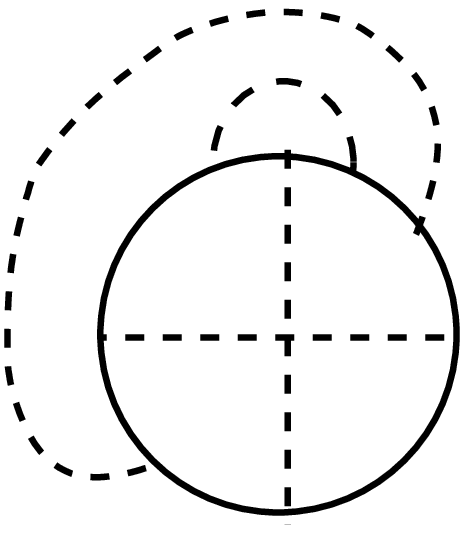} \\
   1 & 2 & 1 & 1
\end{tabular}

\begin{tabular}{cccc}
  \graph{0.77}{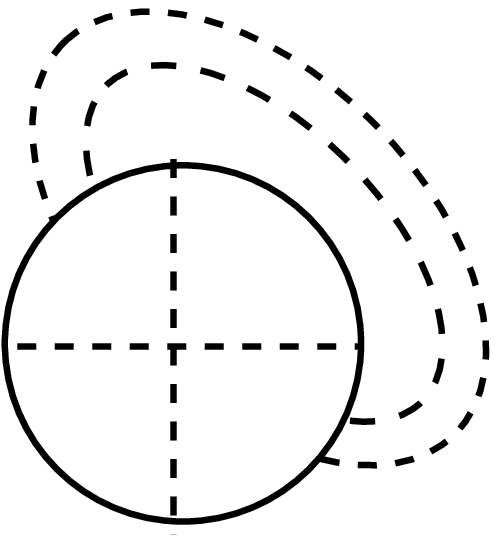} & \graph{0.77}{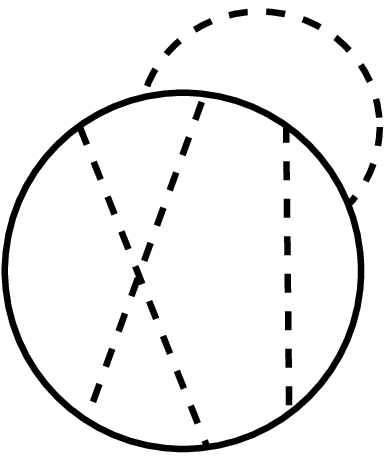} &
  \graph{0.77}{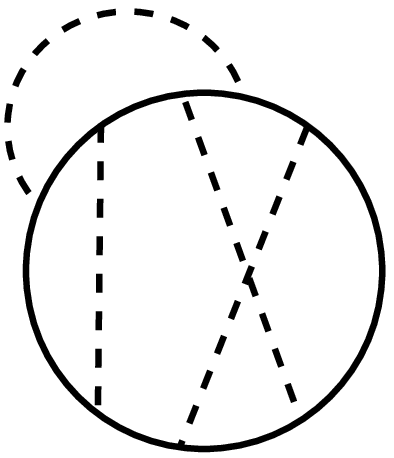} & \graph{0.77}{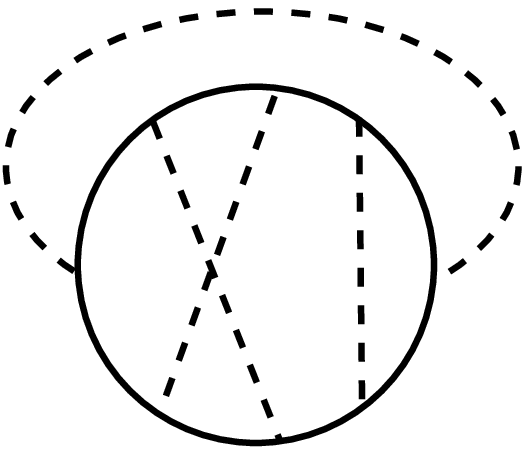} \\
   2 & 1 & 1 & 1
\end{tabular}

\begin{tabular}{cc}
  \graph{0.77}{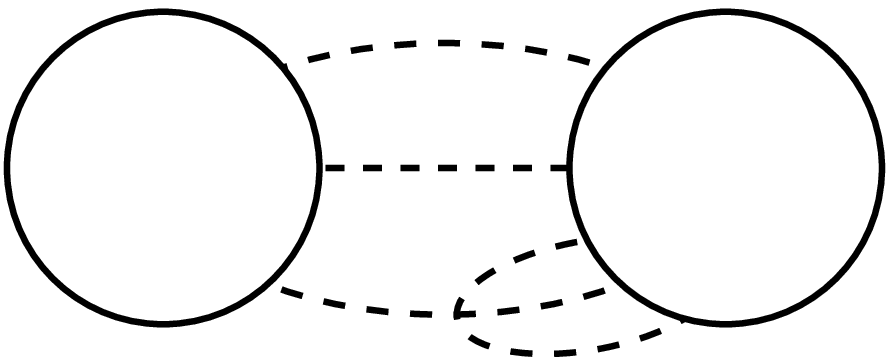} & \graph{0.77}{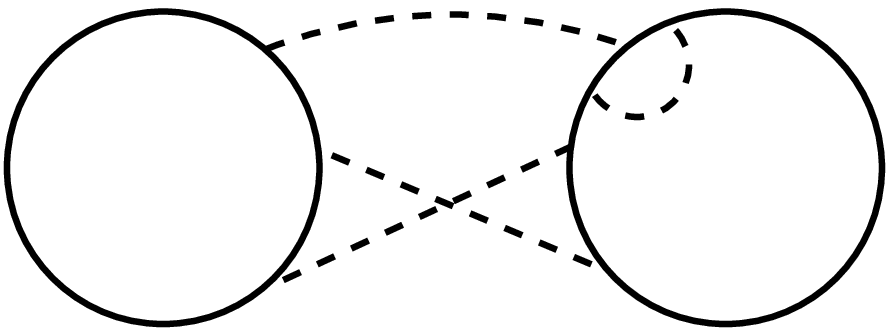} \\
  1 & 1
\end{tabular}

\begin{tabular}{cc}
  \graph{0.77}{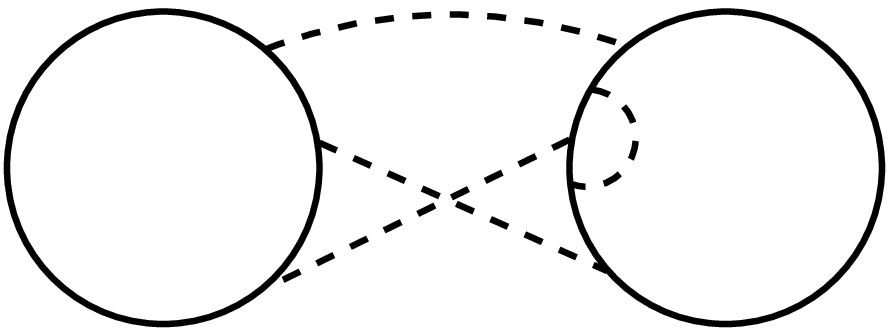} & \graph{0.77}{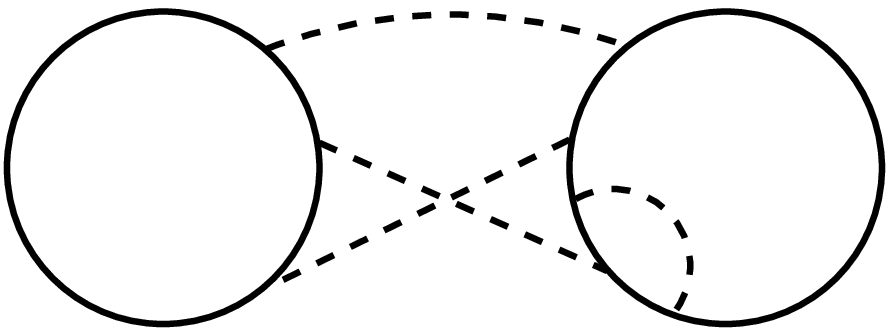} \\
  1 & 1
\end{tabular}

\begin{tabular}{ccc}
  \graph{0.77}{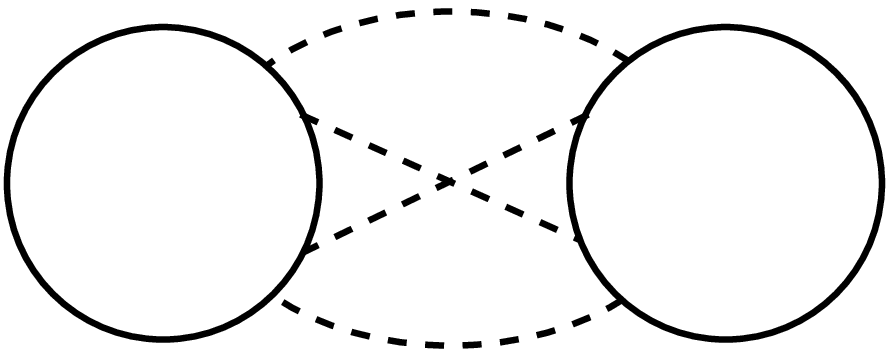} & \graph{0.77}{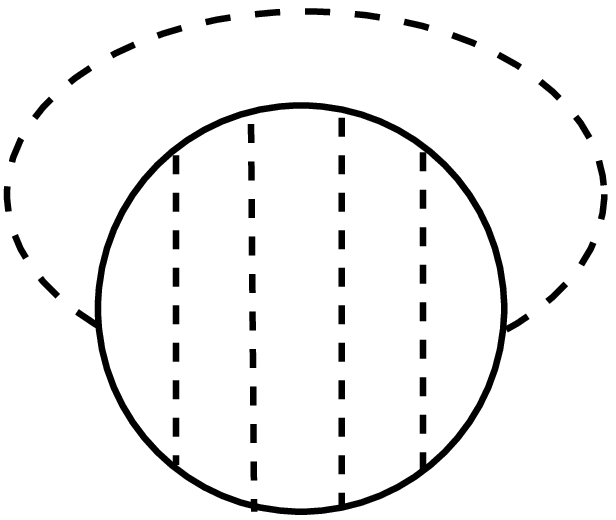} &
  \graph{0.77}{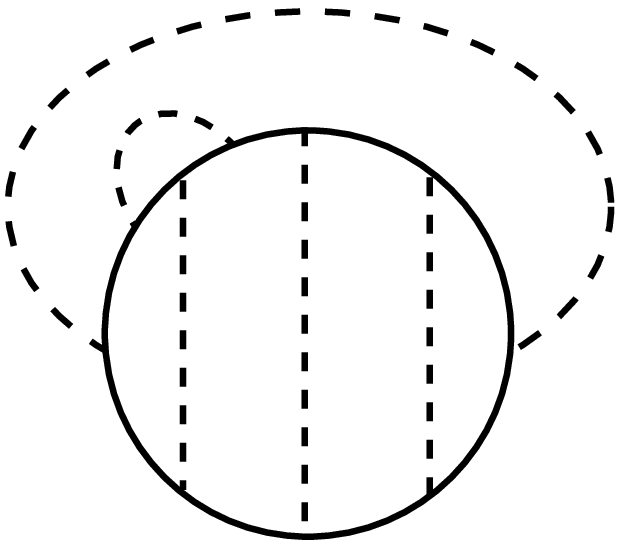} \\
   2 & 2 & 1
\end{tabular}

\begin{tabular}{ccc}
   \graph{0.77}{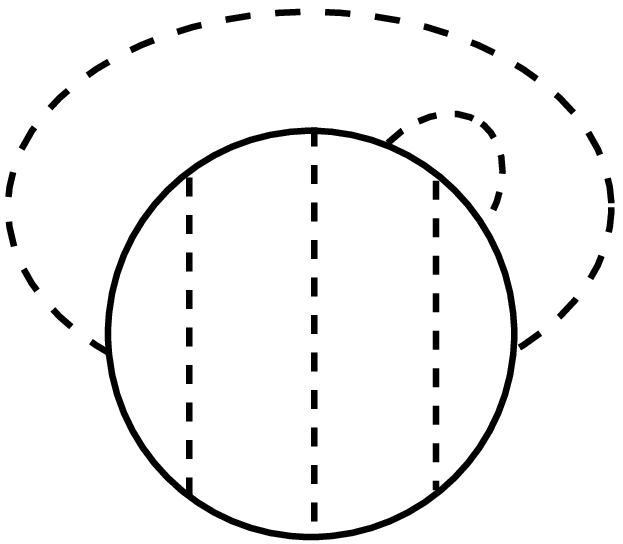} & \graph{0.77}{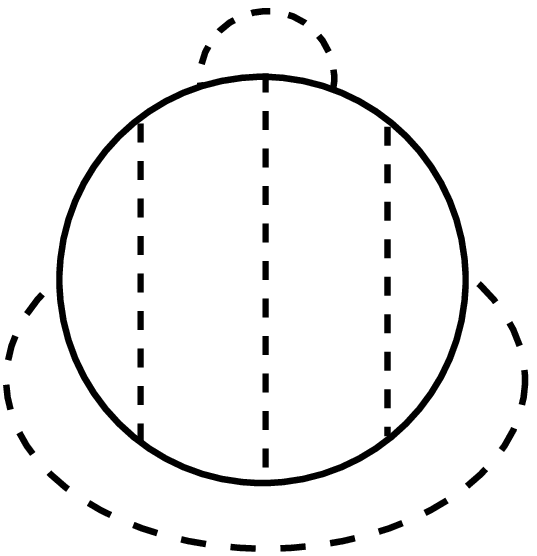} &
   \graph{0.77}{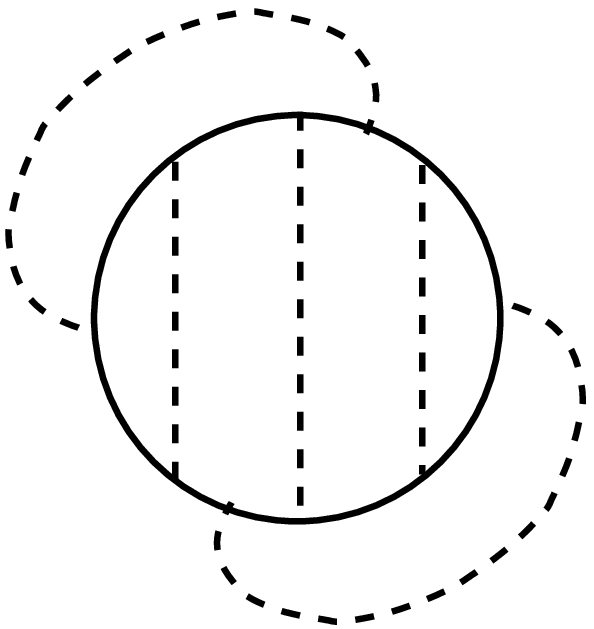}\\
   1  & 1 & 2
\end{tabular}

\begin{tabular}{ccc}
  \graph{0.77}{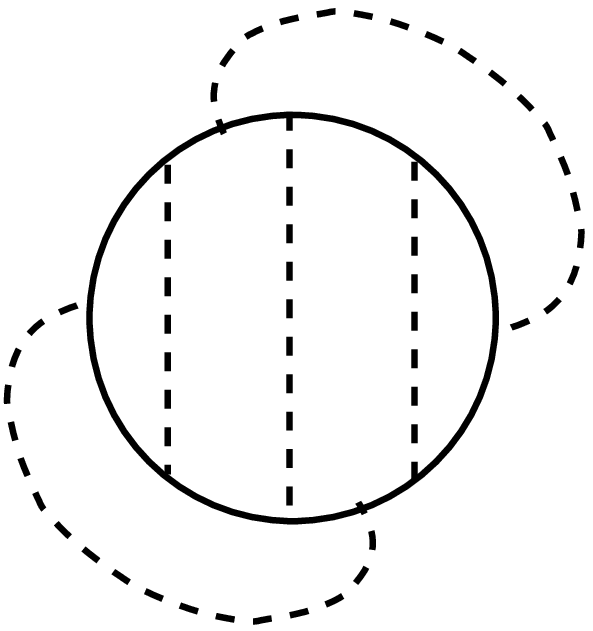} & \graph{0.77}{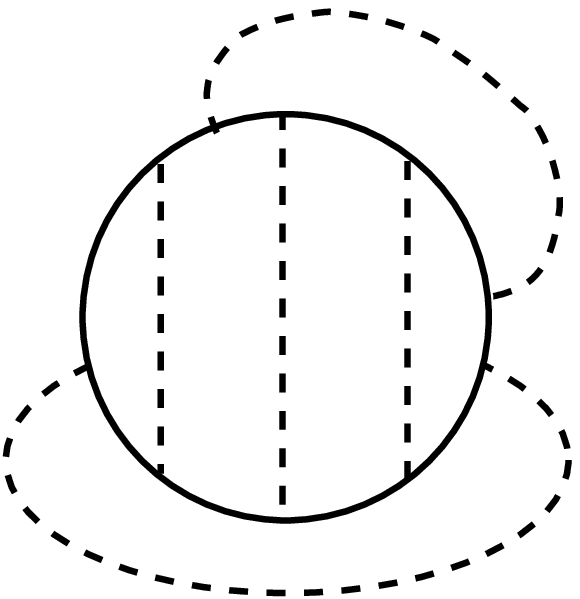} &
  \graph{0.77}{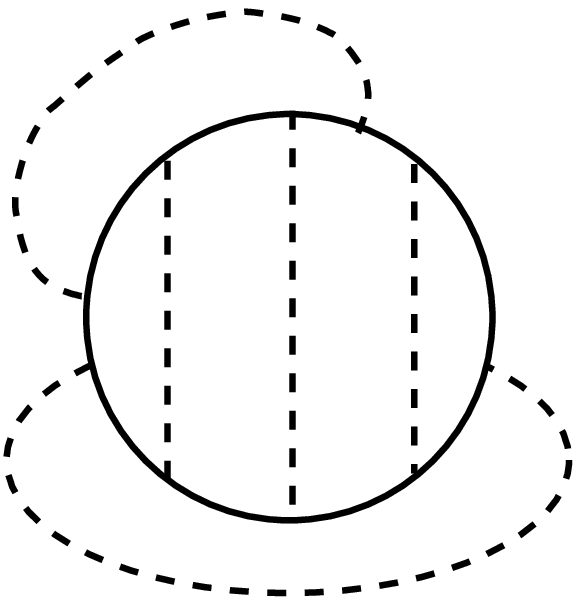}\\
   2 & 1 & 1
\end{tabular}

\begin{tabular}{ccc}
   \graph{0.77}{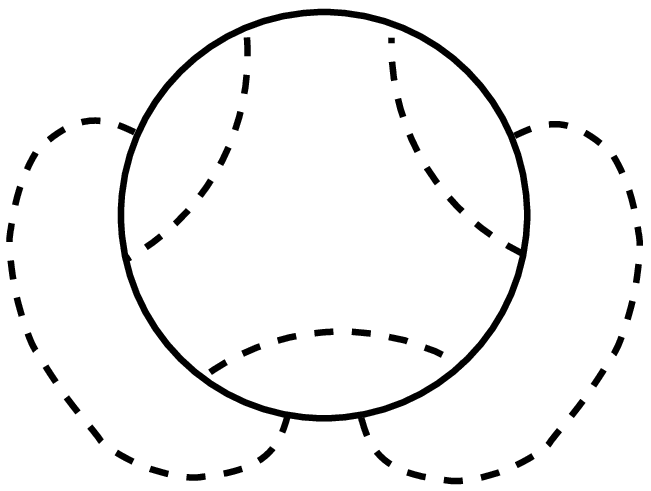} & \graph{0.77}{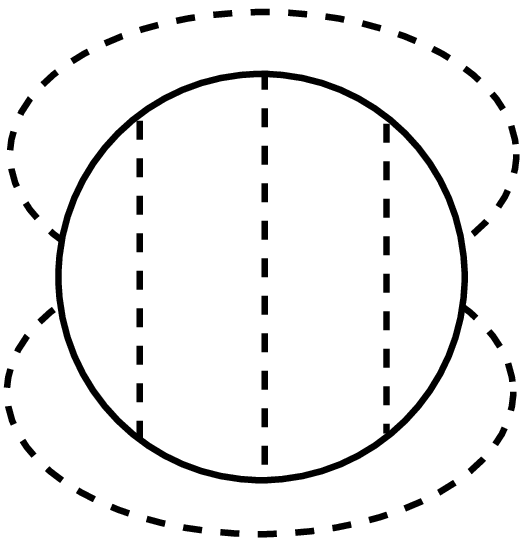} & \graph{0.77}{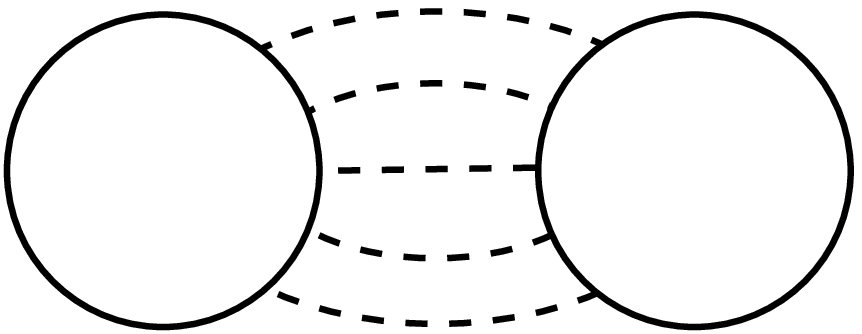} \\
  1 & 2 & 10
\end{tabular}

\begin{tabular}{cc}
  \graph{0.77}{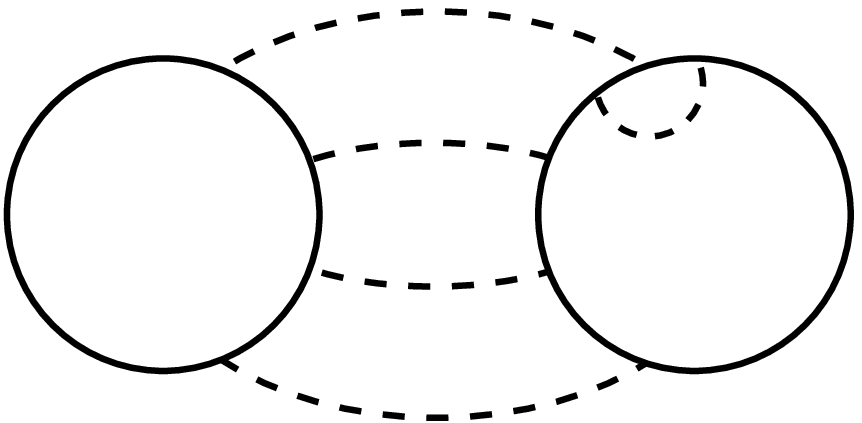} & \graph{0.77}{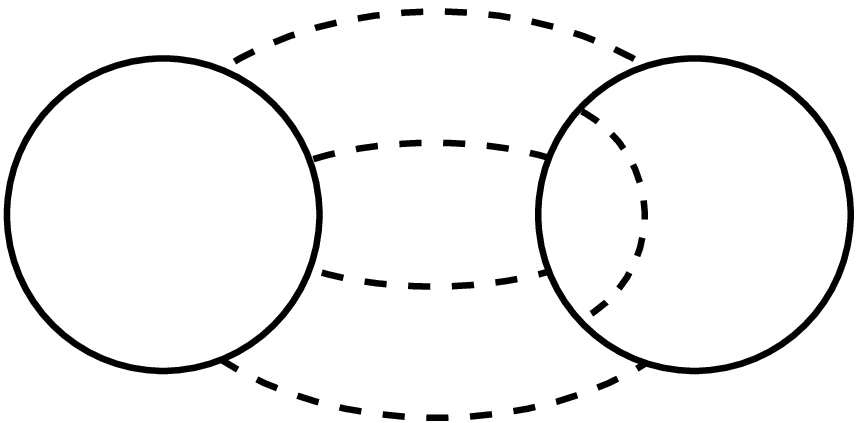} \\
   1 & 2
\end{tabular}

\begin{tabular}{cc}
  \graph{0.77}{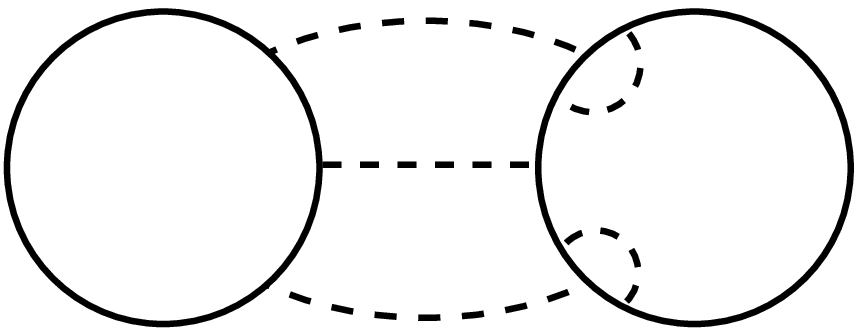} & \graph{0.77}{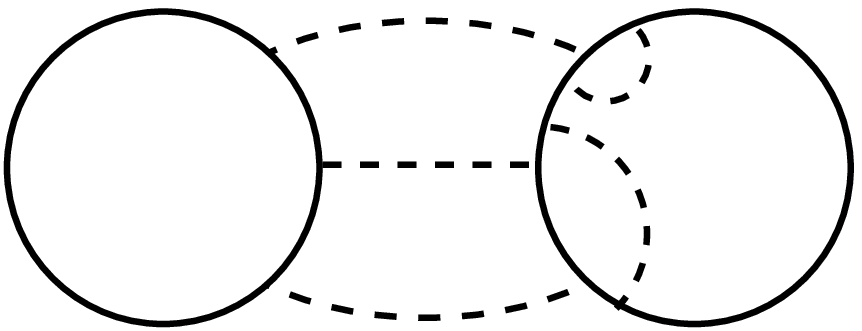} \\
  1 & 1
\end{tabular}

\begin{tabular}{cc}
  \graph{0.77}{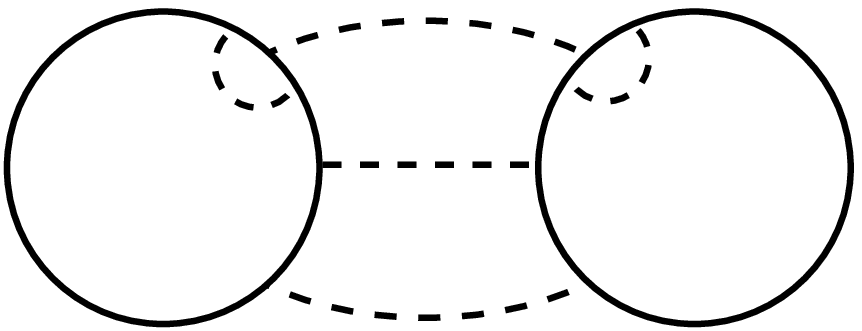} & \graph{0.77}{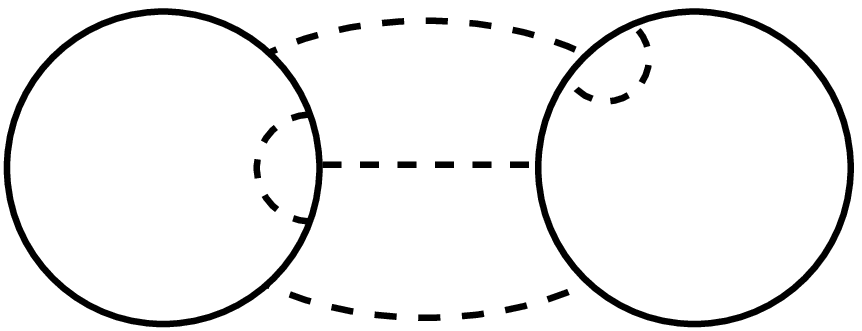} \\
   2 & 2
\end{tabular}

\begin{tabular}{cc}
  \graph{0.77}{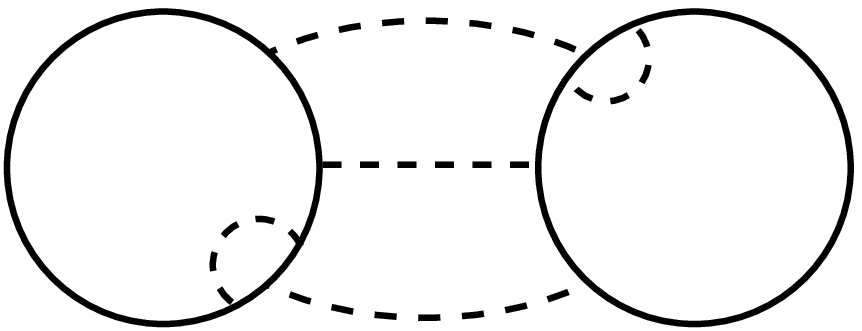} & \graph{0.77}{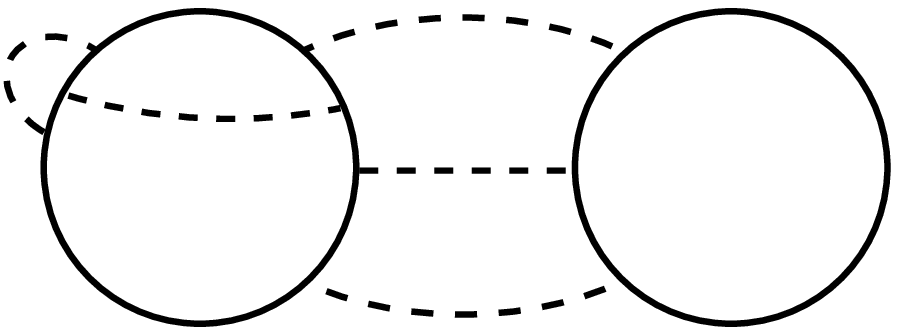}\\
  2 & 1
\end{tabular}

\begin{tabular}{cc}
  \graph{0.77}{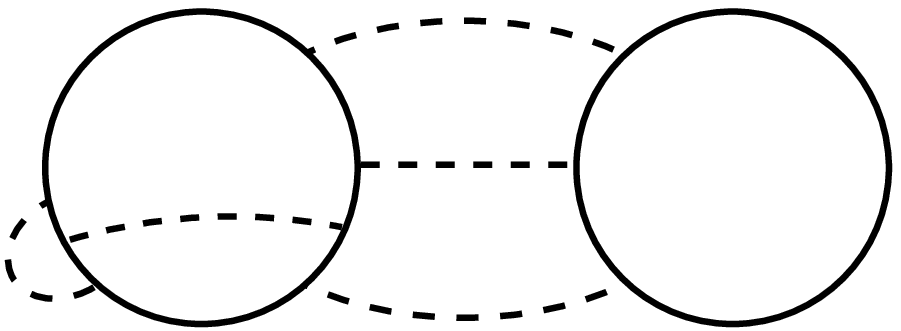} & \graph{0.77}{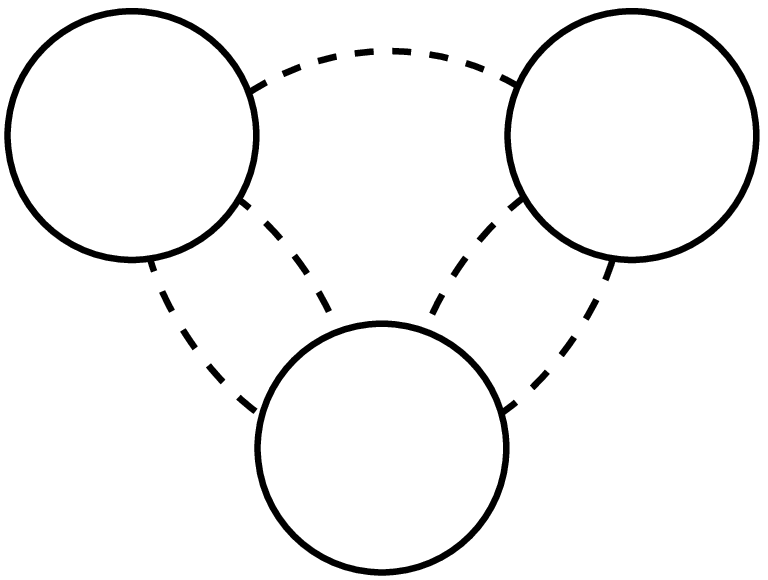} \\
   1 & 2
\end{tabular}

\section{Derivation of SDE for full propagator}
\label{sec:der_SDE}

To begin with, we write the full propagator by the 1PI propagator.
\begin{align}
  \label{full_1PI}
  \raisebox{-8pt}{\scalebox{0.8}{\includegraphics{SDE_1.eps}}} & =\ \
  \raisebox{-2pt}{\scalebox{0.7}{\includegraphics{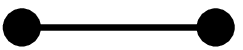}}}\ +\
  \raisebox{-10pt}{\scalebox{0.7}{\includegraphics{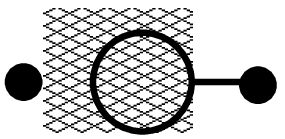}}}\
  +\
  \raisebox{-10pt}{\scalebox{0.7}{\includegraphics{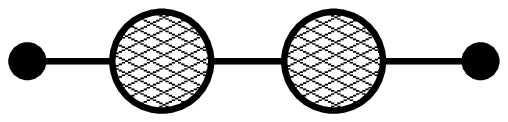}}}\ +\
  \cdots\\
  & = \ \
    \raisebox{-2pt}{\scalebox{0.7}{\includegraphics{bare.eps}}}\ +\
    \raisebox{-10pt}{\scalebox{0.7}{\includegraphics{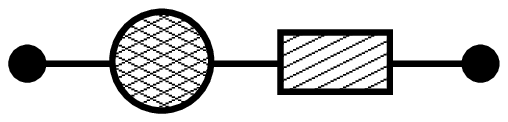}}}.
\end{align}
By multiplying the inverse full propagator from the right and the
inverse bare propagator from the left, we get the equation,
\begin{equation}
  \label{SDE-1}
  \Big(\
   \raisebox{-2pt}{\scalebox{0.7}{\includegraphics{bare.eps}}}\ \Big)^{-1}
  \ = \ \Big(
   \raisebox{-8pt}{\scalebox{0.7}{\includegraphics{SDE_1.eps}}}\Big)^{-1}
  \ + \ \
   \raisebox{-9pt}{\scalebox{0.7}{\includegraphics{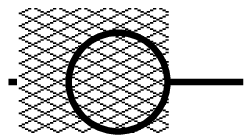}}}.
\end{equation}
After combining eq.(\ref{SDE-1}) with the relation of 1PI propagator
with the full propagator,
\begin{equation}
   \raisebox{-9pt}{\scalebox{0.7}{\includegraphics{1pi.eps}}} \ = \ \
    \raisebox{-0pt}{\scalebox{0.7}{\includegraphics{SDE_2.eps}}}\ + \
    \raisebox{-25pt}{\scalebox{0.7}{\includegraphics{SDE_3.eps}}}\ +
   \ \cdots,
\end{equation}
we are able to take the massless limit (set the inverse bare
propagator to zero).
As a result we get the eq.~(\ref{SDE}).

\section{2PI free energy}
\label{sec:calcG}

Here we present the 2PI free energies for the SO(7) and the SO(4) ansatz.
\begin{align}
\frac{G_{\rm SO(7)}}{N^2} =&  + 3 + 3\,\ln (2)
  - {\displaystyle \frac {1}{2}} \,\ln (
V_1^{7}\,V_2^{3}) + {\displaystyle \frac {1}{2}} 
\,\ln (u^{16}) \nn\\
+ &  g\,(21\,V_1^{2} + 3\,V_2^{2} + 21\,
V_1\,V_2 - 56\,u^{2}\,V_1 + 24
\,u^{2}\,V_2) \nn\\
+ & g^{2}\,( - {\displaystyle \frac {63}{2}} \,V_1
^{4} - {\displaystyle \frac {9}{2}} \,V_2^{4} - 
{\displaystyle \frac {63}{2}} \,V_1^{2}\,V_2^{2}
 - 140\,u^{4}\,V_1^{2} - 168\,u^{4}\,
V_1\,V_2 - 12\,u^{4}\,V_2^{2}) \nn\\
+ & g^{3}(126\,V_1^{6} + 14\,V_2^{6} + 63\,V_1^{4}\,V_2^{2} 
+ 21\,V_2^{4}\,V_1^{2} + 
91\,V_1^{3}\,V_2^{3} - 672\,u^{4}\,
V_1^{4} \nn\\
& - 96\,u^{4}\,V_2^{4} - 1400\,u^{6}\,V_1^{3} 
- 168\,u^{6}\,V_1\,
V_2^{2} - 504\,u^{6}\,V_1^{2}\,V_2 - 232\,u^{6}\,V_2^{3}) \nn\\
+ & g^{4}
( - 672\,V_1^{4}\,V_2^{4} - 168\,V_2^{6}
\,V_1^{2} - 756\,V_1^{6}\,V_2^{2} - 210\,
V_2^{5}\,V_1^{3} - 630\,V_1^{5}\,V_2^{3} - 1134\,V_1^{8} \nn\\
& - 75\,V_2^{8} - 840\,u^{4}\,V_1
^{4}\,V_2^{2} - 168\,u^{4}\,V_2^{4}\,
V_1^{2} - 336\,u^{4}\,V_1^{3}\,V_2^{3} + 5040\,u^{4}\,V_1^{6} \nn\\
& + 528\,u^{4}\,V_2^{6} - 8064\,u^{6}\,V_1^{5} + 2688\,u^{6}\,V_2^{5}
 + 9744\,u^{8}\,V_1\,V_2^{3} + 4032\,
u^{6}\,V_1\,V_2^{4} \nn\\
& - 22372\,u^{8}\,V_1^{4} - 22512\,
u^{8}\,V_1^{3}\,V_2 - 4032\,u
^{6}\,V_1^{3}\,V_2^{2} + 1344\,u^{6}\,
V_2^{3}\,V_1^{2} \nn\\
& + 168\,u^{8}\,V_1^{2}\,V_2^{2}
 - 12096\,u^{6}\,V_2\,V_1^{4} + 924\,
u^{8}\,V_2^{4}) \nn\\
+ & g^{5}(12096\,V_1^{10}\,V_1\,V_2^{4} \nn\\
& + 8316\,V_1^{7}\,V_2^{3} + 
{\displaystyle \frac {70686}{5}} \,V_1^{10} + 9072\,
V_2^{4}\,V_1^{6} - 24672\,u^{6}\,
V_2^{7} - 63504\,u^{4}\,V_1^{8} \nn\\
& + 10206\,V_1^{8}\,V_2^{2} + 
{\displaystyle \frac {2838}{5}} \,V_2^{10} + 
{\displaystyle \frac {35196}{5}} \,V_1^{5}\,V_2^{
5} + 17856\,u^{10}\,V_2^{5} - 392000\,u^{10}\,V_1^{5} \nn\\
& + 22176\,u^{6}\,V_1^{7} - 4080\,
u^{4}\,V_2^{8} + 1764\,V_2^{7}\,
V_1^{3} + 1470\,V_2^{8}\,V_1^{2} + 3696\,
V_1^{4}\,V_2^{6} \nn\\
& + 3072\,u^{8}\,V_2^{6} - 96768\,
u^{8}\,V_1^{6} - 5712\,u^{4}\,
V_1^{5}\,V_2^{3} - 498624\,u^{10}\,
V_1^{4}\,V_2 \nn\\
& - 279552\,u^{10}\,V_1^{3}\,V_2
^{2} - 139776\,u^{10}\,V_1^{2}\,V_2^{3}
 - 22176\,u^{6}\,V_1\,V_2^{6} \nn\\
& + 23184\,u^{6}\,V_1^{5}\,V_2^{2
} - 9744\,u^{6}\,V_2^{5}\,V_1^{2} - 
24528\,u^{6}\,V_2^{4}\,V_1^{3} + 27216
\,u^{6}\,V_1^{4}\,V_2^{3} \nn\\
& + 90720\,u^{6}\,V_2\,V_1^{6} - 
96768\,u^{8}\,V_2^{5}\,V_1 + 2016\,
u^{4}\,V_1^{6}\,V_2^{2} + 11424\,
u^{4}\,V_1^{4}\,V_2^{4} \nn\\
& + 672\,u^{4}\,V_2^{6}\,V_1^{2}
 - 3024\,u^{4}\,V_2^{5}\,V_1^{3} - 
64512\,u^{8}\,V_1^{5}\,V_2 - 134400\,
u^{8}\,V_1^{4}\,V_2^{2} \nn\\
& - 86016\,u^{8}\,V_1^{3}\,V_2^{3} - 59136\,u^{8}\,V_2^{4}\,V_1^{2}),
  \\[1ex]
\frac{G_{\rm SO(4)}}{N^2}   =& - {\displaystyle \frac {1}{2}} \,\ln (
V_1^{4}\,V_2^{6})  + 3 + 7\,\ln (2) + 
{\displaystyle \frac {1}{2}} \,\ln ({\displaystyle \frac {
1}{256}} \,u^{16}) \nn\\
+ & \,g\,(6\,V_1^{2} + 15\,V_2^{2} + 24\,
V_1\,V_2 - 32\,u^{2}\,V_1) \nn\\
+ & \,g^{2}\,( - 9\,V_1^{4} - {\displaystyle \frac {
45}{2}} \,V_2^{4} - 36\,V_1^{2}\,V_2^{2}
 - 32\,u^{4}\,V_1^{2} - 192\,u^{4}\,
V_1\,V_2 + 96\,u^{4}\,V_2^{2}) \nn\\
+ & g^{3}\, (30\,V_1^{6} + 85\,V_2^{6} + 36\,V_1^{4}
\,V_2^{2} + 60\,V_2^{4}\,V_1^{2} + 104\,
V_1^{3}\,V_2^{3} - 192\,u^{4}\,V_1^{4} \nn\\
& - 384\,u^{4}\,V_1^{2}\,V_2^{2}
 + 192\,u^{4}\,V_2^{4} - 128\,u^{6}\,
V_1^{3} - 768\,u^{6}\,V_1^{2}\,V_2 - 1152\,u^{6}\,V_1\,V_2^{2} \nn\\
& - 256\,u^{6}\,V_2^{3})\nn\\
+ & g^{4}(
 - 660\,V_2^{6}\,V_1^{2} - 324\,V_1^{6}\,
V_2^{2} - 600\,V_2^{5}\,V_1^{3} - 360\,
V_1^{5}\,V_2^{3} \nn\\
& - 189\,V_1^{8} - 690\,V_2^{8} - 822\,
V_1^{4}\,V_2^{4} + 1152\,u^{4}\,
V_1^{6} + 768\,u^{4}\,V_1^{4}\,V_2^{2} \nn\\
& + 1536\,u^{4}\,V_1^{3}\,V_2^{3}
 - 768\,u^{4}\,V_2^{6} - 1152\,u^{6}
\,V_1^{5} - 4608\,u^{6}\,V_2\,V_1^{4} \nn\\
& - 3840\,u^{6}\,V_1^{3}\,V_2^{2}
 - 3072\,u^{6}\,V_2^{3}\,V_1^{2} + 384
\,u^{6}\,V_1\,V_2^{4} + 2304\,u^{6}\,V_2^{5} \nn\\
& - 8448\,u^{8}\,V_1^{3}\,V_2 - 
8448\,u^{8}\,V_1^{2}\,V_2^{2} - 16128\,
u^{8}\,V_1\,V_2^{3} - 1792\,u
^{8}\,V_1^{4} \nn\\
& + 3840\,u^{8}\,V_2^{4})\nn\\
+ & g^{5}(
 - 121344\,u^{10}\,V_1\,V_2^{4} + 3456
\,V_1^{7}\,V_2^{3} + {\displaystyle \frac {8496}{
5}} \,V_1^{10} \nn\\
& + 6264\,V_2^{4}\,V_1^{6} - 26880\,
u^{6}\,V_2^{7} - 10368\,u^{4}\,
V_1^{8} + 3240\,V_1^{8}\,V_2^{2} + 7860\,
V_2^{10} \nn\\
& + {\displaystyle \frac {44544}{5}} \,V_1^{5}\,
V_2^{5} - 43008\,u^{10}\,V_2^{5} - 
15872\,u^{10}\,V_1^{5} + 1152\,u^{6}
\,V_1^{7} + 7296\,u^{4}\,V_2^{8} \nn\\
& + 7200\,V_2^{7}\,V_1^{3} + 8160\,V_2^{8}\,V_1^{2} + 9480\,V_1^{4}\,V_2^{6}
 - 84480\,u^{8}\,V_2^{6} + 4608\,u^{8
}\,V_1^{6} \nn\\
& - 12288\,u^{4}\,V_1^{5}\,V_2^{3
} - 122880\,u^{10}\,V_1^{4}\,V_2 - 
245760\,u^{10}\,V_1^{3}\,V_2^{2} \nn\\
& - 153600\,u^{10}\,V_1^{2}\,V_2
^{3} - 7296\,u^{6}\,V_1\,V_2^{6} - 
12288\,u^{6}\,V_2^{5}\,V_1^{2} + 11520
\,u^{6}\,V_2^{4}\,V_1^{3} \nn\\
& + 20736\,u^{6}\,V_1^{4}\,V_2^{3
} + 27648\,u^{6}\,V_2\,V_1^{6} + 102912
\,u^{8}\,V_2^{5}\,V_1 - 9216\,u^{4}\,V_1^{6}\,V_2^{2} \nn\\
& - 12288\,u^{4}\,V_1^{4}\,V_2^{4
} + 3072\,u^{4}\,V_2^{6}\,V_1^{2} - 
7680\,u^{4}\,V_2^{5}\,V_1^{3} - 32256\,
u^{8}\,V_1^{5}\,V_2 \nn\\
& - 113664\,u^{8}\,V_1^{4}\,V_2^{
2} - 92160\,u^{8}\,V_1^{3}\,V_2^{3} - 
49152\,u^{8}\,V_2^{4}\,V_1^{2}).
\end{align}

\section{Improved free energy }
\label{sec:calcF}
\allowdisplaybreaks

We present $F(M^i_0-gM^i_0, m_0-gm_0)$ for the SO(7) and the SO(4) ansatz.
The $F^{\rm improved}_k$ for each ansatz is obtained by neglecting
$O(g^{k+1})$ terms and setting $g=1$ in the following expression.
Here we use variables $P_0^i$ and $q_0$ for notational simplicity,
which are defined by $P_0^1= -7/(2 M_0^1)$, $P_0^2=-3/(2 M_0^2)$ and
$q_0=8/m_0$ for the SO(7) ansatz and $P_0^1= -2/M_0^1$,
$P_0^2=-3/M_0^2$ and $q_0=8/m_0$ for the SO(4) ansatz.
They represent the propagators of the mean field action $S_0$.

\begin{align}
\frac{1}{N^2}F_{\rm SO(7)}&(M^i_0-gM^i_0, m_0-gm_0) = \nonumber \\
-& {\displaystyle \frac {1}{2}} 
\,\ln ((P_0^1)^{7}\,(P_0^2)^{3})
+ 3\,\ln (2) + {\displaystyle \frac {1}{2}} \,
\ln ( q_0^{16}) \nn\\
+ & \,g (24\, q_0^{2}\,(P_0^2) + 3\,(P_0^2)^{2}
 + 21\,(P_0^1)^{2} - 56\, q_0^{2}\,(P_0^1) + 21\,
(P_0^2)\,(P_0^1) + 3) \nn\\
+ & g^2 ( - 
{\displaystyle \frac {483}{2}} \,(P_0^1)^{2}\,(P_0^2)^{2}
 + 42\,(P_0^2)\,(P_0^1) - {\displaystyle \frac {33}{2}} 
\,(P_0^2)^{4} \nn\\
& - 96\, q_0^{2}\,(P_0^2)^{3} + 336\, q_0
^{2}\,(P_0^1)^{2}\,(P_0^2) - 84\,(P_0^2)^{3}\,
(P_0^1) - 168\, q_0^{2}\,(P_0^1) + 42\,(P_0^1)
^{2} + 6\,(P_0^2)^{2} + {\displaystyle \frac {3}{2}}  \nn\\
& - 60\, q_0^{4}\,(P_0^2)^{2} + 196\, q_0
^{4}\,(P_0^1)^{2} - {\displaystyle \frac {567}{2}} \,(P_0^1)^{4} 
- 252\,(P_0^1)^{3}\,(P_0^2) + 72\, q_0^{2}
\,(P_0^2) + 672\, q_0^{2}\,(P_0^1)^{3} \nn\\
& - 336\, q_0^{2}\,(P_0^2)^{2}\,(P_0^1) - 
840\, q_0^{4}\,(P_0^1)\,(P_0^2)) \nn\\
+ & \,g^3 (1 - 18144\, q_0^{2}\,(P_0^1)^{5} + 1680\,
 q_0^{2}\,(P_0^1)^{2}\,(P_0^2) + 6615\,(P_0^1)
^{4}\,(P_0^2)^{2} + 4032\, q_0^{2}\,(P_0^2)^{4}\,
(P_0^1) \nn\\
& - 5040\, q_0^{4}\,(P_0^1)\,(P_0^2) - 12096
\, q_0^{2}\,(P_0^1)^{4}\,(P_0^2) +
11424\,(P_0^1)^{2}\,(P_0^2)^{2}\,q_0^{4}
 + 5040\, q_0^{4}
\,(P_0^2)^{3}\,(P_0^1) \nn\\
& + 15624\, q_0^{6}\,(P_0^1)^{2}\,(P_0^2) - 
1680\, q_0^{2}\,(P_0^2)^{2}\,(P_0^1) + 6384\,
 q_0^{2}\,(P_0^2)^{3}\,(P_0^1)^{2} - 3696\,
 q_0^{2}\,(P_0^1)^{3}\,(P_0^2)^{2} \nn\\
& - 168\, q_0^{6}\,(P_0^1)\,(P_0^2)^{2} + 
1152\, q_0^{4}\,(P_0^2)^{4} + 7728\, q_0^{4}\,
(P_0^1)^{3}\,(P_0^2) + 2541\,(P_0^2)^{4}\,(P_0^1)
^{2} + 3976\, q_0^{6}\,(P_0^1)^{3} \nn\\
& + 3360\, q_0^{2}\,(P_0^1)^{3} + 5747\,(P_0^1)^{3}\,(P_0^2)^{3} 
- 336\, q_0^{2}\,(P_0^1) + 
1176\, q_0^{4}\,(P_0^1)^{2} + 6804\,(P_0^1)^{5}\,
(P_0^2) \nn\\
& + 63\,(P_0^2)\,(P_0^1) - 966\,(P_0^1)^{2}\,
(P_0^2)^{2} + 1056\, q_0^{2}\,(P_0^2)^{5} - 336\,
(P_0^2)^{3}\,(P_0^1) + 144\, q_0^{2}\,(P_0^2)
 - 480\, q_0^{2}\,(P_0^2)^{3} \nn\\
& - 360\, q_0^{4}\,(P_0^2)^{2} - 1256\,q_0^{6}\,(P_0^2)^{3} + 924\,(P_0^1)\,(P_0^2)^{5} - 
1008\,(P_0^1)^{3}\,(P_0^2) + 166\,(P_0^2)^{6} + 6678
\,(P_0^1)^{6} \nn\\
& + 9\,(P_0^2)^{2} - 66\,(P_0^2)^{4} - 1134\,
(P_0^1)^{4} + 63\,(P_0^1)^{2}) \nn\\
+ & g^4 (64176\, q_0^{2}\,(P_0^1)^{4}\,(P_0^2)^{3} - 127008\, q_0^{2}\,(P_0^1)^{5} + 314496\,
 q_0^{2}\,(P_0^1)^{5}\,(P_0^2)^{2} \nn\\
& - 313488\, q_0^{6}\,(P_0^1)^{4}\,(P_0^2)
 + 5040\, q_0^{2}\,(P_0^1)^{2}\,(P_0^2) + 39690\,
(P_0^1)^{4}\,(P_0^2)^{2} + 28224\, q_0^{2}\,
(P_0^2)^{4}\,(P_0^1) \nn\\
& - 17640\, q_0^{4}\,(P_0^1)\,(P_0^2) - 
255528\,(P_0^1)^{4}\,(P_0^2)^{2}\, q_0^{4} - 
182448\, q_0^{6}\,(P_0^1)^{3}\,(P_0^2)^{2} \nn\\
& - 191688\, q_0^{4}\,(P_0^2)^{4}\,(P_0^1)^{
2} - 84672\, q_0^{2}\,(P_0^1)^{4}\,(P_0^2) + 91392
\,(P_0^1)^{2}\,(P_0^2)^{2}\, q_0^{4} + 40320\,
 q_0^{4}\,(P_0^2)^{3}\,(P_0^1) \nn\\
& - 194544\, q_0^{2}\,(P_0^1)^{3}\,(P_0^2)^{
4} + 140616\, q_0^{6}\,(P_0^1)^{2}\,(P_0^2) + 
108360\, q_0^{8}\,(P_0^1)^{2}\,(P_0^2)^{2} - 96642
\,(P_0^2)^{5}\,(P_0^1)^{3} \nn\\
& - 272160\,(P_0^2)\,(P_0^1)^{5}\, q_0^{4}
 - 5040\, q_0^{2}\,(P_0^2)^{2}\,(P_0^1) + 544320\,
 q_0^{2}\,(P_0^1)^{6}\,(P_0^2) - 100212\,q_0^{8}\,(P_0^1)^{4} \nn\\
& + 9744\, q_0^{6}\,(P_0^1)\,(P_0^2)^{4} + 
44688\, q_0^{2}\,(P_0^2)^{3}\,(P_0^1)^{2} - 25872
\, q_0^{2}\,(P_0^1)^{3}\,(P_0^2)^{2} - 1512\,
 q_0^{6}\,(P_0^1)\,(P_0^2)^{2} \nn\\
& - 13104\, q_0^{8}\,(P_0^1)^{3}\,(P_0^2) + 
169680\, q_0^{8}\,(P_0^1)\,(P_0^2)^{3} + 9216\,
 q_0^{4}\,(P_0^2)^{4} - 73920\, q_0^{2}\,
(P_0^2)^{6}\,(P_0^1) \nn\\
& + 61824\, q_0^{4}\,(P_0^1)^{3}\,(P_0^2) - 
101472\, q_0^{4}\,(P_0^2)^{5}\,(P_0^1) - 261072\,
 q_0^{4}\,(P_0^1)^{3}\,(P_0^2)^{3} - 231840\,
 q_0^{6}\,(P_0^1)^{5} \nn\\
& + 14220\, q_0^{8}\,(P_0^2)^{4} - 109200\,
 q_0^{6}\,(P_0^2)^{3}\,(P_0^1)^{2} - 147840\,
 q_0^{2}\,(P_0^2)^{5}\,(P_0^1)^{2} + 15246\,
(P_0^2)^{4}\,(P_0^1)^{2} \nn\\
& + 35784\, q_0^{6}\,(P_0^1)^{3} + 10080\,
 q_0^{2}\,(P_0^1)^{3} + 34482\,(P_0^1)^{3}\,
(P_0^2)^{3} - 560\, q_0^{2}\,(P_0^1) - 26448\,
 q_0^{4}\,(P_0^2)^{6} \nn\\
& + 4116\, q_0^{4}\,(P_0^1)^{2} - 187110\,
(P_0^1)^{4}\,(P_0^2)^{4} + 40824\,(P_0^1)^{5}\,
(P_0^2) + 84\,(P_0^2)\,(P_0^1) + 13920\, q_0^{
6}\,(P_0^2)^{5} \nn\\
& - 2415\,(P_0^1)^{2}\,(P_0^2)^{2} + 7392\,q_0^{2}\,(P_0^2)^{5} - 840\,(P_0^2)^{3}\,(P_0^1) + 
240\, q_0^{2}\,(P_0^2) - 1440\, q_0^{2}\,
(P_0^2)^{3} \nn\\
& - 1260\, q_0^{4}\,(P_0^2)^{2} - 42672\,
(P_0^2)^{6}\,(P_0^1)^{2} - 11304\, q_0^{6}\,
(P_0^2)^{3} + 5544\,(P_0^1)\,(P_0^2)^{5} - 178416\,
(P_0^1)^{6}\, q_0^{4} \nn\\
& - 236502\,(P_0^1)^{5}\,(P_0^2)^{3} - 240408\,
(P_0^1)^{7}\,(P_0^2) + 641088\, q_0^{2}\,(P_0^1)^{7} - 247212\,(P_0^1)^{6}\,(P_0^2)^{2} \nn\\
& - 13944\,(P_0^2)^{7}\,(P_0^1) - 2520\,(P_0^1)
^{3}\,(P_0^2) - 15936\,(P_0^2)^{7}\, q_0^{2} + 996
\,(P_0^2)^{6} + 40068\,(P_0^1)^{6} + 12\,(P_0^2)^{2}
 \nn\\
& - 165\,(P_0^2)^{4} - 2835\,(P_0^1)^{4} + 84\,
(P_0^1)^{2} - 202986\,(P_0^1)^{8} - 2199\,(P_0^2)^{8}
 + {\displaystyle \frac {3}{4}} ) \nn\\
+ & g^5(577584\, q_0^{2}\,(P_0^1)^{4}\,
(P_0^2)^{3} - 508032\, q_0^{2}\,(P_0^1)^{5} + 
2830464\, q_0^{2}\,(P_0^1)^{5}\,(P_0^2)^{2} \nn\\
& - 11009376\,(P_0^1)^{6}\,(P_0^2)^{3}\, q_0
^{2} - 3448368\, q_0^{6}\,(P_0^1)^{4}\,(P_0^2) + 
11760\, q_0^{2}\,(P_0^1)^{2}\,(P_0^2) \nn\\
& + 1878912\, q_0^{6}\,(P_0^2)^{5}\,(P_0^1)
^{2} + 138915\,(P_0^1)^{4}\,(P_0^2)^{2} + 3873408\,
(P_0^1)^{2}\,(P_0^2)^{7}\, q_0^{2} \nn\\
& + 112896\, q_0^{2}\,(P_0^2)^{4}\,(P_0^1)
 - 780864\, q_0^{8}\,(P_0^1)^{3}\,(P_0^2)^{3} + 
8285760\, q_0^{4}\,(P_0^1)^{4}\,(P_0^2)^{4} \nn\\
& - 47040\, q_0^{4}\,(P_0^1)\,(P_0^2) - 
3321696\, q_0^{8}\,(P_0^2)^{5}\,(P_0^1) - 4235616
\,(P_0^2)\, q_0^{8}\,(P_0^1)^{5} \nn\\
& + 11690784\,(P_0^1)^{6}\, q_0^{4}\,(P_0^2)
^{2} - 2555280\,(P_0^1)^{4}\,(P_0^2)^{2}\, q_0^{4}
 - 2006928\, q_0^{6}\,(P_0^1)^{3}\,(P_0^2)^{2} \nn\\
& + 1561728\,(P_0^2)^{8}\, q_0^{2}\,(P_0^1)
 + 9767520\,(P_0^2)\, q_0^{6}\,(P_0^1)^{6} - 
19885824\,(P_0^1)^{7}\,(P_0^2)^{2}\, q_0^{2} \nn\\
& - 4906944\, q_0^{8}\,(P_0^1)^{4}\,(P_0^2)
^{2} - 1916880\, q_0^{4}\,(P_0^2)^{4}\,(P_0^1)^{2}
 - 338688\, q_0^{2}\,(P_0^1)^{4}\,(P_0^2) \nn\\
& + 411264\,(P_0^1)^{2}\,(P_0^2)^{2}\, q_0^{
4} - 6339648\, q_0^{8}\,(P_0^1)^{2}\,(P_0^2)^{4}
 + 181440\, q_0^{4}\,(P_0^2)^{3}\,(P_0^1) \nn\\
& - 622944\,(P_0^1)^{5}\,(P_0^2)^{4}\, q_0^{
2} + 5608512\, q_0^{6}\,(P_0^2)^{2}\,(P_0^1)^{5}
 - 26925696\,(P_0^2)\,(P_0^1)^{8}\, q_0^{2} \nn\\
& - 1750896\, q_0^{2}\,(P_0^1)^{3}\,(P_0^2)
^{4} + 703080\, q_0^{6}\,(P_0^1)^{2}\,(P_0^2) + 
16039296\,(P_0^2)\, q_0^{4}\,(P_0^1)^{7} \nn\\
& + 1300320\, q_0^{8}\,(P_0^1)^{2}\,(P_0^2)
^{2} - 773136\,(P_0^2)^{5}\,(P_0^1)^{3} - 2721600\,
(P_0^2)\,(P_0^1)^{5}\, q_0^{4} \nn\\
& - 11760\, q_0^{2}\,(P_0^2)^{2}\,(P_0^1) - 
570528\, q_0^{10}\,(P_0^1)\,(P_0^2)^{4} - 8170848
\,(P_0^2)\, q_0^{10}\,(P_0^1)^{4} \nn\\
& + 4898880\, q_0^{2}\,(P_0^1)^{6}\,(P_0^2)
 + 2779392\, q_0^{4}\,(P_0^2)^{7}\,(P_0^1) - 
1202544\, q_0^{8}\,(P_0^1)^{4} \nn\\
& + 5450592\, q_0^{4}\,(P_0^2)^{6}\,(P_0^1)
^{2} + 107184\, q_0^{6}\,(P_0^1)\,(P_0^2)^{4} + 
178752\, q_0^{2}\,(P_0^2)^{3}\,(P_0^1)^{2} \nn\\
& + 8013600\, q_0^{4}\,(P_0^1)^{3}\,(P_0^2)
^{5} + 9508128\, q_0^{4}\,(P_0^1)^{5}\,(P_0^2)^{3}
 - 103488\, q_0^{2}\,(P_0^1)^{3}\,(P_0^2)^{2} \nn\\
& - 7560\, q_0^{6}\,(P_0^1)\,(P_0^2)^{2} - 
157248\, q_0^{8}\,(P_0^1)^{3}\,(P_0^2) - 1912512\,
 q_0^{10}\,(P_0^1)^{3}\,(P_0^2)^{2} \nn\\
& + 2036160\, q_0^{8}\,(P_0^1)\,(P_0^2)^{3}
 + 5197920\, q_0^{6}\,(P_0^1)^{3}\,(P_0^2)^{4} + 
41472\, q_0^{4}\,(P_0^2)^{4} \nn\\
& + 4110624\, q_0^{6}\,(P_0^1)^{4}\,(P_0^2)
^{3} - 665280\, q_0^{2}\,(P_0^2)^{6}\,(P_0^1) + 
278208\, q_0^{4}\,(P_0^1)^{3}\,(P_0^2) \nn\\
& + 364896\, q_0^{6}\,(P_0^2)^{6}\,(P_0^1)
 - 1014720\, q_0^{4}\,(P_0^2)^{5}\,(P_0^1) - 
2610720\, q_0^{4}\,(P_0^1)^{3}\,(P_0^2)^{3} \nn\\
& - 2550240\, q_0^{6}\,(P_0^1)^{5} - 9748032\,
 q_0^{10}\,(P_0^1)^{2}\,(P_0^2)^{3} + 170640\,
 q_0^{8}\,(P_0^2)^{4} \nn\\
& - 1201200\, q_0^{6}\,(P_0^2)^{3}\,(P_0^1)
^{2} + 7336224\,(P_0^1)^{4}\,(P_0^2)^{5}\, q_0^{2}
 - 1330560\, q_0^{2}\,(P_0^2)^{5}\,(P_0^1)^{2} \nn\\
& + 53361\,(P_0^2)^{4}\,(P_0^1)^{2} + 6365856\,
(P_0^1)^{3}\,(P_0^2)^{6}\, q_0^{2} + 178920\,
 q_0^{6}\,(P_0^1)^{3} + 23520\, q_0^{2}\,
(P_0^1)^{3} \nn\\
& + 120687\,(P_0^1)^{3}\,(P_0^2)^{3} - 840\,
 q_0^{2}\,(P_0^1) - 264480\, q_0^{4}\,(P_0^2)^{6} + 10976\, q_0^{4}\,(P_0^1)^{2} \nn\\
& - 1496880\,(P_0^1)^{4}\,(P_0^2)^{4} + 142884\,
(P_0^1)^{5}\,(P_0^2) + 105\,(P_0^2)\,(P_0^1) + 
153120\, q_0^{6}\,(P_0^2)^{5} \nn\\
& - 4830\,(P_0^1)^{2}\,(P_0^2)^{2} + 281472\,
(P_0^2)^{9}\, q_0^{2} + 246288\,(P_0^2)^{9}\,
(P_0^1) + 29568\, q_0^{2}\,(P_0^2)^{5} - 1680\,
(P_0^2)^{3}\,(P_0^1) \nn\\
& + 360\, q_0^{2}\,(P_0^2) - 3360\, q_0^{
2}\,(P_0^2)^{3} - 3360\, q_0^{4}\,(P_0^2)^{2} - 
341376\,(P_0^2)^{6}\,(P_0^1)^{2} - 56520\, q_0^{6}
\,(P_0^2)^{3} \nn\\
& + 19404\,(P_0^1)\,(P_0^2)^{5} - 1784160\,(P_0^1)^{6}\, q_0^{4} - 1892016\,(P_0^1)^{5}\,(P_0^2)
^{3} - 1923264\,(P_0^1)^{7}\,(P_0^2) \nn\\
& + 5769792\, q_0^{2}\,(P_0^1)^{7} - 1977696\,
(P_0^1)^{6}\,(P_0^2)^{2} - 25982208\,(P_0^1)^{9}\,
 q_0^{2} - {\displaystyle \frac {476448}{5}} \, q_0
^{10}\,(P_0^1)^{5} \nn\\
& + 14270256\, q_0^{4}\,(P_0^1)^{8} + 3342528\,
 q_0^{8}\,(P_0^1)^{6} + 10644480\, q_0^{6}\,
(P_0^1)^{7} + 9743328\,(P_0^2)\,(P_0^1)^{9} \nn\\
& - 111552\,(P_0^2)^{7}\,(P_0^1) - 5040\,(P_0^1)^{3}\,(P_0^2) - 143424\,(P_0^2)^{7}\, q_0^{2} + 
4224276\,(P_0^1)^{4}\,(P_0^2)^{6} \nn\\
& - 18432\, q_0^{6}\,(P_0^2)^{7} + 
{\displaystyle \frac {1239072}{5}} \, q_0^{10}\,(P_0^2)^{5} + 10877328\,(P_0^1)^{7}\,(P_0^2)^{3} + 627792\,
 q_0^{4}\,(P_0^2)^{8} \nn\\
& + 10724238\,(P_0^1)^{8}\,(P_0^2)^{2} + 
{\displaystyle \frac {37165968}{5}} \,(P_0^2)^{5}\,(P_0^1)^{5} + 9821196\,(P_0^1)^{6}\,(P_0^2)^{4} + 2083536\,
(P_0^1)^{3}\,(P_0^2)^{7} \nn\\
& - 604608\, q_0^{8}\,(P_0^2)^{6} + 853230\,
(P_0^2)^{8}\,(P_0^1)^{2} + 3486\,(P_0^2)^{6} + 140238
\,(P_0^1)^{6} + 15\,(P_0^2)^{2} - 330\,(P_0^2)^{4}
 \nn\\
& - 5670\,(P_0^1)^{4} + 105\,(P_0^1)^{2} - 1623888
\,(P_0^1)^{8} - 17592\,(P_0^2)^{8} + {\displaystyle 
\frac {170934}{5}} \,(P_0^2)^{10} + {\displaystyle \frac {3}{
5}}  \nn\\
& + {\displaystyle \frac {35778078}{5}} \,(P_0^1)^{10}) ,
       \\[1ex]
\frac{1}{N^2} F_{\rm SO(4)}&(M^i_0-gM^i_0, m_0-gm_0) = \nonumber \\
& 3
\,\ln (2)
  + {\displaystyle \frac {1}{2}} \,\ln ( q_0
^{16}) - {\displaystyle \frac {1}{2}} \,\ln ((P_0^1)^{
4}\,(P_0^2)^{6}) \nn\\
+ & g (3 - 32\, q_0^{2}\,(P_0^1) + 6\,(P_0^1)
^{2} + 15\,(P_0^2)^{2} + 24\,(P_0^2)\,(P_0^1)) \nn\\
+ & g^2 (12\,
(P_0^1)^{2} \nn\\
& - 240\,(P_0^2)^{3}\,(P_0^1) - 45\,(P_0^1)^{4}
 - 32\, q_0^{4}\,(P_0^1)^{2} - {\displaystyle \frac {
345}{2}} \,(P_0^2)^{4} + 384\, q_0^{2}\,(P_0^1)^{2
}\,(P_0^2) + 96\, q_0^{4}\,(P_0^2)^{2} \nn\\
& - 192\, q_0^{4}\,(P_0^1)\,(P_0^2) + 192\,
 q_0^{2}\,(P_0^1)^{3} + {\displaystyle \frac {3}{2}} 
 - 276\,(P_0^1)^{2}\,(P_0^2)^{2} - 144\,(P_0^1)^{3}\,
(P_0^2) - 96\, q_0^{2}\,(P_0^1) \nn\\
& + 48\,(P_0^2)\,(P_0^1) + 30\,(P_0^2)^{2}) \nn\\
+ & g^3 (1 - 2880\, q_0^{2}\,(P_0^1)^{5} + 1920\,
 q_0^{2}\,(P_0^1)^{2}\,(P_0^2) + 4212\,(P_0^1)
^{4}\,(P_0^2)^{2} - 1152\, q_0^{4}\,(P_0^1)\,
(P_0^2) \nn\\
& - 6912\, q_0^{2}\,(P_0^1)^{4}\,(P_0^2) + 
3456\,(P_0^1)^{2}\,(P_0^2)^{2}\, q_0^{4} + 384\,
 q_0^{4}\,(P_0^2)^{3}\,(P_0^1) + 2304\, q_0
^{6}\,(P_0^1)^{2}\,(P_0^2) \nn\\
& - 3840\, q_0^{2}\,(P_0^2)^{3}\,(P_0^1)^{2}
 - 8832\, q_0^{2}\,(P_0^1)^{3}\,(P_0^2)^{2} - 4224
\, q_0^{6}\,(P_0^1)\,(P_0^2)^{2} - 1728\,q_0^{4}\,(P_0^2)^{4} \nn\\
& + 1920\, q_0^{4}\,(P_0^1)^{3}\,(P_0^2) + 
6540\,(P_0^2)^{4}\,(P_0^1)^{2} + 1920\, q_0^{6}\,
(P_0^1)^{3} + 960\, q_0^{2}\,(P_0^1)^{3} + 6952\,
(P_0^1)^{3}\,(P_0^2)^{3} \nn\\
& - 192\, q_0^{2}\,(P_0^1) - 192\, q_0^{4
}\,(P_0^1)^{2} + 2160\,(P_0^1)^{5}\,(P_0^2) + 72\,
(P_0^2)\,(P_0^1) - 1104\,(P_0^1)^{2}\,(P_0^2)^{2}
 \nn\\
& - 960\,(P_0^2)^{3}\,(P_0^1) + 576\, q_0^{4
}\,(P_0^2)^{2} - 256\, q_0^{6}\,(P_0^2)^{3} + 5520
\,(P_0^1)\,(P_0^2)^{5} - 576\,(P_0^1)^{3}\,(P_0^2) + 3485\,(P_0^2)^{6} \nn\\
& + 606\,(P_0^1)^{6} + 45\,(P_0^2)^{2} - 690\,
(P_0^2)^{4} - 180\,(P_0^1)^{4} + 18\,(P_0^1)^{2} + 
1728\, q_0^{4}\,(P_0^1)^{4}) \nn\\
+ & g^4 (
333696\, q_0^{2}\,(P_0^1)^{4}\,(P_0^2)^{3} - 20160
\, q_0^{2}\,(P_0^1)^{5} + 269568\, q_0^{2}\,
(P_0^1)^{5}\,(P_0^2)^{2} - 82944\, q_0^{6}\,
(P_0^1)^{4}\,(P_0^2) \nn\\
& + 5760\, q_0^{2}\,(P_0^1)^{2}\,(P_0^2) + 
25272\,(P_0^1)^{4}\,(P_0^2)^{2} - 4032\, q_0^{4}\,
(P_0^1)\,(P_0^2) - 110592\,(P_0^1)^{4}\,(P_0^2)^{
2}\, q_0^{4} \nn\\
& - 27648\, q_0^{6}\,(P_0^1)^{3}\,(P_0^2)^{2
} - 43008\, q_0^{4}\,(P_0^2)^{4}\,(P_0^1)^{2} - 
48384\, q_0^{2}\,(P_0^1)^{4}\,(P_0^2) + 27648\,
(P_0^1)^{2}\,(P_0^2)^{2}\, q_0^{4} \nn\\
& + 3072\, q_0^{4}\,(P_0^2)^{3}\,(P_0^1) + 
209280\, q_0^{2}\,(P_0^1)^{3}\,(P_0^2)^{4} + 20736
\, q_0^{6}\,(P_0^1)^{2}\,(P_0^2) + 80640\,q_0^{8}\,(P_0^1)^{2}\,(P_0^2)^{2} \nn\\
& - 246600\,(P_0^2)^{5}\,(P_0^1)^{3} - 100224\,
(P_0^2)\,(P_0^1)^{5}\, q_0^{4} + 172800\,q_0^{2}\,(P_0^1)^{6}\,(P_0^2) + 2304\, q_0^{8}\,
(P_0^1)^{4} \nn\\
& + 84864\, q_0^{6}\,(P_0^1)\,(P_0^2)^{4} - 
26880\, q_0^{2}\,(P_0^2)^{3}\,(P_0^1)^{2} - 61824
\, q_0^{2}\,(P_0^1)^{3}\,(P_0^2)^{2} - 38016\,
 q_0^{6}\,(P_0^1)\,(P_0^2)^{2} \nn\\
& + 34560\, q_0^{8}\,(P_0^1)^{3}\,(P_0^2) - 
28416\, q_0^{8}\,(P_0^1)\,(P_0^2)^{3} - 13824\,
 q_0^{4}\,(P_0^2)^{4} + 15360\, q_0^{4}\,
(P_0^1)^{3}\,(P_0^2) \nn\\
& + 11136\, q_0^{4}\,(P_0^2)^{5}\,(P_0^1) - 
99840\, q_0^{4}\,(P_0^1)^{3}\,(P_0^2)^{3} - 35712
\, q_0^{6}\,(P_0^1)^{5} + 6912\, q_0^{8}\,
(P_0^2)^{4} \nn\\
& + 67584\, q_0^{6}\,(P_0^2)^{3}\,(P_0^1)^{2
} + 88320\, q_0^{2}\,(P_0^2)^{5}\,(P_0^1)^{2} + 
39240\,(P_0^2)^{4}\,(P_0^1)^{2} + 17280\, q_0^{6}
\,(P_0^1)^{3} \nn\\
& + 2880\, q_0^{2}\,(P_0^1)^{3} + 41712\,
(P_0^1)^{3}\,(P_0^2)^{3} - 320\, q_0^{2}\,(P_0^1) + 45312\, q_0^{4}\,(P_0^2)^{6} - 672\, q_0
^{4}\,(P_0^1)^{2} \nn\\
& - 238374\,(P_0^1)^{4}\,(P_0^2)^{4} + 12960\,
(P_0^1)^{5}\,(P_0^2) + 96\,(P_0^2)\,(P_0^1) + 
9984\, q_0^{6}\,(P_0^2)^{5} - 2760\,(P_0^1)^{2}\,
(P_0^2)^{2} \nn\\
& - 2400\,(P_0^2)^{3}\,(P_0^1) + 2016\, q_0
^{4}\,(P_0^2)^{2} - 214260\,(P_0^2)^{6}\,(P_0^1)^{2}
 - 2304\, q_0^{6}\,(P_0^2)^{3} + 33120\,(P_0^1)\,
(P_0^2)^{5} \nn\\
& - 61056\,(P_0^1)^{6}\, q_0^{4} - 163224\,
(P_0^1)^{5}\,(P_0^2)^{3} - 43632\,(P_0^1)^{7}\,
(P_0^2) + 58176\, q_0^{2}\,(P_0^1)^{7}  + 60\,(P_0^2)^{2} \nn\\
& - 94500\,(P_0^1)^{6}\,(P_0^2)^{2} - 167280\,
(P_0^2)^{7}\,(P_0^1) - 1440\,(P_0^1)^{3}\,(P_0^2)
 + 20910\,(P_0^2)^{6} + 3636\,(P_0^1)^{6} \nn\\
& - 1725\,(P_0^2)^{4} - 450\,(P_0^1)^{4} + 24\,
(P_0^1)^{2} - 10665\,(P_0^1)^{8} - 91140\,(P_0^2)^{8}
 + {\displaystyle \frac {3}{4}}  + 13824\, q_0^{4}\,
(P_0^1)^{4}) \nn\\
+ & g^5 (3003264\, q_0^{2}\,(P_0^1)^{4}\,
(P_0^2)^{3} - 80640\, q_0^{2}\,(P_0^1)^{5} + 
2426112\, q_0^{2}\,(P_0^1)^{5}\,(P_0^2)^{2} \nn\\
& - 13057920\,(P_0^1)^{6}\,(P_0^2)^{3}\, q_0
^{2} - 912384\, q_0^{6}\,(P_0^1)^{4}\,(P_0^2) + 
13440\, q_0^{2}\,(P_0^1)^{2}\,(P_0^2) \nn\\
& - 2666496\, q_0^{6}\,(P_0^2)^{5}\,(P_0^1)
^{2} + 88452\,(P_0^1)^{4}\,(P_0^2)^{2} - 2676480\,
(P_0^1)^{2}\,(P_0^2)^{7}\, q_0^{2} \nn\\
& - 2239488\, q_0^{8}\,(P_0^1)^{3}\,(P_0^2)
^{3} + 4918272\, q_0^{4}\,(P_0^1)^{4}\,(P_0^2)^{4}
 - 10752\, q_0^{4}\,(P_0^1)\,(P_0^2) \nn\\
& + 476160\, q_0^{8}\,(P_0^2)^{5}\,(P_0^1)
 - 1244160\,(P_0^2)\, q_0^{8}\,(P_0^1)^{5} + 
5985792\,(P_0^1)^{6}\, q_0^{4}\,(P_0^2)^{2} \nn\\
& - 1105920\,(P_0^1)^{4}\,(P_0^2)^{2}\, q_0
^{4} - 304128\, q_0^{6}\,(P_0^1)^{3}\,(P_0^2)^{2}
 + 2059776\,(P_0^2)\, q_0^{6}\,(P_0^1)^{6} \nn\\
& - 9072000\,(P_0^1)^{7}\,(P_0^2)^{2}\, q_0
^{2} - 2276352\, q_0^{8}\,(P_0^1)^{4}\,(P_0^2)^{2}
 - 430080\, q_0^{4}\,(P_0^2)^{4}\,(P_0^1)^{2} \nn\\
& - 193536\, q_0^{2}\,(P_0^1)^{4}\,(P_0^2)
 + 124416\,(P_0^1)^{2}\,(P_0^2)^{2}\, q_0^{4} - 
602112\, q_0^{8}\,(P_0^1)^{2}\,(P_0^2)^{4} \nn\\
& + 13824\, q_0^{4}\,(P_0^2)^{3}\,(P_0^1) - 
15255936\,(P_0^1)^{5}\,(P_0^2)^{4}\, q_0^{2} + 
2366208\, q_0^{6}\,(P_0^2)^{2}\,(P_0^1)^{5} \nn\\
& - 4886784\,(P_0^2)\,(P_0^1)^{8}\, q_0^{2}
 + 1883520\, q_0^{2}\,(P_0^1)^{3}\,(P_0^2)^{4} + 
103680\, q_0^{6}\,(P_0^1)^{2}\,(P_0^2) \nn\\
& + 4745088\,(P_0^2)\, q_0^{4}\,(P_0^1)^{7}
 + 967680\, q_0^{8}\,(P_0^1)^{2}\,(P_0^2)^{2} - 
1972800\,(P_0^2)^{5}\,(P_0^1)^{3} \nn\\
& - 1002240\,(P_0^2)\,(P_0^1)^{5}\, q_0^{4}
 - 723456\, q_0^{10}\,(P_0^1)\,(P_0^2)^{4} - 
958464\,(P_0^2)\, q_0^{10}\,(P_0^1)^{4} \nn\\
& + 1555200\, q_0^{2}\,(P_0^1)^{6}\,(P_0^2)
 - 836736\, q_0^{4}\,(P_0^2)^{7}\,(P_0^1) + 27648
\, q_0^{8}\,(P_0^1)^{4} + 663552\, q_0^{4}\,
(P_0^2)^{6}\,(P_0^1)^{2} \nn\\
& + 933504\, q_0^{6}\,(P_0^1)\,(P_0^2)^{4}
 - 107520\, q_0^{2}\,(P_0^2)^{3}\,(P_0^1)^{2} + 
2390016\, q_0^{4}\,(P_0^1)^{3}\,(P_0^2)^{5} \nn\\
& + 6532608\, q_0^{4}\,(P_0^1)^{5}\,(P_0^2)
^{3} - 247296\, q_0^{2}\,(P_0^1)^{3}\,(P_0^2)^{2}
 - 190080\, q_0^{6}\,(P_0^1)\,(P_0^2)^{2} \nn\\
& + 414720\, q_0^{8}\,(P_0^1)^{3}\,(P_0^2)
 + 233472\, q_0^{10}\,(P_0^1)^{3}\,(P_0^2)^{2} - 
340992\, q_0^{8}\,(P_0^1)\,(P_0^2)^{3} \nn\\
& - 1826304\, q_0^{6}\,(P_0^1)^{3}\,(P_0^2)
^{4} - 62208\, q_0^{4}\,(P_0^2)^{4} + 567552\,q_0^{6}\,(P_0^1)^{4}\,(P_0^2)^{3} + 69120\, q_0^{4
}\,(P_0^1)^{3}\,(P_0^2) \nn\\
& - 2517120\, q_0^{6}\,(P_0^2)^{6}\,(P_0^1)
 + 111360\, q_0^{4}\,(P_0^2)^{5}\,(P_0^1) - 998400
\, q_0^{4}\,(P_0^1)^{3}\,(P_0^2)^{3} \nn\\
& - 392832\, q_0^{6}\,(P_0^1)^{5} + 1812480\,
 q_0^{10}\,(P_0^1)^{2}\,(P_0^2)^{3} + 82944\,
 q_0^{8}\,(P_0^2)^{4} + 743424\, q_0^{6}\,
(P_0^2)^{3}\,(P_0^1)^{2} \nn\\
& - 11836800\,(P_0^1)^{4}\,(P_0^2)^{5}\, q_0
^{2} + 794880\, q_0^{2}\,(P_0^2)^{5}\,(P_0^1)^{2}
 + 137340\,(P_0^2)^{4}\,(P_0^1)^{2} \nn\\
& - 6856320\,(P_0^1)^{3}\,(P_0^2)^{6}\, q_0
^{2} + 86400\, q_0^{6}\,(P_0^1)^{3} + 6720\,q_0^{2}\,(P_0^1)^{3} + 145992\,(P_0^1)^{3}\,(P_0^2)^{3}
 \nn\\
& - 480\, q_0^{2}\,(P_0^1) + 453120\, q_0
^{4}\,(P_0^2)^{6} - 1792\, q_0^{4}\,(P_0^1)^{2} - 
1906992\,(P_0^1)^{4}\,(P_0^2)^{4} + 45360\,(P_0^1)^{5
}\,(P_0^2) \nn\\
& + 120\,(P_0^2)\,(P_0^1) + 109824\, q_0^{6}
\,(P_0^2)^{5} - 5520\,(P_0^1)^{2}\,(P_0^2)^{2} + 
5832960\,(P_0^2)^{9}\,(P_0^1) - 4800\,(P_0^2)^{3}\,
(P_0^1) \nn\\
& + 5376\, q_0^{4}\,(P_0^2)^{2} - 1714080\,
(P_0^2)^{6}\,(P_0^1)^{2} - 11520\, q_0^{6}\,
(P_0^2)^{3} + 115920\,(P_0^1)\,(P_0^2)^{5} \nn\\
& - 610560\,(P_0^1)^{6}\, q_0^{4} - 1305792\,
(P_0^1)^{5}\,(P_0^2)^{3} - 349056\,(P_0^1)^{7}\,
(P_0^2) + 523584\, q_0^{2}\,(P_0^1)^{7} \nn\\
& - 756000\,(P_0^1)^{6}\,(P_0^2)^{2} - 1365120\,
(P_0^1)^{9}\, q_0^{2} - {\displaystyle \frac {2274816
}{5}} \, q_0^{10}\,(P_0^1)^{5} + 2039040\, q_0
^{4}\,(P_0^1)^{8} \nn\\
& - 622080\, q_0^{8}\,(P_0^1)^{6} + 374400\,
 q_0^{6}\,(P_0^1)^{7} + 1023840\,(P_0^2)\,(P_0^1)^{9} - 1338240\,(P_0^2)^{7}\,(P_0^1) \nn\\
& - 2880\,(P_0^1)^{3}\,(P_0^2) + 10791720\,(P_0^1)^{4}\,(P_0^2)^{6} - 395520\, q_0^{6}\,(P_0^2)^{
7} - 67584\, q_0^{10}\,(P_0^2)^{5} \nn\\
& + 4617216\,(P_0^1)^{7}\,(P_0^2)^{3} - 1434624\,
 q_0^{4}\,(P_0^2)^{8} + 2490696\,(P_0^1)^{8}\,
(P_0^2)^{2} + {\displaystyle \frac {49990752}{5}} \,(P_0^2)^{5}\,(P_0^1)^{5} \nn\\
& + 7313976\,(P_0^1)^{6}\,(P_0^2)^{4} + 9979200\,
(P_0^1)^{3}\,(P_0^2)^{7} - 311808\, q_0^{8}\,
(P_0^2)^{6} + 8160000\,(P_0^2)^{8}\,(P_0^1)^{2} \nn\\
& + 73185\,(P_0^2)^{6} + 12726\,(P_0^1)^{6} + 75\,
(P_0^2)^{2} - 3450\,(P_0^2)^{4} - 900\,(P_0^1)^{4} + 
30\,(P_0^1)^{2} - 85320\,(P_0^1)^{8} \nn\\
& - 729120\,(P_0^2)^{8} + 2769360\,(P_0^2)^{10} + 
{\displaystyle \frac {3}{5}}  + {\displaystyle \frac {1096488}{5}
} \,(P_0^1)^{10} + 62208\, q_0^{4}\,(P_0^1)^{4}) .
\end{align}

\section{Numerical values of extrema}
\label{sec:Position_of_plateau}

The positions of the extrema having the lowest free energy are
shown in Table \ref{tab:Position_of_plateau}.
Here $P_0^1$, $P_0^2$ and $q_0$ are related to $M_0^1$, $M_0^2$ and
$m_0$ as in appendix \ref{sec:calcF}.
Since we consider lower orders here, we can not expect clear plateaus.
However at lower orders we expect that
the extremum with the lowest
free energy lies in the plateau and approximates the
free energy well.
We can check it is indeed the case for the $\phi^4$ toy model.
\begin{table}[htbp]
  \begin{center}
    \leavevmode
    \begin{tabular}{c|c||l|l|l} \hline
      ansatz & parameters &
         \multicolumn{1}{c|}{1st order} &
         \multicolumn{1}{c|}{3rd order} &
         \multicolumn{1}{c}{5th order} \\
       \hline \hline
       SO(7) & $P_0^1$ &\hspace{3pt} 0.4105507905  &\hspace{3pt}
         0.4518139994  &\hspace{3pt} 0.4400228898 \\ \cline{2-5}
            & $P_0^2$ &\hspace{3pt} 0.1073836480 &\hspace{3pt}
         0.07040135462  &\hspace{3pt} 0.06351041755  \\ \cline{2-5}
            & $q_0$ &\hspace{3pt} 0.4426595159 & \hspace{3pt}
         0.5126612791  &\hspace{3pt}  0.5082289370 \\ \hline \hline
      SO(4) & $P_0^1$ &\hspace{3pt} 0.5625802884  &\hspace{3pt}
         0.6844310859 &\hspace{3pt} 0.6695144044\\ \cline{2-5}
            & $P_0^2$ &\hspace{3pt} 0.1630908715 &\hspace{3pt}
         0.08577036668 &\hspace{3pt} 0.02122284256\\ \cline{2-5}
            & $q_0$ &\hspace{3pt} 0.4713708814  &\hspace{3pt}
         0.5390149536 &\hspace{3pt} 0.5174809839 \\ \hline
    \end{tabular}
    \caption{The positions of the extrema for each order.}
    \label{tab:Position_of_plateau}
  \end{center}
\end{table}


\end{document}

%% file: b_pro.latex
\setlength{\unitlength}{3947sp}%
\begingroup\makeatletter\ifx\SetFigFont\undefined%
\gdef\SetFigFont#1#2#3#4#5{%
  \reset@font\fontsize{#1}{#2pt}%
  \fontfamily{#3}\fontseries{#4}\fontshape{#5}%
  \selectfont}%
\fi\endgroup%
\begin{picture}(1560,703)(946,-1394)
\put(1696,-1336){\makebox(0,0)[lb]{\smash{\SetFigFont{12}{14.4}{\familydefault}{\mddefault}{\updefault}{\color[rgb]{0,0,0}$j$}%
}}}
\thicklines
{\color[rgb]{0,0,0}\multiput(1201,-1036)(218.18182,0.00000){6}{\line( 1, 0){109.091}}
}%
\put(1711,-856){\makebox(0,0)[lb]{\smash{\SetFigFont{12}{14.4}{\familydefault}{\mddefault}{\updefault}{\color[rgb]{0,0,0}$i$}%
}}}
\put(2506,-1081){\makebox(0,0)[lb]{\smash{\SetFigFont{12}{14.4}{\familydefault}{\mddefault}{\updefault}{\color[rgb]{0,0,0}$\nu$}%
}}}
\put(946,-1081){\makebox(0,0)[lb]{\smash{\SetFigFont{12}{14.4}{\familydefault}{\mddefault}{\updefault}{\color[rgb]{0,0,0}$\mu$}%
}}}
\end{picture}

%% file: f_pro.latex
\setlength{\unitlength}{3947sp}%
\begingroup\makeatletter\ifx\SetFigFont\undefined%
\gdef\SetFigFont#1#2#3#4#5{%
  \reset@font\fontsize{#1}{#2pt}%
  \fontfamily{#3}\fontseries{#4}\fontshape{#5}%
  \selectfont}%
\fi\endgroup%
\begin{picture}(1560,703)(946,-1394)
\put(1696,-1336){\makebox(0,0)[lb]{\smash{\SetFigFont{12}{14.4}{\familydefault}{\mddefault}{\updefault}{\color[rgb]{0,0,0}$j$}%
}}}
\thicklines
{\color[rgb]{0,0,0}\put(1201,-1036){\line( 1, 0){1200}}
}%
\put(1711,-856){\makebox(0,0)[lb]{\smash{\SetFigFont{12}{14.4}{\familydefault}{\mddefault}{\updefault}{\color[rgb]{0,0,0}$i$}%
}}}
\put(2506,-1081){\makebox(0,0)[lb]{\smash{\SetFigFont{12}{14.4}{\familydefault}{\mddefault}{\updefault}{\color[rgb]{0,0,0}$\beta$}%
}}}
\put(946,-1081){\makebox(0,0)[lb]{\smash{\SetFigFont{12}{14.4}{\familydefault}{\mddefault}{\updefault}{\color[rgb]{0,0,0}$\alpha$}%
}}}
\end{picture}

%% file: b_vert.latex
\setlength{\unitlength}{3947sp}%
\begingroup\makeatletter\ifx\SetFigFont\undefined%
\gdef\SetFigFont#1#2#3#4#5{%
  \reset@font\fontsize{#1}{#2pt}%
  \fontfamily{#3}\fontseries{#4}\fontshape{#5}%
  \selectfont}%
\fi\endgroup%
\begin{picture}(1560,1620)(946,-1801)
\thicklines
{\color[rgb]{0,0,0}\multiput(1816,-1591)(0.00000,218.18182){6}{\line( 0, 1){109.091}}
}%
\put(946,-1081){\makebox(0,0)[lb]{\smash{\SetFigFont{12}{14.4}{\familydefault}{\mddefault}{\updefault}{\color[rgb]{0,0,0}$\mu$}%
}}}
\put(2506,-1081){\makebox(0,0)[lb]{\smash{\SetFigFont{12}{14.4}{\familydefault}{\mddefault}{\updefault}{\color[rgb]{0,0,0}$\rho$}%
}}}
\put(1771,-1801){\makebox(0,0)[lb]{\smash{\SetFigFont{12}{14.4}{\familydefault}{\mddefault}{\updefault}{\color[rgb]{0,0,0}$\lambda$}%
}}}
{\color[rgb]{0,0,0}\multiput(1201,-1036)(218.18182,0.00000){6}{\line( 1, 0){109.091}}
}%
\put(1456,-826){\makebox(0,0)[lb]{\smash{\SetFigFont{12}{14.4}{\familydefault}{\mddefault}{\updefault}{\color[rgb]{0,0,0}$i$}%
}}}
\put(2071,-1366){\makebox(0,0)[lb]{\smash{\SetFigFont{12}{14.4}{\familydefault}{\mddefault}{\updefault}{\color[rgb]{0,0,0}$k$}%
}}}
\put(1471,-1351){\makebox(0,0)[lb]{\smash{\SetFigFont{12}{14.4}{\familydefault}{\mddefault}{\updefault}{\color[rgb]{0,0,0}$j$}%
}}}
\put(1771,-346){\makebox(0,0)[lb]{\smash{\SetFigFont{12}{14.4}{\familydefault}{\mddefault}{\updefault}{\color[rgb]{0,0,0}$\nu$}%
}}}
\put(2101,-826){\makebox(0,0)[lb]{\smash{\SetFigFont{12}{14.4}{\familydefault}{\mddefault}{\updefault}{\color[rgb]{0,0,0}$l$}%
}}}
\end{picture}

%% file: b-f_vert.latex
\setlength{\unitlength}{3947sp}%
\begingroup\makeatletter\ifx\SetFigFont\undefined%
\gdef\SetFigFont#1#2#3#4#5{%
  \reset@font\fontsize{#1}{#2pt}%
  \fontfamily{#3}\fontseries{#4}\fontshape{#5}%
  \selectfont}%
\fi\endgroup%
\begin{picture}(1560,1273)(946,-1394)
\thicklines
{\color[rgb]{0,0,0}\put(1201,-1036){\line( 1, 0){1200}}
}%
{\color[rgb]{0,0,0}\multiput(1771,-1066)(0.00000,276.00000){3}{\line( 0, 1){138.000}}
}%
\put(1351,-856){\makebox(0,0)[lb]{\smash{\SetFigFont{12}{14.4}{\familydefault}{\mddefault}{\updefault}{\color[rgb]{0,0,0}$i$}%
}}}
\put(946,-1081){\makebox(0,0)[lb]{\smash{\SetFigFont{12}{14.4}{\familydefault}{\mddefault}{\updefault}{\color[rgb]{0,0,0}$\alpha$}%
}}}
\put(2131,-841){\makebox(0,0)[lb]{\smash{\SetFigFont{12}{14.4}{\familydefault}{\mddefault}{\updefault}{\color[rgb]{0,0,0}$k$}%
}}}
\put(2506,-1081){\makebox(0,0)[lb]{\smash{\SetFigFont{12}{14.4}{\familydefault}{\mddefault}{\updefault}{\color[rgb]{0,0,0}$\beta$}%
}}}
\put(1711,-286){\makebox(0,0)[lb]{\smash{\SetFigFont{12}{14.4}{\familydefault}{\mddefault}{\updefault}{\color[rgb]{0,0,0}$\mu$}%
}}}
\put(1726,-1336){\makebox(0,0)[lb]{\smash{\SetFigFont{12}{14.4}{\familydefault}{\mddefault}{\updefault}{\color[rgb]{0,0,0}$j$}%
}}}
{\color[rgb]{0,0,0}\put(1771,-1051){\circle*{90}}
}%
\end{picture}